\documentclass[twocolumn]{aastex62} 

\received{2019}
\revised{2019}
\accepted{2019}

\submitjournal{ApJ}
\shorttitle{Space densities and emissivities of AGNs at $z>4$}
\shortauthors{Giallongo et al.}

\begin{document} 

\title{THE SPACE DENSITIES AND EMISSIVITIES OF AGNs AT $z> 4$}

\author{E. Giallongo}
\affiliation{INAF-Osservatorio Astronomico di Roma, via Frascati 33, 00078, Monteporzio, Italy}
\author{A. Grazian}
\affiliation{INAF-Osservatorio Astronomico di Roma, via Frascati 33, 00078, Monteporzio, Italy}
\author{F. Fiore}
\affiliation{INAF-Osservatorio Astronomico di Trieste, via G.B. Tiepolo 11, 34131, Trieste, Italy}
\author{D. Kodra}
\affiliation{Department of physics and Astronomy and PITT PACC, University of Pittsburgh, Pittsburgh, PA 15260, USA}
\author{T. Urrutia}
\affiliation{Leibniz Institut f\"{u}r Astrophysik, An der Sternwarte 16, D-14482 Potsdam, Germany}
\author{M. Castellano}
\affiliation{INAF-Osservatorio Astronomico di Roma, via Frascati 33, 00078, Monteporzio, Italy}
\author{S. Cristiani}
\affiliation{INAF-Osservatorio Astronomico di Trieste, via G.B. Tiepolo 11, 34131, Trieste, Italy}
\author{M. Dickinson}
\affiliation{National Optical Astronomy Observatory, Tucson, AZ 85719, USA}
\author{A. Fontana}
\affiliation{INAF-Osservatorio Astronomico di Roma, via Frascati 33, 00078, Monteporzio, Italy}
\author{N. Menci}
\affiliation{INAF-Osservatorio Astronomico di Roma, via Frascati 33, 00078, Monteporzio, Italy}
\author{L. Pentericci}
\affiliation{INAF-Osservatorio Astronomico di Roma, via Frascati 33, 00078, Monteporzio, Italy}
\author{K. Boutsia}
\affiliation{Carnegie Observatories, Las Campanas Observatory, Casilla 601, La Serena, Chile}
\author{J. A. Newman}
\affiliation{Department of physics and Astronomy and PITT PACC, University of Pittsburgh, Pittsburgh, PA 15260, USA}
\author{S. Puccetti}
\affiliation{Agenzia Spaziale Italiana-Unit\`a di Ricerca Scientifica, via del Politecnico, 00133 Roma, Italy}

\begin{abstract}
The study of the space density of bright AGNs at $z>4$ has been subject to extensive effort  given its importance  for the estimate of the cosmological ionizing emissivity and  growth of supermassive black holes. In this context  we have recently derived high space densities of AGNs at $z\sim 4$ and $-25<M_{1450}<-23$ in the COSMOS field from a spectroscopically complete sample. In the present paper we attempt to extend the knowledge of the AGN space density at fainter magnitudes ($-22.5<M_{1450}<-18.5$) in the  $4<z<6.1$ redshift interval by means of a multiwavelength sample of galaxies in the CANDELS GOODS-South, GOODS-North and EGS fields. We use an updated   criterion to extract faint AGNs from a population of NIR (rest-frame UV) selected galaxies at  photometric $z>4$ showing X-ray detection in deep Chandra images available for the three CANDELS fields. We have collected a photometric sample of 32 AGN candidates in the selected redshift interval, six of which having spectroscopic redshifts.  Including  our COSMOS sample  as well as other bright QSO samples allows a first guess on the shape of the  UV  luminosity function at $z\sim 4.5$.  The resulting emissivity and photoionization rate  appear consistent with that derived from the photoionization level of the intergalactic medium at $z\sim 4.5$. An extrapolation  to $z\sim 5.6$ suggests an important  AGN contribution to the IGM ionization if there are no significant changes in the shape of the UV luminosity function.

\end{abstract}


\section{Introduction}

The cosmological evolution of the AGN population is a key property to understand the growth of supermassive black holes in galaxies  and to assess the contribution of this population to the expected ionizing photon budget of the integalactic medium (IGM)
as a function of redshift. Concerning the latter aspect,  galaxies hosting  an AGN  are known to ionize their surrounding IGM especially when the nuclear emission overwhelms that from the hosting stellar population by order of magnitudes as in the most powerful quasars. Their ionizing flux with an escape fraction $f_{esc}\lesssim 100$\% propagates by  $\sim 10$ Mpc at $z\sim 5.5$ as observed in quasar spectra (e.g., \citet{prochaska09},\citet{worseck14}).
Recent results derived from $z\sim 4$ fainter AGNs with luminosities comparable to that of local Seyferts  are showing that the escape fraction does not change strongly from $M_{1450}\sim -29$ (\citet{cristiani16}) down to $M_{1450}\sim -24.5$ keeping average values $\langle f_{esc}\rangle \sim 80$\% (\citet{grazian18}).

Their strong ionizing capabilities are the required  precondition  for a significant contribution of   AGNs  to the expected photoionization rate at any redshift, provided that their
volume density at intermediate and faint luminosities is sufficiently high. There is general consensus about the shape and evolution of the AGN luminosity function
at $z<4$ which is able to provide the required  photoionization rate (\citet{haardt12}). Moreover, direct HST far UV measures in a sub-area of the GOODS-North field
in the redshift interval $z\sim 2.5-3$ suggest that AGNs can dominate the contribution to the photoionization rate at $z<4$ (\citet{jones18}). The estimate of  the AGN ionizing emissivity at $z>4$ is more controversial and probably affected by systematics in the adopted AGN selection functions.
Even at $z\sim 4-4.5$ there are contrasting results on the AGN space density. On one hand AGN surveys based on standard color selection coupled with point-like appearance suggest at these redshifts relatively low space densities at $-24<M_{1450}<-26$ resulting in a shallow and progressive bending of the luminosity function toward the faint end. Indeed the faint extrapolations of the SDSS and  the Subaru quasar surveys all agree toward a minor contribution of AGNs to the IGM photoionization rate at $z>4$ (e.g., \citet{akiyama18}, \citet{onoue17}, \citet{matsuoka18}, \citet{mcgreer18}).

On the other hand, multiwavelength deep surveys at $z\gtrsim 4$ are progressively discovering an increasing number of previously unknown faint AGNs able to produce a rather steep luminosity function down to  $M_{1450}\sim -24$ (\citet{glikman11}, \citet{civano11}, \citet{fiore12}). In particular the Glikman survey was based on a reasonably complete spectroscopic sample of optically selected AGNs but the disagreement with the SDSS and Subaru faint-extended surveys is significant. The Glikman points in fact appear higher by a factor up to $\sim 5$ at $M\sim -23.5$ with respect to e.g.  \citet{akiyama18}. Very recently we have obtained preliminary results from an ongoing spectroscopic survey of AGNs in the COSMOS field (\citet{boutsia18}). The AGN candidates have been selected by different criteria based e.g. on color or X-ray selections and photometric redshifts. Adding data collected from the literature we ended with a corrected space density of $1.6\times  10^{-6}$ Mpc$^{-3}$ at $z\sim 4$ and $M_{1450}\sim -23.5$ even a bit higher than found by \citet{glikman11}, suggesting serious incompleteness at the faint end of the bright large area surveys based on standard color selections. This new result leaves open the possibility of a significant contribution by the AGN population to the ionizing photon budget at $z\sim 4.5$.

In light of our recent results in the COSMOS field we try to extend in the present paper the knowledge of the luminosity function to fainter magnitudes adopting an improved analysis on a larger and deeper dataset with respect to the one  used in our previous works  (\citet{fiore12}, \citet{giallongo15}, hereafter G15). In G15 we have shown that the inclusion of the X-ray detection in the multiwavelength analysis  of galaxies at $4<z<6.5$ can allow to probe  the faint-end  of the AGN UV luminosity function down to $M_{1450}\sim -18$. In particular, the method proposed by \citet{fiore12}  is effective in discovering faint high-redshift AGN candidates among the high $z$ galaxies selected in NIR H-band images (UV rest-frame)  which show any detection in deep and high resolution Chandra X-ray images. We also made a first attempt to estimate the global shape of the UV luminosity function in the redshift interval $z=4-6.5$ and gave the corresponding estimate of the predicted photoionization rate provided by the global AGN population. We concluded that a scenario where AGNs can give a significant contribution to the reionization was consistent with the performed analysis. \citet{madau15} have then evaluated the global evolutionary scenario for a reionization driven by the AGN population while \citet{finkelstein19} have proposed a scenario where both AGNs and star forming galaxies compete at various redshifts to provide the required emissivity to keep the IGM highly ionized. The G15 results have been questioned more recently by e.g., \citet{cappelluti16}, \citet{vito18} and \citet{parsa18}  on the basis of different evaluations of the significance of some X-ray detections and of different estimates of photometric redshifts which could result in a significant contamination of the G15 sample by low redshift sources.

In the present paper we give an improved analysis on a larger dataset based on the new 7Msec Chandra image in the CANDELS GOODS-South (GDS) coupled with shallower X-ray images in the CANDELS GOODS-North (GDN) and EGS fields. In the new analysis we benefit from the higher signal-to-noise ratio (S/N)  in the deepest X-ray regions coupled with an area more than 3 times larger and updated estimates of CANDELS photometric redshifts. In Sect. 2 we describe the CANDELS catalog of high redshift AGN candidates including the X-ray detection and photometric redshift estimates  making at the same time a critical analysis of the G15 and   \citet{parsa18}  results.
In Sect. 3 we derive the  AGN UV luminosity function in two redshift intervals $z=4-5$ and $z=5-6.1$. In  Sect. 4 we show predictions on the expected AGN UV emissivities and photoionization rates in the same redshift intervals. Finally, in Sect. 5 we draw our conclusions.

Throughout the paper we adopt round cosmological parameters  $\Omega_{\Lambda}=0.7$, $\Omega_{0}=0.3$, and Hubble constant $h=0.7$ in units of 100 km/s/Mpc. Apparent magnitudes are in the AB photometric system.

\section{The AGN sample in the CANDELS fields}

The CANDELS catalog including the GOODS and EGS fields covers an area of about 0.15 deg$^2$ at an average NIR depth $H\sim 27$ at the HST resolution.
As in G15 we select galaxies in the $H$ band which samples the rest-frame UV emission  ($\lambda<3000$ {\AA}) at $z>4$. Thus our sample is an UV selected sample of AGNs at $z>4$. The CANDELS optical-NIR photometric catalog of galaxies in the GOODS-South area is the same as used in G15 (\citet{guo13}). In the present paper we add  the CANDELS GOODS-North (\citet{barro19}) and EGS fields (\citet{stefanon17}) selecting galaxies in the H band down  to $H=27-27.5$ depending on the exposure maps in all the fields. The three fields cover an area of $\sim 170, 176, 205.5$  arcmin$^2$ at a mean depth of $H\sim 27$. The availability of deep IRAC images from the {\sl Spitzer Space Telescope}   covering the CANDELS fields to 26 AB mag ($3\sigma$) at both 3.6 and 4.5 $\mu$m (\citet{ashby13}) is also important for the accuracy of the photometric redshifts.

Catalogs of photometric redshifts have been provided by the CANDELS team for the galaxies in the used fields. \citet{dahlen13} have made a first comparison analysis among the various codes. Nine independent redshift estimates have been statistically combined to produce a best estimate. These estimates published in \citet{santini15} have been used in G15 to derive the redshift distribution of faint AGN candidates. In the present work we use an updated estimate by the CANDELS team which further develops the \citet{dahlen13} analysis optimally combining four redshift probability distribution functions (PDFs)  by four groups within CANDELS. The method involves a combination of the different PDF(z) based on the minimum Frechet distance (Kodra et al. 2019 in prep.) which provides more reliable confidence intervals when compared with spectroscopic redshifts. 
More specifically Kodra et al. (2019) calculate the distance of the PDF of each participant team from the other participants (also called the Frechet Distance), by taking the difference of the PDFs at each point, and then summing up all these differences for the entire redshift interval where the PDFs are evaluated.
For each source they  identify the participant team which has the smallest sum of differences from the other participants (i.e.  the minimum Frechet Distance) and adopt its best redshift estimate and PDF. The method has been checked using sources with spectroscopic redshifts and four teams have been selected as those giving the best global agreement.

We adopt in general the resulting best photometric redshifts  or spectroscopic redshifts where available. We also note in this context that  the spectroscopic redshifts available in our sample are in  good agreement with the photometric estimates in all the cases.

The best fit solutions for the photometric redshifts have been derived fitting to the observed SEDs the spectral energy distributions predicted by stellar population synthesis models without considering any dominant contribution from AGN emission.  This choice is supported by the fact that at $z>4$, the photometric estimate of the redshift  is mainly based on the dropout of the SED due to neutral absorption by the IGM shortward of the Ly$\alpha$ and Lyman limit wavelengths  independently of the specific intrinsic spectral shape (G15).  Moreover, in faint AGNs which are partly obscured, the host galaxy contribution is expected to be significant in the rest-frame optical band  (e.g. \citet{bongiorno12}, \citet{bongiorno14}, \citet{lamassa17}).

\subsection{X-ray data analysis}

As in G15 the AGN selection from the parent  $z>4$  catalog is based on the detection of significant X-ray emission in the source position measured
in the HST/WFC3 H-band image. In this updated analysis we benefit from the new 7 Msec GOODS-South  Chandra image as well as from the shallower 2 Msec GOODS-North and 0.7 Msec EGS Chandra images. The data reduction was done reprocessing all the observations using the {\it Chandra} Interactive Analysis Observations (CIAO) software (v4.8; \citet{fruscione06}) and CALDB v4.8. Intervals of high background were determined by creating 0.3--10 keV background light curves in intervals of 100 s. We rejected time intervals where the background count rate was 3$\sigma$ above the mean value of the background count rate in the observation.  For each observation, we produced energy filtered events files and exposure maps in several energy bands. We registered and refined X-ray astrometry of each observation adopting a sample of bright point-like X-ray sources selected from the  {\it Chandra} catalogs available for the three fields (Xue et  al. 2011, Alexander et al. 2003, Xue et al. 2016 and Nandra et al. 2015). The positions of these bright sources have been recovered in each single observation providing the needed rototranslation. The cleaned and astrometrically corrected event files have then been  generated for each frame and coadded with a residual positional error  of 0.02 arcsec. The astrometric solutions found for the full mosaics following this procedure may not be  consistent with the CANDELS astrometric solution. Shifts of  $\approx0.1-0.2$ arcsec are seen between the X-ray and the CANDELS bright sources (see also Luo et al. 2017). Since we are interested in collecting X-ray counts at the position of the CANDELS sources, we realigned the mosaics to the CANDELS reference frames using bright X-ray and optical sources.

We search  for X-ray counts in the final coadded X-ray images at the position of each CANDELS source in the NIR H band. This is the key to reach the lowest possible flux limit, since analyzing the X-ray emission at the position of known sources allows one to use a less conservative threshold for source detection than in a blind search. We do not correct for possible offsets betwen the X-ray centoid and the CANDELS position.
The X-ray detection strategy and photometry are based on the {\it ephot} software and are extensively discussed in Fiore et al. (2012).
To reach the faintest X-ray flux limits we chose  the (circular) photometric apertures and energy bands that minimize the background and maximize the SNR.  We
use apertures from one to seven arcsec (diameter). Because the Chandra Point Spread Function (PSF) varies strongly with the offaxis
angle (and also with the energy band, although less strongly), we use apertures allowing to collect at least 35$\div$40\% of the counts at each
offaxis angle. This is important to obtain reliable fluxes, limiting the uncertainty on the PSF aperture correction, which is obviously larger for larger PSF shapes. On the other hand, large apertures may be affected by  contamination of foreground sources close to the position of the main target.
To limit the contribution of contaminants we carefully checked all detections based on large apertures and excluded from our sample 
all the cases where the spurious X-ray flux comes from adjacent bright sources. As we will discuss in the following, this is the main difference with respect to the analysis reported in G15. 

The sky coverage and associated incompleteness correction is estimated by using MonteCarlo simulations (see \citet{fiore12}  for details). 
The study of faint X-ray source population requires the most careful possible characterization of the background. Following \citet{fiore12}, we extracted average
backgrounds  by excluding regions of 10, 15 and 20 arcsec of radii around bright sources. The corresponding background spectra were very similar and the background obtained with a 10 arcsec exclusion region has been adopted  to estimate the background at the position of each CANDELS source.  We first extracted spectra at the position of each CANDELS source. We then normalized the average background to the counts detected in the X-ray spectra at the position of the CANDELS sources in the 7-11 keV energy band, where the contribution of the X-ray sources with respect to the internal Chandra ACIS background is negligible (for faint sources) or small (for bright sources), due to the sharp decrease in the Chandra effective area.  This procedure allows better count statistics, minimizes systematic errors due to e.g. source crowding, varying exposure times etc. It also allows a direct estimate of the Poisson probability for background fluctuations at the position of the CANDELS sources.

We estimated the Poisson probability that the counts extracted from a given energy band and a given aperture at the position of each CANDELS source were a fluctuation of the background (estimated as described above). We finally chose the aperture and energy bands producing the smallest probability.

To associate a reliable probability to each X-ray search we resorted again to simulations. We first simulated between $10^5$ and $10^6$ background X-ray spectra at the position of each CANDELS source, to use exactly the same exposure time, vignetting, and PSF. We then applied the {\it ephot} detection tool to these simulations and studied the number of spurious sources as a function of the parameters used: 1) the Poisson probability that the simulated counts are indeed a background fluctuation; 2) the size of the source extraction region; 3) the background subtracted counts in the broad 0.3-4 keV band; 4) the energy band width $E_{Max}/E_{min}$. In this multidimensional space we chose a combination of parameters providing a number of spurious detections smaller than one every 5000 spectra.  The number of candidates with $z>4$ in the three fields analyzed  is 4084 (1489 in  GDS, 1341 in  GDN and 1254 in the EGS) and we expect about 1 spurious detection in the overall sample. The estimates  of spurious detection probabilities are given in Table 1.

{\it e-phot} was also run on the galaxy samples with fixed energy bands ($0.5-2$ keV), thus optimizing only for the size of the source extraction region. The X-ray fluxes in the band $0.5-2$ keV were estimated from {\it e-phot} count rates in the $0.5-2$ keV band, after PSF correction, if the S/N ratio in this band is higher than 2.5 or from the count-rates in the band which optimizes the detection otherwise. To convert count-rates into fluxes we used a spectrum with a photon index estimated from the ratio of the counts in the $0.5-2$ keV and $2-4$ keV bands when a source was detected in both bands, or with a fixed photon index of 1.4 otherwise. To test our photometry, we compared our $0.5-2$ keV flux with those published by \citet{luo17} for the CDFS, \citet{xue16} for the CDFN and \citet{nandra15}  for the EGS.  The agreement is generally good, with the median of our samples in agreement with those of the published samples within 2\% for the CDFS sources, 6\% for the CDFN sources and 4\% for the EGS sources.\\ \\


\subsection{Result from the X-ray data analysis}

We have detected in the X-ray images 32 AGN candidates with $4<z<6.1$ whose positions, with relative astrometric accuracy of $\sim 1$ arcsec, redshifts, and H magnitudes are given in Table 1. This table also lists the results from the X-ray detection procedure: X-ray best detection energy band, photometric  aperture (diameter), probability of spurious detection and $0.5-2$ keV flux.

Of the 32 AGN candidates, 19 come from the GOODS-South field (11 in common with \citet{luo17}), 8 from GOODS-North (6 in common with \citet{xue16}), and 5 from the EGS field (1 in common with \citet{nandra15}). The X-ray contours of the  32 sources  overlaid with the WFC3 H-band images are shown in the Appendix (Figures 12, 13, 14).

Of the 19 AGN candidates in the GOODS-South 15 are in common with G15. In 4 of the 7 candidates removed from the original G15 catalog (GDS4285, GDS4952, GDS5501, GDS31334) the X-ray detection remains confirmed at the chosen probability threshold, using a typical detection cell of 2-3 arcsec around the CANDELS sources. However, the comparison of the Chandra map with the HST image suggests that most of the X-ray emission is actually produced by contaminating sources within 1-2 arcsec from the CANDELS high-z target. This contamination was underestimated in G15 because of the shallower X-ray images with respect to the present Chandra 7Msec images. Thus the X-ray association in the present sample is now more robust. The statistical significance associated to the remaining 3 of the 7 removed candidates (GDS12130, GDS9713, GDS33073) was just above threshold in the 4 Msec analysis (G15), but it is just below threshold in the 7Msec analysis.  This may be due to either source variability or background fluctuation. These 7 sources were also removed by the \citet{parsa18} analysis of GOODS-South field. On the other hand there are  4 new candidates at $z>4$ in the new 7Ms GOODS-South image which increase to 19 the total number.

We have also produced an X-ray stack of  all the new 14 sources  not included in the previous X-ray selected catalogs (see Figure 15 in the Appendix). The X-ray stack image in the {\it fixed} 0.8-3 keV energy band shows a significant $S/N\sim 10$ implying that the bulk of our new sources are X-ray emitters. More details are in the Appendix. 

Here we remind that the associated X-ray luminosities are in general $L_X\gtrsim 10^{43}$ erg s$^{-1}$ in the $2-10$ keV band.  These luminosities are more typical of Seyfert-like and QSO sources rather than  starburst galaxies for which these high X-ray luminosities would correspond to very high star formation rates $\gtrsim 2000$ M$_{\odot}$ yr$^{-1}$ (\citet{ranalli03}) well above the average SFRs derived from stellar SED fits to our composite sample. Moreover  selecting high redshift X-ray sources would imply sampling high rest-frame spectral energies $>7$ keV,  where the galaxy contribution is progressively decreasing. We can not exclude however a significant contribution from stars to the X-ray luminosity of few peculiar sources in our sample. 

As in G15 we note that the possible presence of significant X-ray absorption does not imply a similar absorption shortward of the Lyman continuum, because the physical regions responsible for the UV and X-ray emissions are different in size and ionization state. Indeed X-ray absorption is in general close to the emitting region and originated by metals associated to a highly ionized hydrogen.

\begin{table*}
\caption{The  NIR/X AGN candidates catalog}
\begin{center}
\begin{tabular}{r r c c c l l  c r r}
\hline\hline
 & ID & RA & Dec & z  & H &  X-ray$^a$  & X-ray$^b$   &  Spurious Det. Prob.  &$F_X$ (0.5-2keV)  \\
 &    &    &     &    &    &  $\Delta E$ keV             & arcsec   & $10^{-5}$  &       $10^{-17}$  erg s$^{-1}$ cm$^{-2}$   \\
\hline
GDS & 273 &53.1220463 &-27.9387409 & 4.49, 4.762$^s$ & 23.98 & 1.1-1.8 & 6 & $<0.1$  & $6.5$ \\ 
&$2527^*$ & 53.1392984 &-27.8907090 & 4.10  $ $   &25.85           & 0.6-2.2 & 5 &$1$ & $3.4$ \\ 
&4356 &53.1465968 &-27.8709872  & 5.21  &26.39               & 1.4-5.2 & 2 &$<0.1$ & $3.2$ \\ 
&5248 & 53.1480453 &-27.8618602 & 4.66  &26.48               & 0.8-4.2 & 2  &5-10 & $1.0$\\
&5375 & 53.1026292 &-27.8606307 & 4.22  &25.18               & 0.8-2.0 & 2 &1-2 & $2.1$ \\ 
&6131 & 53.0916055 &-27.8533421 & 4.55 &24.20                & 0.4-6.3 & 2 &$<0.1$ & $3.5$\\ 
&8687 & 53.0868634 &-27.8295859 & 4.17, 4.400$^s$  &26.92    & 0.3-2.6 & 4 &10-20 & $1.9$\\ 
&8884 & 53.1970699 &-27.8278566 & 4.51  &25.77               & 0.8-3.4 & 7 &10-20  & $3.4$ \\ 
&$9945^*$ & 53.1619508 &-27.8190897 & 4.34, 4.497$^s$  &25.01   & 0.3-4.2 & 3  &0.2-1  & $1.1$\\  
&11287 & 53.0689924 &-27.8071692 & 4.93  &25.08              & 1.1-1.8 & 4 &2-5  & $1.2$\\ 
&11847 & 53.1040317 &-27.8023590 & 5.01  $ $  &24.53              & 0.4-3.9 & 4  &0.2  & $2.2$\\ 
&14800 & 53.0211735 &-27.7823645 & 4.92, 4.823$^s$  &23.46  & 0.4-1.6 & 5  &5-10  & $2.9$\\  
&16822 & 53.1115637 &-27.7677714 & 4.50  &25.70              & 0.7-5.2 & 2  &$<0.1$  & $9.3$\\ 
&19713 & 53.1198898 &-27.7430349 & 5.18  &25.33              & 0.5-2.0 & 2  &$<0.1$  & $3.9$\\ 
&20765 & 53.1583449 &-27.7334854 & 5.60  &24.46              & 0.8-1.8 & 5  &$<0.1$  & $6.3$\\ 
&23757 & 53.2036444 &-27.7143907 & 4.01 &24.57               & 1.1-2.4 & 5 &10-20  & $4.8$\\ 
&28476 & 53.0646867 &-27.8625539 & 5.72  &26.79              & 0.9-2.0 & 4  &0.2-1  & $2.8$ \\
&$29323^*$ & 53.0409764 &-27.8376619 & 5.37  &26.35              & 0.5-5.2 & 4  &$<0.1$  & $9.8$ \\
&33160 & 53.0062504 &-27.7340678 & 6.05  &25.92              & 0.5-4.5 & 5  &$<0.1$  & $6.6$\\  
\hline
GDN & 3326 & 189.14362635 &62.16167882 & 4.87 &  25.18       & 0.7-5.5 & 4 &$<0.1$ & $12.5$\\ 
& 3333 & 189.19983697 &62.16148604 & 4.99, 5.186$^s$& 23.65  & 0.4-4.8 & 4  &$<0.1$  & $27.2$\\
& $4333^*$ & 189.05890459 &62.17155019 & 4.55  & 26.31           & 1.1-6.3 & 4  &$<0.1$ & $8.5$\\
& 4572 & 189.32906181 &62.17385428 & 4.15  & 25.20           & 0.6-4.2 & 4  &$<0.1$ & $21.5$\\
& 5986 & 189.18805775 &62.18521547 & 4.22  & 25.09           & 0.8-1.7 & 4  &0.01-0.1 & $2.1$\\
& 15188 & 189.19076305 &62.24677265 & 5.80  & 25.00          & 0.4-1.7 & 6  &$<0.1$ & $5.7$\\
& 24110 & 189.29924275 &62.37008003 & 4.31, 4.055$^s$& 23.85 & 0.4-6.8 & 7  &0.2-1 &  $9.8$\\
& 28055 & 189.18933832 & 62.13845277 & 5.18  & 26.09        & 0.4-4.5 & 2  &0.2-1 & $3.0$\\
\hline 
EGS &7454 & 215.2492254 &53.0681778 & 4.87 &  26.80              &  0.8-2.0 & 6  &0.1-1 & $9.1$ \\
&8046 & 214.8608139 &52.7967059 & 4.11 &  24.00              &  0.6-3.9 & 5  &0.1-1 & $5.4$\\
&20415 & 215.0341520 &52.9844549 & 4.31 &  25.42            &  0.4-3.9 & 2 &10-20 & $3.4$\\ 
&23182 & 214.9485284 &52.9381169 & 4.85 &  25.59            &  0.3-3.4 & 4 &0.1-1 & $7.9 $\\
&40754 & 214.6202004 &52.7525725 & 4.01 &  25.94           &  1.1-6.8 & 6  &$<0.1$ & $11.0$\\ 
\hline
\end{tabular}
\end{center}
\textsuperscript{\normalsize a}{X-ray detection band; \textsuperscript{\normalsize b}X-ray photometric aperture; \textsuperscript{\normalsize s}spectroscopic redshift; 

 }

\tablecomments{IDs and H-band CANDELS coordinates for GOODS-South, GOODS-North and EGS are from
\cite{guo13}, \citet{barro19}  and \cite{stefanon17},
respectively. Errors in the X-ray flux estimate  range from 10\% to 30\%.  \\
 Sources with spectroscopic redshifts: GDS273
from \cite{vanzella08}, GDS8687 (this paper), GDS9945 from
\cite{herenz17}, GDS14800 from \cite{balestra10} GDN3333 
from \cite{barger08} and references therein, GDN24110 from
\cite{barger14} and references therein.\\
Objects marked with an asterisk are not used for the estimate of the luminosity function.  GDS9945 has an uncertain X-ray association. The other  sources have  $M_{1450}>-18.5$}
\end{table*}

\subsection{Photometric redshifts}

Most sources in the HST H-band image are compact  or too faint for any morphological classification and only a few spectroscopic redshifts are available from the literature  (see Table 1).  

Thus, a critical issue for these sources is related to the photometric estimate of  redshifts. In spite of the fact that we have adopted a combination of different redshift estimates provided by different authors, the redshifts of a few sources remain uncertain due to their faintness, the almost power-law shape of the SED and the contamination of their UV/blue ground-based photometry  by nearby sources.
Few AGN candidates  suffer from these large uncertainties, namely GDS2527, GDS11847, GDS33160.  
 It is worth noting however that GDS2527 has $M_{1450}>-18$ and it is not used for the computation of the $z=4-5$ LF. GDS11847 and GDS33160 are at $z>5$ and their broad PDF(z) could bias the high-$z$, LF  estimate. We quantify this bias when computing volume densities in Section 3.1.

The SEDs and PDFs(z)  of all the AGN candidates together with their X-ray, optical and IR multiwavelength images are given in the Appendix. In particular the PDFs(z) show the uncertainties associated with the redshift estimates which however still depend on the SED templates adopted. From the images it is possible to check the faintness of the $z>4$ AGN candidates in the optical bands and the need for deep IR and X-ray detections for any effective multiwavelength selection. 

A recent re-evaluation of the G15 sample by Parsa et al. (2018)  resulted in lower photometric redshifts for a significant number of sources  and consequently  lower luminosity functions, especially at $z>5$. In all the discrepant cases (except one) this is due to low photometric redshifts obtained by means of the adoption by Parsa et al. (2018) of AGN templates coupled with large dust reddening (see also the Appendix). First of all we note they adopt Calzetti extinction curve which is more suitable for starburst galaxies rather than for AGNs which typically show  an SMC extinction curve  especially at $z< 4$ (e.g. \citet{richards03}). Second, they remove  from the fit IRAC data at 5.8 and 8 $\mu m$ which are important especially for red sources. This introduces further degeneracy favouring in some cases                 low-redshift, dusty solutions. 

To check the robustness of their criticism we have also included pure AGN templates in our library following the recipe adopted by \citet{hsu14} in the extended GOODS-S field adding possible dust reddening. In Figure 8 in the Appendix we show the difference between the two redshift estimates.  The average difference is very small with  essentially no offset between the two estimates. We also note there is not significant difference between the average dust reddening derived by redshift best fit solutions obtained using galaxy or AGN templates  ($E(B-V)\sim 0.27$ vs. $E(B-V)\sim 0.15$, respectively). However, there is a fraction of 20\% of objects with  low redshift solutions by AGN templates with strong reddening. These solutions however would imply {\it pure} AGN spectral energy distributions in sources with low luminosities which in the extreme cases are more typical of dwarf galaxies $M_B\sim -12,-15$ without assuming any contamination by the UV light of the host galaxy.  We consider unlikely these low-z solutions.

Dusty and reddened AGN templates from the UV to the NIR giving low redshift solutions   appear moreover  in contrast with that resulting from  recent multicomponent analyses performed on a sample of reddened AGNs by \citet{bongiorno14} and \citet{lamassa17}. Indeed their  multicomponent analyses performed on reddened broad-line AGNs with spectroscopic redshifts are showing optical spectra dominated by the SEDs of the  host galaxy. The AGN continuum appears to mainly shape the NIR rest-frame region (\citet{bongiorno14}, \citet{lamassa17}) and not the rest frame optical region as assumed by Parsa et al. (2018).   Moreover, the fainter high redshift AGNs at $z\sim 5-6$ seem less dusty with respect to the bright QSOs observed in the same redshift interval. Indeed  recent ALMA observations seem to support a scenario where faint AGNs appear to inhabit normal main-sequence or quiescent galaxies (Izumi et al. 2018) in contrast with the brightest QSOs at similar redshifts. 

For all these reasons we keep the photometric redshift solutions  obtained by a combination of the probability redshift distributions derived by  galaxy templates as described above.

\subsection{Two spectroscopic redshifts from the MUSE-Wide survey}

We have extracted MUSE spectra at the position of our 12 candidates falling in the  MUSE-wide survey in GOODS-South (\citet{herenz17},  \citet{urrutia18}).
The MUSE-Wide survey is a GTO programme  covering a relatively large area in GOODS-South with a relatively shallow exposure time of 1h.  MUSE is the integral field unit at ESO-VLT with a FoV of 1
arcmin$^2$ covered with a spatial sampling at 0.2 arcsec and a spectral resolution of 2.5 {\AA}  from 4750 {\AA} to 9350 {\AA} (\cite{bacon09}). Details on the data reduction analysis are provided in \citet{herenz17}  where the detection algorithm LSDCat has been used to select emission lines by means of a matched filtering procedure. This analysis has provided the detection of GDS9945 as a clear  Lyman $\alpha$ emitter at $z=4.50$ with a possible AGN signature of weak NV emission. This source was already known as Lyman $\alpha$ emitter by means of previous unpublished spectroscopic information (see G15), however since
its X-ray position could be contaminated by a close brighter galaxy (see Figure 9 in the Appendix) the source was not used for the computation of the LF. We have also visually checked for weak emission line features finding another tentative detection in GDS8687 at $z=4.40$ in broad agreement with our photometric redshift but in contrast with the \citet{parsa18}  estimate which puts GDS8687 at $z=3.57$ so out of the $z>4$ sample. The emission lines of the two sources are shown in Figure 1. The lines have a $S/N\sim 34, 4$, fluxes $f\sim  40,
9 \times 10^{-18}$ erg s$^{-1}$ cm$^{-2}$, luminosities $L\sim 7, 1.8 \times 10^{42}$ erg s$^{-1}$. which are typical of Lyman $\alpha$ emitters at $z\sim 5-6$. In particular luminosities are confined between 0.1$L^*$ and $L^*$ at $z\sim 4-4.5$   at the faint end of the luminosity interval where AGN activity could be present in Lyman $\alpha$ emitters (e.g., \citet{sobral19}). 

\begin{figure}
\centering
\plotone{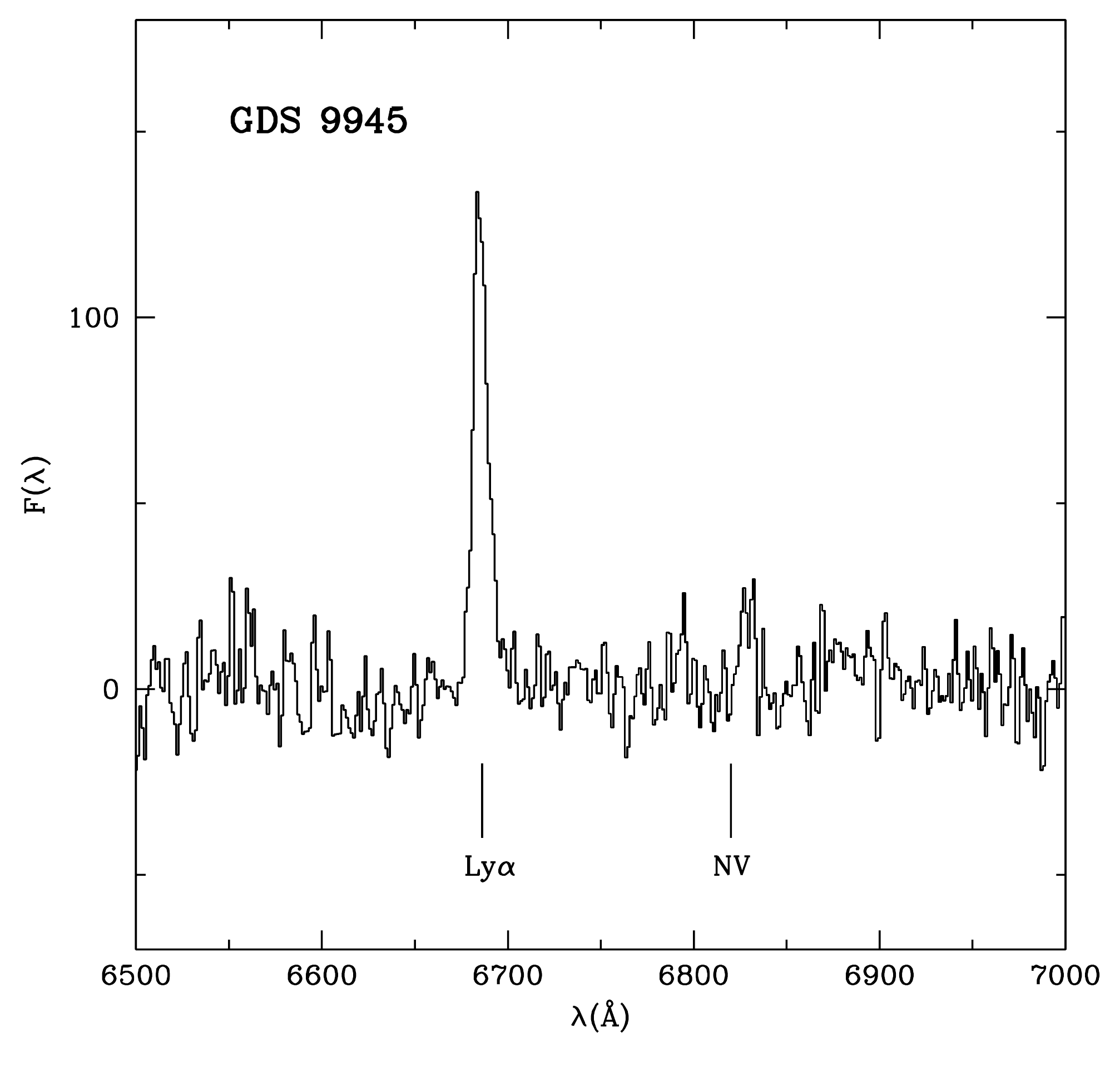}
\plotone{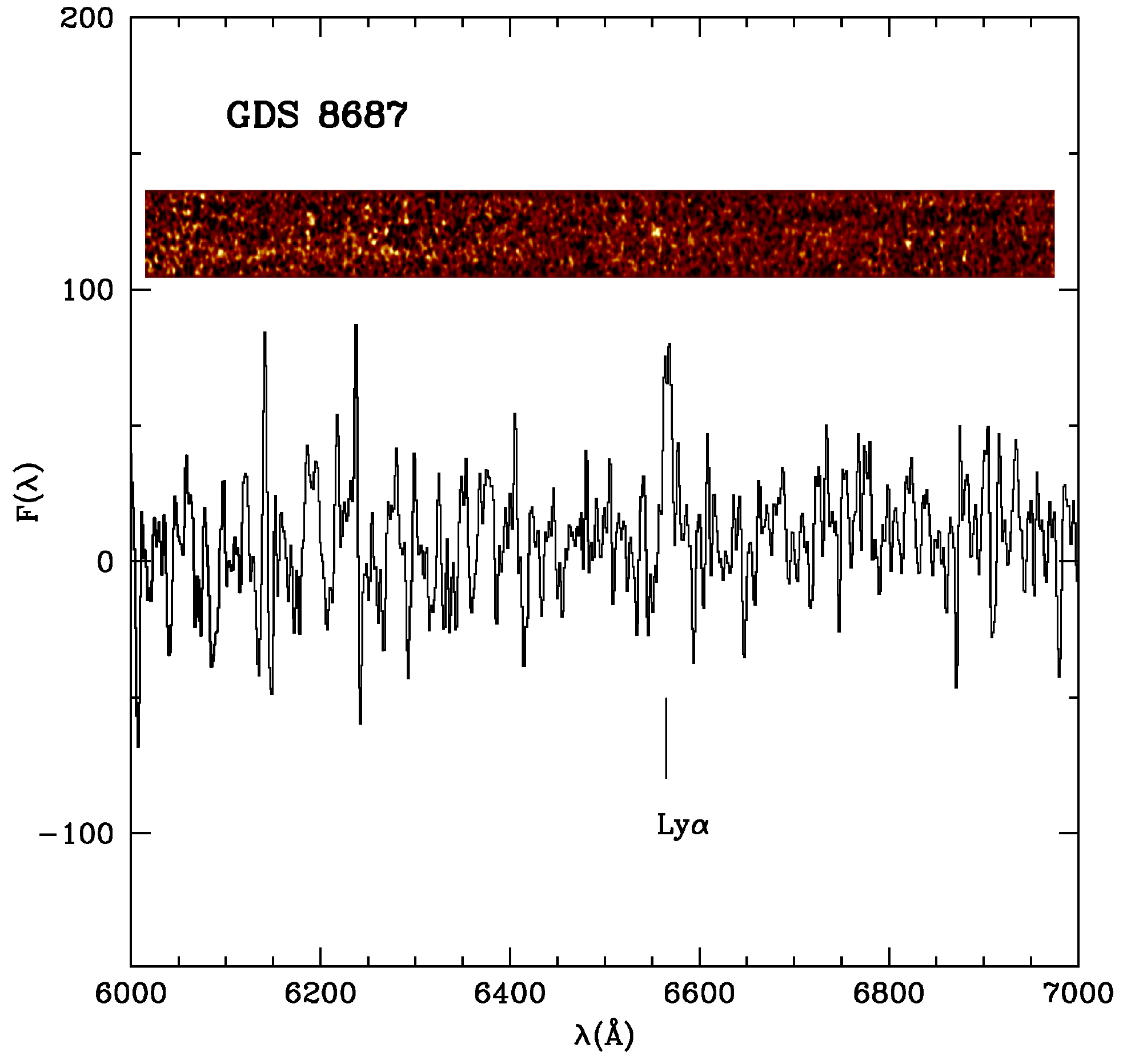} 
\caption{Spectra of two AGNs with emission line detection from the data cube by \citet{urrutia18}. Fluxes are in arbitrary units. The Lyman $\alpha$ emitter GDS9945 (top panel)  is included in the  \cite{herenz17} paper. Here we show a slightly offsetted spectrum showing a flux-reduced, slightly asymmetric Ly$\alpha$ line.  A weak NV possible emission feature 1500 km s$^{-1}$ redward of the expected position is present. Tentative identification of Lyman $\alpha$ emission is also provided for GDS8687. The 2D extracted spectrum is shown in the middle panel and the 1D spectrum around the emission line in the bottom panel.  The vertical lines mark the wavelength positions of the assumed Lyman $\alpha$ emission.}

\end{figure}

\subsection{Rest-frame UV luminosities}

Although the photometric redshifts have been estimated from galaxy templates we are assuming that the AGN luminosity is driving the UV SED of our sources with  a minor contamination by the host galaxy. To check the validity of our assumption we show in Figure 2 the 2 keV Luminosity vs monochromatic UV luminosity at 2500 {\AA} for our 32 sources.  Luminosities at 2500 {\AA}  have been derived from the observed apparent magnitudes closer to $2500\times (1+z)$ {\AA}, namely J band for $4<z<5$ sources and H band for $5<z<6$. Luminosities at 2 keV have been derived from the 0.3-2 keV flux assuming a photon spectral index $\Gamma=1.4$ for these faint sources which is more similar to the background spectrum and takes into account a possible absorption.

We have compared the distribution of our sources in the $L_X-L_{UV}$ plane to the  relation measured by \citet{lusso10} for the   brighter, type 1, COSMOS AGNs. It appears that most of the objects are within 1$\sigma$ from the correlation without presence of any significant bias. We conclude that in our sample the AGN contribution to the UV luminosity is on average prevailing  on the host galaxy contribution.  This is consistent with the SED analysis performed by \citet{bongiorno12} on  the type 1 COSMOS sample (their Figure 4, right panel) where a two-component model (AGN+galaxy templates) has been adopted. An AGN contribution at least by 50-60\% of the total UV flux has been found within a 25th percentile of the unobscured AGN sample for $\lambda<3000$ {\AA}.  Thus our sample in Figure 2 simply represents the fainter end  of the COSMOS $L_X-L_{UV}$ correlation but at higher redshifts.

Also for the computation of the UV luminosity function in the next section, rest frame UV 1450 {\AA} absolute magnitudes $M_{1450}$  are  derived directly from apparent magnitudes in those filters with  effective rest frame wavelengths closer  to $1450\times (1+z)$ {\AA}.  This minimizes uncertainties in k-corrections.

\begin{figure}
\centering
\plotone{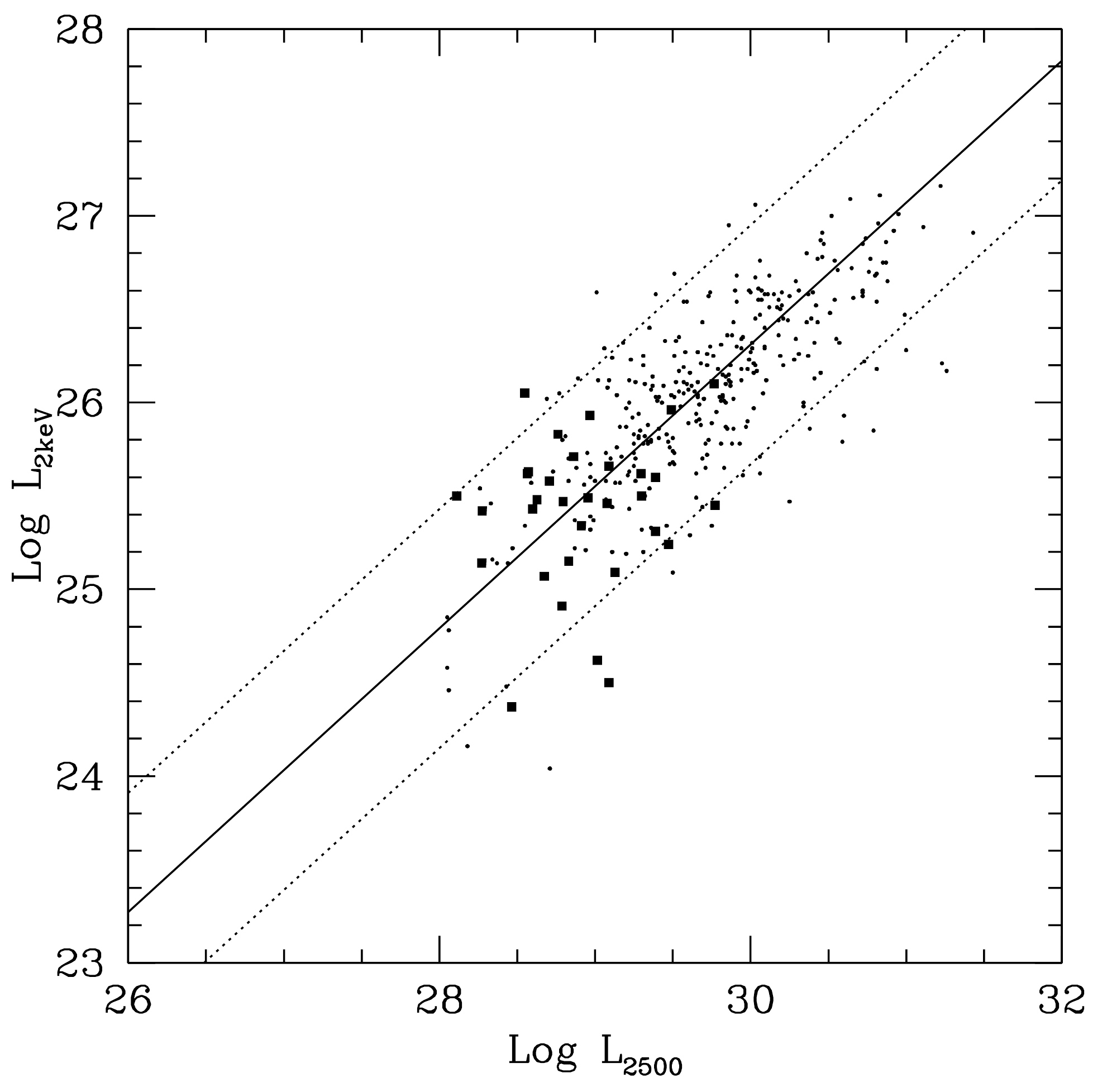}
\caption{ $\text{Log} (L(\nu)_{2keV})$ vs. $\text{Log} (L(\nu)_{2500})$  for our AGN candidates (filled squares). The \citet{lusso10} AGN COSMOS sample is also shown  for comparison as small dots. The  continuous and dotted lines represent the \citet{lusso10} best fit correlation with $1\sigma$ uncertainties in the offset parameter.}
\end{figure}

\section{The UV AGN luminosity function}

\subsection{$1/V_{max}$ analysis}

Our  magnitude-selected galaxy sample in the UV rest-frame band is then used to estimate the 1450 {\AA} luminosity function  following G15. The extended version of the  standard $1/V_{max}$ algorithm (\citet{schmidt68}) is adopted where different regions  with different magnitude limits are combined together in the volume estimate of each object (e.g., \citet{avni80}).  

For a given redshift interval $(z_{low}, z_{up})$, these volumes are confined between $z_{low}$ and $z_{lim} ( j)$, the latter being defined as the minimum between $z_{up}$ and the maximum redshift at which the object could have been observed within the magnitude limit of the $j-th$ region.  Indeed, the complexity of the exposure map in the H-band image of the CANDELS fields requires a composition of several sub-areas with  associated magnitude limits (see G15).

As in G15 the galaxy volume density $\phi(M, z)$ in a given $(\Delta z, \Delta M)$ bin is:
\begin{equation}
\phi(M,z)=\frac{1}{\Delta M} \sum_{i=1}^n\left[ \sum _j \omega (j)  \int _{z_{low}} ^{z_{lim}(i,j)} \frac{dV}{dz} dz  \right]^{-1} 
\end{equation}
where $\omega( j)$ is the area in units of steradians corresponding to the different regions, $ n$ is the number of objects in the redshift-magnitude bin and $dV/dz$ is the comoving volume element per steradian.

\begin{figure}
\centering
\plotone{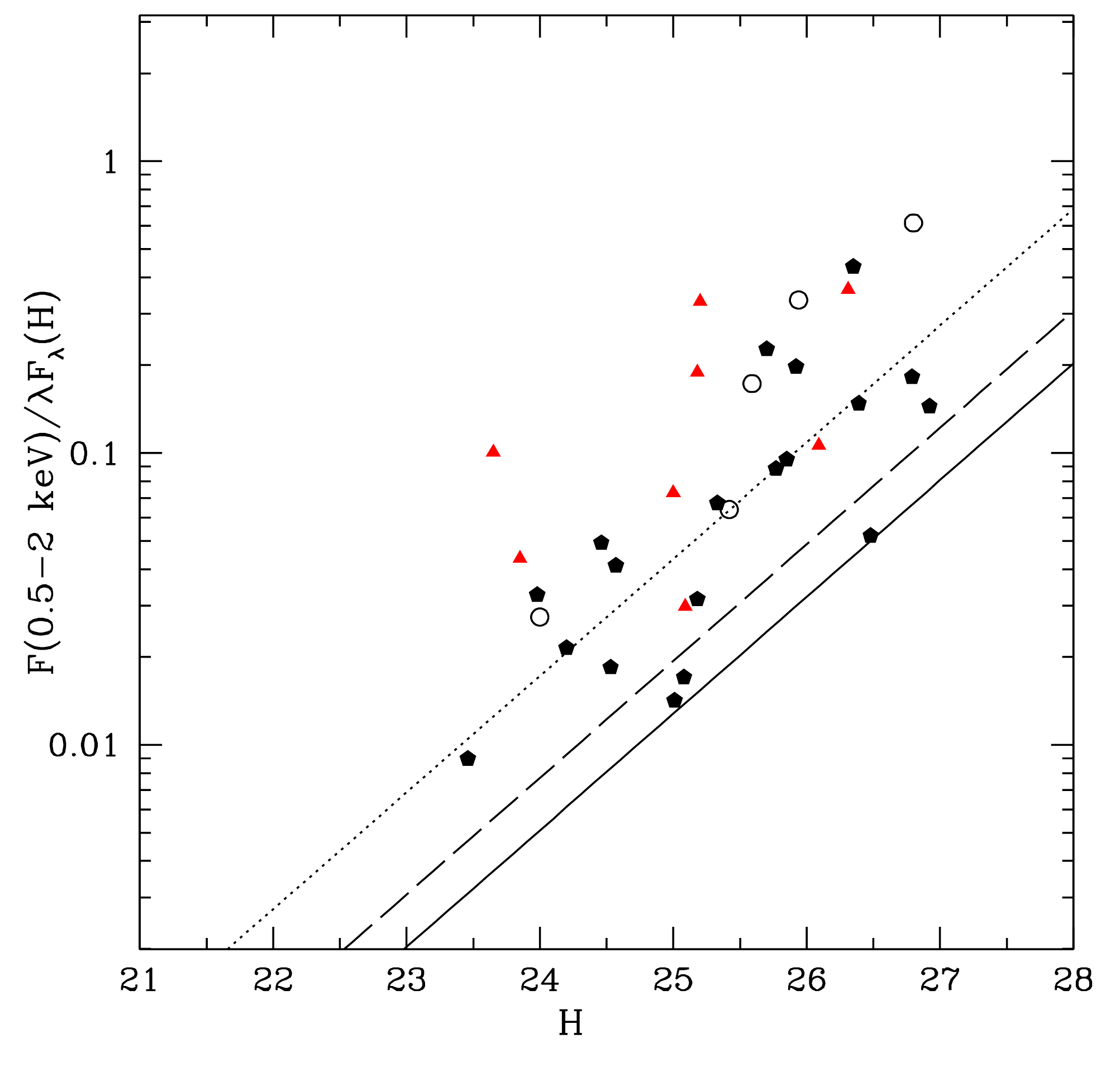}
\caption{Adimensional X/H flux ratios as a function of the H-band magnitude for the GOODS-South (pentagons), GOODS-North (red triangles) and EGS (empty circles)  AGN candidates. Straight lines represent the adopted X-ray  flux limits  of the three fields at 40\% completeness level, $10^{-17}$ (continuous), $1.5\times 10^{-17}$ (dashed) and  $3.4\times 10^{-17}$ (dotted) in erg cm$^{-2}$ s$^{-1}$ in the $0.5-2$ keV band for GDS, GDN and EGS respectively.}
\end{figure}

The estimates of the UV luminosity functions of AGN candidates selected by their X-ray flux are subject to significant correction for incompleteness since at faint NIR  (rest-frame UV) fluxes only sources which are relatively bright in the X-ray band can be detected even in the deepest GOODS-South field.  Figure 3 shows the X/H flux ratio as a function of the H band magnitude for our AGN candidates in the three fields (GDS 9945 left out). The straight lines indicate the locus of the constant X-ray flux limits adopted for the three fields. It is clear for example that at $H\gtrsim 26$ sources with  $F_X/(\lambda f_H)\lesssim 0.1$ can not be detected at the X-ray flux limit of the EGS survey. Overall our AGN sampe is thus biased against relatively weak X-ray emitters as pointed out in G15. This aspect has been taken into account when computing the faint-end UV luminosity function.

\begin{figure*}
\centering
\plotone{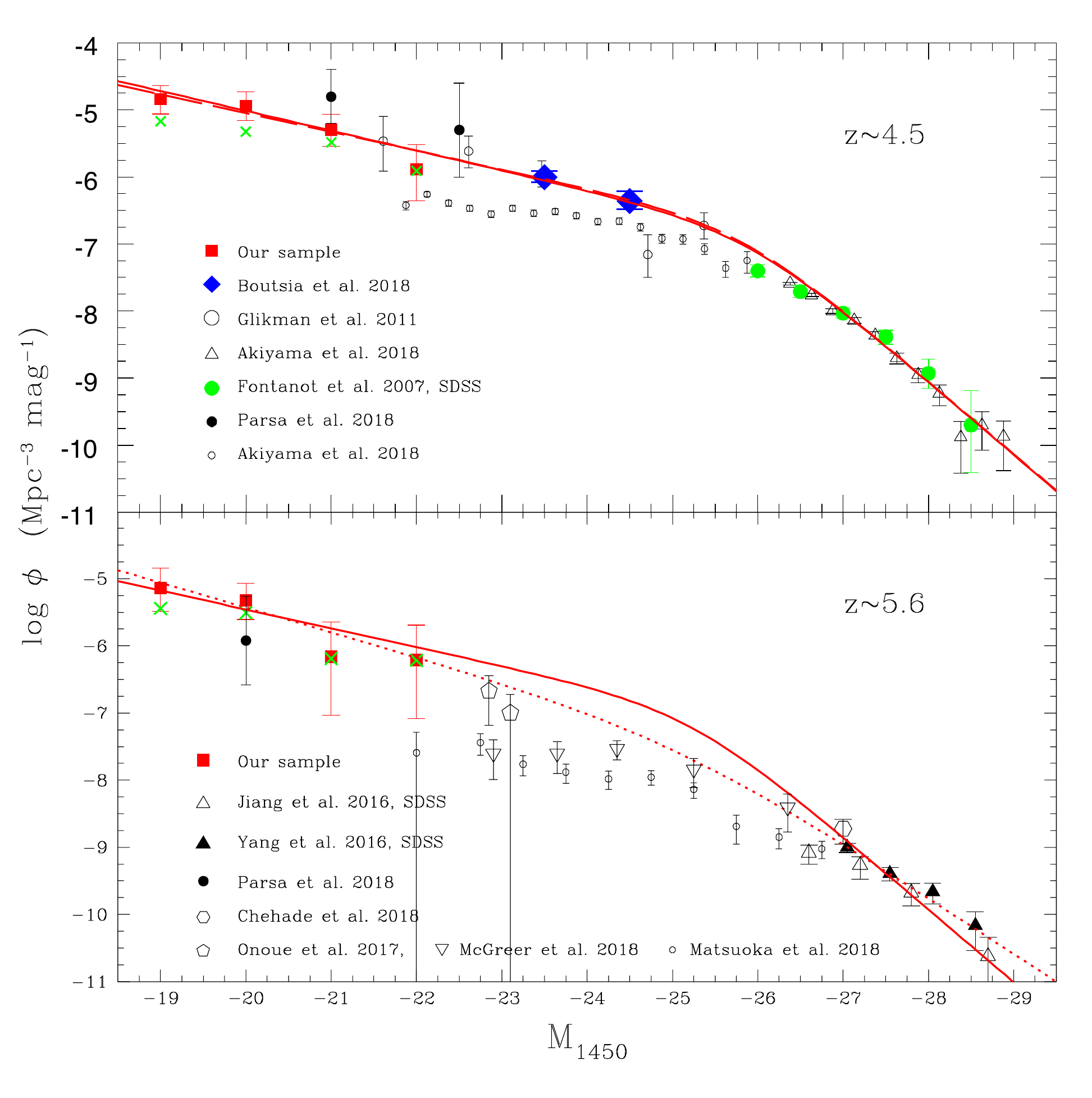}
\caption{UV 1450 {\AA} AGN luminosity functions in two redshift intervals. Different symbols represent $1/V_{max}$ data points from different surveys as explained in the figure box. Green crosses are CANDELS points only corrected for incompleteness in the H band counts while red squares are the same  points corrected for incompleteness in the X-ray detection due to the $X/H$ distribution (see Sect. 3.1).\\
Upper panel:  dashed red curve  (model 1) is the best fit LF derived  at $z=4.5$ connecting CANDELS data with our recent COSMOS spectroscopic sample (\citet{boutsia18} and with the SDSS data as analyzed by \citet{fontanot07} (green bullets) and   \citet{akiyama18}  (open triangles); continuous curve (model 2) is the best fit obtained including also the NOAO sample (\citet{glikman11}); only SDSS quasar densities at $M_{1450}<-27$ have been considered for the fit.  Other volume densities derived by color selected surveys at intermediate magnitudes are shown for comparison.\\
Lower panel: dotted curve  (model 3) is the best fit solution obtained at $z=5.6$ leaving free all the LF parameters; continuous curve (model 4) is the best fit solution  obtained  fixing the two slopes to the best fit values obtained at $z\sim 4.5$  in model 2. Only the CANDELS  and SDSS with $M_{1450}<-27$ samples have been used for the analysis. Other volume densities derived by color selected surveys at intermediate magnitudes are shown for comparison.}
\end{figure*}

The resulting volume densities as a function of $M_{1450}, z$  are shown in Table 2 and Figure 4 in two redshift intervals $\Delta z=4-5$ and $\Delta z =5-6.1$. 
Volume densities have been first corrected for incompleteness in the H-band galaxy counts at the faintest limits showing e.g.,  a 50\% drop at $H\sim 27$ in GOODS-South and at $H\sim 26.5$ in the EGS field. The second correction takes into account  the loss of AGN candidates with X-ray flux below the X-ray flux limit for AGNs with a given H magnitude. The incompleteness fraction is derived from the same $X/H$ distribution observed above the X-ray flux threshold for a given H-band flux of each source. The objects have been weighted considering the shape of the corrected H counts. It is clear that the adopted X/H distribution could be biased by selection effects (e.g., volume effects in small area surveys) and could be different from the intrinsic X/H distribution of faint AGNs at $z>4$, which is essentially unknown at the redshifts and luminosities probed here. Nevertheless this is an  attempt to estimate any correction for incompleteness due to  the X/H properties of the faint AGN population.
Again, as in G15,  corrections by 10-20\% have also been  applied to volume densities due to spatial fluctuations of the X-ray flux limits  for each X-ray position.

\begin{table*}
\caption{AGN luminosity functions from 1/V$_{max}$ analysis}
\centering
\begin{tabular}{c l r r c r}
\hline\hline
$\Delta z$ & M$_{1450}$ & $\phi_{obs}$ & $\phi_{corr}$ \ \ \ & N$_{obj}$ & $\phi_{MC}$  \ \ \ \\
\hline\hline
$4-5$ & $-19$ & $6.81$ & $14.54_{-5.81}^{+8.72}$ & 6  & $18.05\pm4.27$\\
 & $-20$ & $4.74$ & $11.47_{-4.59}^{+6.88}$ & 6  & $8.03\pm 3.34$\\
 & $-21$ & $3.29$ & $5.08_{-2.21}^{+3.45}$ & 5  & $4.52\pm 1.15$\\
 & $-22$ & $1.24$ & $1.31_{-0.87}^{+1.74}$ & 2  & $1.33\pm 0.11$\\

\hline
$5-6.1$ & $-19$ & $3.62$ & $7.27_{-4.02}^{+7.12}$ & 3 &  $6.27\pm 3.42$\\
 & $-20$ & $3.12$ & $4.77_{-2.31}^{+3.79}$ & 4  & $2.91\pm 1.84$\\
 & $-21$ & $0.65$ & $0.69_{-0.60}^{+1.61}$ & 1  & $1.13\pm 0.70$\\
 & $-22$ & $0.61$ & $0.62_{-0.54}^{+1.44}$ & 1  & $0.80\pm 0.33$\\

\hline
\end{tabular}\\
$\phi_{corr}$ is $\phi_{obs}$ volume corrected for incompleteness in the X/H distribution. $\phi$, $\phi_{corr}$ and $\phi_{MC}$ are in units of  $10^{-6}$ Mpc$^{-3}$ mag$^{-1}$.\\
$\phi_{MC}$ are average volume densities derived from 1000 simulated catalogs where random photometric redshifts have been extracted from the PDF(z) of each source. See details in Sect. 3.1.
\end{table*}


\begin{table*}
\caption{Parametric AGN luminosity functions, emissivities and photoionization rates$^1$}
\begin{center}
\begin{tabular}{c l l l c l c c l}
\hline\hline
Model & \ \ $\Delta z$ & \ \ $\beta$ & \ \ $\gamma$ & M$_{break}$ & $ \ log \phi^*$ & $ \epsilon_{1450}$ & $ \epsilon_{912}$ & $ \ \ \Gamma$ \\
\hline\hline
1 & $4-5$  & 1.70 & 3.71 & $-25.81$ & $-6.68$ & $6.50_{-3.77}^{+24.26}$ & $3.89_{-2.26}^{+14.52}$ & $0.56_{-0.33}^{+2.09}$ \\
2 & $4-5$  & 1.74 & 3.72 & $-25.89$ & $-6.76$ & $6.35_{-3.51}^{+8.53}$ & $3.80_{-2.10}^{+5.10}$ & $0.55_{-0.30}^{+0.73}$ \\     
\hline
3 & $5-6.1$  & 1.92 & 3.09 & $-25.06$ & $-7.29$ & 1.33 & 0.80 & 0.07\\
4 & $5-6.1$  & 1.74$^a$ & 3.72$^a$ & $-25.37$ & $-7.05$ & $1.94_{-1.59}^{+7.37}$ & $1.16_{-0.95}^{+4.41}$ & $0.11_{-0.09}^{+0.41}$ \\  

\hline
\end{tabular}\\
\end{center}
$^1\phi^*$ in units Mpc$^{-3}$ Mag$^{-1}$, $\epsilon$ in units of $10^{24}$ ergs s$^{-1}$ Hz$^{-1}$ Mpc$^{-3}$ , $\Gamma$ in units of $10^{-12}$ s$^{-1}$;   model 2 also includes the NOAO   data points (\citet{glikman11}).   \\
1$\sigma$ errors in $\beta,\gamma,$M$_{break},\log\phi^*$ are $\sim 0.4,0.7,1.3,1.0$ respectively for both  models 1,2\\ 
1$\sigma$ errors in M$_{break},\log\phi^*$ are $\sim 0.7,0.7$ for model 4\\
$\beta,\gamma,$M$_{break},\log\phi^*$ for model 3 are essentially  not constrained by the present data.\\
Errors for $\epsilon$ and $\Gamma$ are derived computing the 68\% joint probability distribution for M$_{break}$ and $\log \phi^*$ having fixed the two slopes\\ 
a) fixed value. \\

\end{table*}

Poisson errors  in  LF bins have been derived  adopting the recipe by \citet{gehrels86} which is also valid for a small number of sources.  Systematic errors as for example  due  to cosmic variance in small area surveys are expected to be smaller than in G15 due to the three fields used in the present  analysis. 

The LFs are shown in Table 2 and Figure 4 with (red filled squares) or without (green crosses) corrections for incompleteness with respect to the expected X/H distribution. They amount to a factor $\sim 2$ at the faintest magnitude bins. It is to note that the small corrections applied to the brightest points at $M_{1450}\sim -22$ in our sample are probably  lower limits since even at these relatively bright magnitudes there could be  AGNs not detected in X-ray. We know for example there is a relatively bright AGN ($R\sim 24.8$)  with spectroscopic redshift at $z\sim 2$ in GOODS-North which lacks X-ray detection  in the 2Msec X-ray image (see e.g., Figure 2 in \citet{steidel02}). Figure 4 shows a rather flat luminosity function in the UV luminosity interval typical of the local Seyfert population $-21\lesssim  M_{1450}\lesssim -18.5$ with corrected densities $\sim 10^{-5 }$ Mpc$^{-3}$ mag$^{-1}$ at $z\sim 4.5$. 

We note that the number of candidates in the higher redshift bin selected $\Delta z=5-6.1$ is small and prevents any  preliminary estimate of a specific shape. In this context  the X-ray detection of sources at $z\gtrsim 6.5$ is strongly disfavoured by sampling the spectral energy distribution at progressively higher rest-frame energies going from 10 keV to 16 keV  in the redshift interval $\Delta z=4-7$ for observations taken at $\sim 2$ keV.  In this range of energies a typical AGN spectrum drops by a factor $\lesssim 2$. Thus  a typical AGN with a fixed bolometric luminosity (fixed H magnitude) which would be barely detected at 2 keV at $z\sim 4$ will be undetected if at $z\sim 7$.

As already mentioned, a few AGN candidates  suffer from large uncertainties in the estimate of the photometric redshift. 
Thus to check in general  how the uncertainties in the photometric redshifts affect the LF estimates we have extracted randomly by a Monte Carlo technique the photometric redshift for each source according to its PDF(z)  shown in Figure 7. In this procedure we have also included AGN candidates at $3<z<4$ some of which having a significant probability of being at $z>4$. These lower $z$ sources have been selected adopting the same procedure used for the $z>4$ AGN candidates.
 In case a spectroscopic redshift is available for an AGN we fix the photometric redshift to the spectroscopic one.  We then processed the simulated catalog with the same software adopted for the observed catalog. We did  not vary the H-band magnitudes and the weight assigned to each object to take into account the X/optical flux ratio incompleteness. We produced  1000 simulated catalogs used to  compute the mean values of the simulated $\phi_{MC}$ for the same  magnitude bins adopted for the LF derived by the best fit photometric $z$. The scatter of the LF has been derived for each absolute magnitude and redshift interval as half of the difference between the 84th and 16th percentiles.
  
Form Table 2 it is possible to conclude that the resulting $\phi_{MC}$ mean values are consistent with the volume densities derived from the  best estimates of the photometric redshifts,  indicating that the few broad PDF(z) distributions present in our sample do not significantly bias the estimate of the luminosity function.
The resulting scatter of the simulated  $\phi_{MC}$   results  lower than the errors computed  according to the \citet{gehrels86} recipe, probably due to the small number of sources per bin.  Thus the poissonian errors shown in Table 2 and Figure 4  do not represent a significant underestimate of the true errors.
 
For these reasons the volume densities  based on the redshift best estimates are therefore  used  in the next sections  to estimate the global shape of the LF and its associated emissivity.

Our new corrected volume densities at $z\sim 4.5$ are e.g. a factor $\sim 2$ lower with respect to those derived in G15 at  $M_{1450}=-20$. These lower values are due to fluctuations of number densities of AGN candidates  in the three fields, to the different incompleteness correction derived from  the new X/H distribution
and  to the cleaning of uncertain X-ray sources, as described in Sect. 2.2.  Our volume  densities are consistent with those derived by \citet{parsa18} at $z<5$ and $M_{1450}\sim -20$ while are a factor 3 higher at $z=5.6$. This discrepancy is mainly due to the lower redshifts derived   by Parsa et al. (2018)  for a significant fraction of our GOODS-South sample, probably due to the reddened AGN templates adopted for the estimate of the photometric redshifts, as discussed in the previous section.

\subsection{On the global shape of the AGN luminosity function at $z>4$}

The prediction of the ionizing AGN emissivity critically depends on the global shape of the luminosity function and on the escape fraction of ionizing photons from the AGN host galaxy. 

In this context a homogeneous UV sample of $z>4$ AGNs selected on the basis of the source X-ray detection and extended
on a sufficiently large magnitude interval ($-29\lesssim M_{1450}\lesssim -18$) is not available with the present instrumentation. For this reason we have connected our data points of the luminosity function to that derived by different optical selected surveys under various assumptions about their possible incompleteness. 

More specifically, to derive a first guess on the shape of the UV luminosity function we have compared in Figure 4 our volume densities derived at $M_{1450}>-22$ with that of the brightest QSOs selected in the SDSS survey where selection effects with respect to the morphological appearance and X-ray properties are thought to be small.  In other words, at the brightest absolute magnitudes sampled by the SDSS survey $M_{1450}<-27$  no  X-ray QSOs with strong absorption in the rest-frame optical/UV are expected and the optically selected sample should be representative of the overall AGN population.
Volume densities derived at slightly different  redshifts have been scaled to our average redshifts adopting a rescaling in normalization.
We adopted a redshift decrease  $\Delta log \phi = -k \Delta z$ where $k=0.34$ in the redshift interval  $z=4-5$ (\citet{schindler18}) and $k=0.5$ at higher $z$
(\citet{fan01}). At $z\sim 4.5$ QSO volume densities derived from the same SDSS sample  by different authors  differ by more than a factor of two at $M_{1450}>-27$ due to different procedures adopted to derive the selection function and consequently  the correction for incompleteness (e.g., \citet{richards06}, \citet{fontanot07}).  For this reason we considered only QSO volume densities derived at $M_{1450}<-27$.

We have also connected our CANDELS densities with the ones derived by us in the COSMOS field  based on an extensive spectroscopic campaign of candidates
selected on a multiwavelength criterion (\citet{boutsia18}). This  includes not only the standard color selection but also the use of phtotometric redshift catalogs and especially X-ray detection. In this respect it is the most similar selection criterion to the one adopted in G15 and in the present work. For this reason the derived volume densities are particularly high at intermediate absolute magnitudes ($M_{1450}\sim -24$) much higher than found by e.g. \citet{akiyama18} in their color selected QSO sample but closer  to the ones derived by \citet{glikman11} in their multicolor survey (NOAO deep wide-field survey).

We have therefore included both \citet{boutsia18} and \citet{glikman11} in the best fit analysis of the luminosity function  scaling the published densities from $z=3.9$ and $z=4.2$ to  $z=4.5$, respectively.  Model 1 includes our COSMOS sample and the SDSS data. Model 2 adds the NOAO sample to the data of model 1.

To check how the inclusion of brighter surveys can alter the faint end slope of the LF  we have first derived a best fit power-law slope  $\beta=1.74$ for our CANDELS data only. The slope is definitely flatter than found at the bright end suggesting the presence of a significant break at intermediate magnitudes. For this reason we adopted a two power-law shape for the LF of the type
\begin{equation}
\phi=\frac{\phi^*}{10^{0.4(M_{break}-M)(\beta-1)}+10^{0.4(M_{break}-M)(\gamma-1)} }
\end{equation}

Two best fit solutions are shown in Table 3 and Figure 4 (upper panel) at $z\sim 4.5$. The first solution (model 1)  shows a flat faint-end slope consistent with that derived by the CANDELS data alone. A sharp break is present at intermediate magnitudes ($M_{1450}\sim -25.8$) followed at the bright end by a 
 steep  power-law $\gamma \sim 3.7$. Such a steep slope at bright magnitudes ($M_{1450}<-27$) seems supported by the recent evaluations based on the ELQS-N QSO survey at $z\lesssim 4$ (\citet{schindler18}).

The second solution (model 2) includes the NOAO survey in the analysis. The resulting densities are lower but closer to our COSMOS data and the 
LF shape derived from the analysis of the four samples is very similar to the one derived in model 1. 
Our CANDELS data thus represent the natural extension of the LF after the bending shown in the \citet{boutsia18} and \citet{glikman11} data. 

At $z\sim 5.6$ given the poor and uncertain statistics we can not derive with a similar accuracy a global luminosity function  and the uncertainties associated to the LF parameters are much larger. Indeed, in this redshift interval our sample is statistically confined to $M_{1450}>-20$ since each of the two brightest bins includes only one object  although the source in the brightest CANDELS bin at $M_{1450}\sim -22$ is a secure AGN (GDN3333) at the spectroscopic redshift $z=5.186$.  Therefore a best fit LF derived from the joint analysis of  the bright $M_{1450}<-27$ SDSS sample and the very faint CANDELS sample resulted in a steeper faint-end slope $\beta\sim 1.9$ and a shallower bright-end slope $\gamma \sim 3.1$, as shown in table 3 (model 3). If we assume that the shape of the LF does not change appreciably from $z=4.5$ to $z=5.6$  we can fix the two LF slopes found at $z=4.5$ in model 2 and adapt normalization and break magnitude to follow the slightly different density evolutions of the bright and faint sides of the LF. For this model 4 we derived a break magnitude $M_{1450}\simeq -25.2$ and a density decline respect to the previous redshift bin by  factors 2-5 from the faint to the bright-end, respectively. These factors however are sensitive to  the uncertain break position of the luminosity function.

The very uncertain model 3 predicts volume densities at intermediate magnitudes $-23\lesssim M_{1450}\lesssim -25$ which are consistent with or slightly higher than the ones derived from standard optical color selected QSO surveys by \citet{mcgreer18} at $z\sim 5$ and by \citet{onoue17} and \citet{matsuoka18} at $z\sim 6$, all scaled at $z=5.6$.  Model 4,  with its slopes invariance, predicts a number of sources in the same magnitude interval which is a factor 3-10 higher.  While this appears as a larger factor we note  that a factor 3--4 is also present at $z\sim 4.5$ between the Subaru color survey by \citet{akiyama18} and the COSMOS spectroscopic sample by \citet{boutsia18}. This systematic discrepancy between spectroscopic samples of color selected AGNs  and AGNs selected by a broader multiwavelength criterion as in the COSMOS field could depend on the progressive increase of the incompleteness at fainter magnitudes  in color and morphological selected AGNs due to various selection effects. For example, in the COSMOS field about half of the X-ray spectroscopically confirmed AGNs lies outside the standard optical color selection. Moreover, the Subaru surveys  seem to rely on tighter color selections and appear more conservative to avoid large contamination by Galactic stars, for this reason the final estimated volume densities are the results of a non straightforward balance between contamination and incompleteness corrections. Thus either a large incompleteness is still present at intermediate absolute magnitudes in optical color-selected, point-like surveys or a strong and quick change in the shape of the luminosity function should be envisaged.

In summary, a sharp break of the LF  with  slope values changing  by $\sim 2$  is required at $z\sim 4.5$ to follow the volume density decrease from the CANDELS sample down to the COSMOS and SDSS samples. This sharp change implies a major role of AGNs with intermediate luminosity to the ionizing emissivity of the global AGN population up to  $z\sim 5$ as outlined in the next section. With the present multiwavelength CANDELS dataset it is not possible to derive similar constraints in the redshift interval $5<z<6.1$ and  predictions on the global AGN  ionizing emissivity rely on extrapolations about the LF shape, as already pointed out in G15.

\section{AGN Hydrogen Ionizing Emissivity and Photoionization rate}

The global ionizing emissivity has been computed adopting the same AGN SED as in G15.  It is  represented by a double power law from $\lambda=1450$ {\AA} to $\lambda=912$ {\AA}  (\citet{schirber03,telfer02} and \citet{vandenberk01}). It could be noted that in these faint AGNs the stellar contribution could in principle steepen the spectral shape shortward of $\lambda=1450$ {\AA}. However given the limited wavelength range involved in the interpolation changing the average slope from e.g., -0.44 to -1.4 shortward of 1450 {\AA} reduces the photoionization rates by about 10 percent only. Moreover, the X/H distribution and especially the $L_X$ vs. $L_{UV}$ correlation found in our sample and shown in Sect. 2.5 and 3.1 are similar to that found in the brighter COSMOS AGNs (\citet{fiore12}) suggesting a modest contribution to the UV flux by the stellar populations of the galaxies  hosting  AGN X-ray activity at levels $L_X\gtrsim 10^{43}$ erg s$^{-1}$. It is also to note that the multicomponent analysis of the SED of the best studied local Seyferts, where the stellar contribution is more important, seems to indicate that the continuum spectral shape is modeled by a galaxy spectrum in the optical region down to about 3000 {\AA} where an AGN component becomes progressively important at shorter wavelengths (e.g., \citet{marin18}). Considering also the recent multicomponent SED analysis performed on red relatively faint QSOs by \citet{lamassa17}, the predicted UV shape shortward of 1000 {\AA} seems in many cases driven by the AGN component while the galaxy component appears significant mainly in the optical bands. 

We assume an escape fraction $ \langle f_{esc}\rangle =1$ as a reference value as observed in bright quasars at high redshifts (\citet{prochaska09}, \citet{worseck14}). Recently we have evaluated the escape fraction in a small lower luminosity AGN sample at $z\sim 4$ with average absolute magnitudes  as low as $M_{1450}\sim -24$, i.e. where we already expect a significant contribution to the AGN emissivity. We derived $ \langle f_{esc}\rangle \sim 0.8$ (\citet{grazian18}). Since we obtained this measure neglecting the IGM contribution to the Lyman limit absorption,  the derived value should be considered in this context as a lower limit to the intrinsic one. Thus, although a more robust estimate waits for a complete and unbiased sample of low luminosity AGNs, our pilot sample suggests little evolution in the escape fraction with decreasing luminosities from $M_{1450}\lesssim -27$ down to $M_{1450}\sim -24.5$.

We consider the contribution by sources as faint as  $M_{1450}= -18$ up to the brightest limit of our sample $M_{1450}= -29$. It is to note that sources fainter than our low luminosity limit do not provide a significant contribution given the rather flat faint-end slope suggested by our sample. Moreover the significant change of the LF slope around the break magnitude implies a major contribution to the ionizing production rate  just by AGNs near the expected break, as extensively discussed in \cite{giallongo12}.

\begin{figure}
\epsscale{1.2}
\plotone{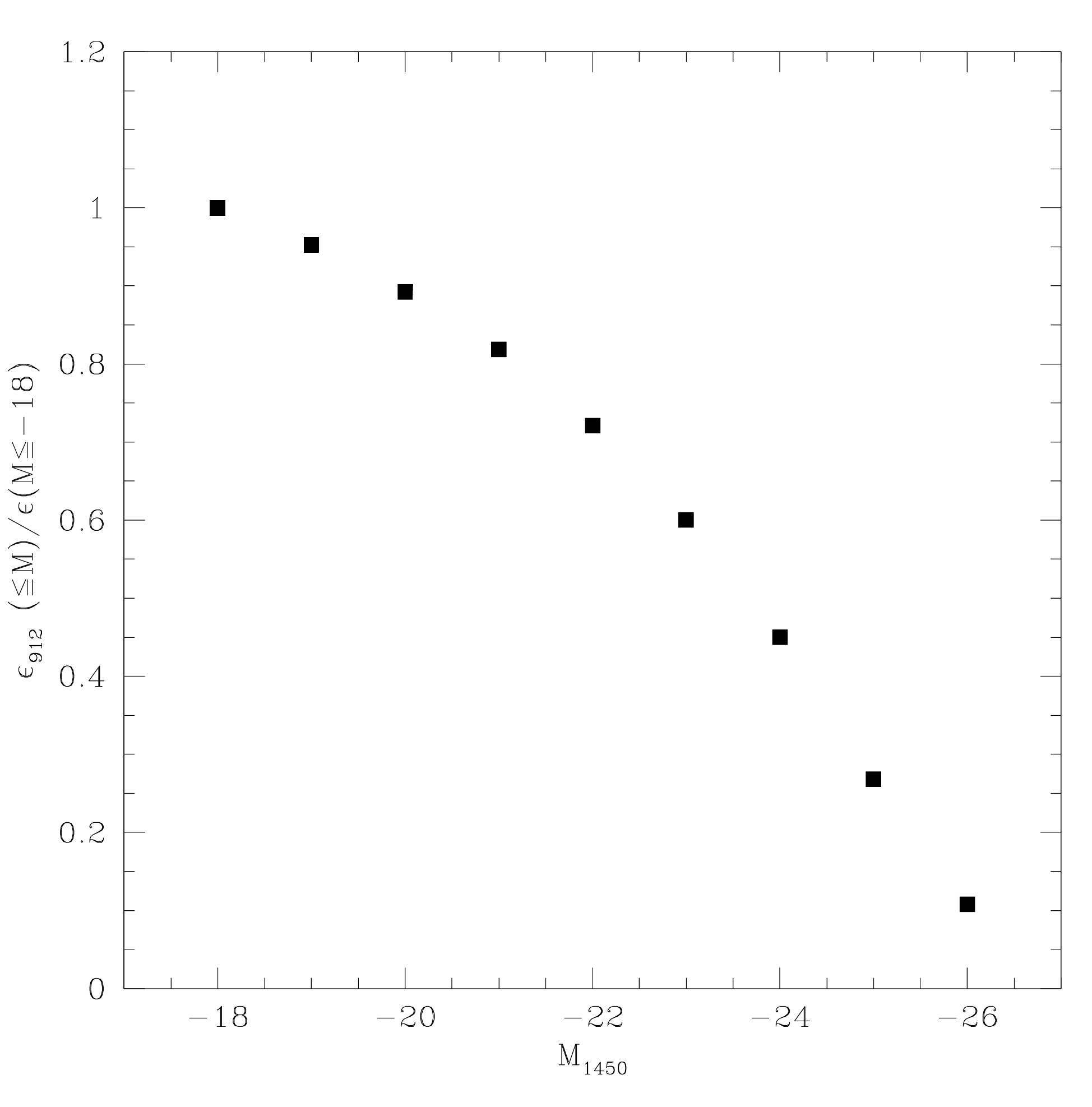}
\caption{Fractional ionizing emissivity at $z=4.5$ for model 2. The fraction is relative to the emissivity computed for $M_{1450}\leq -18$. Similar results can be derived for model 1.}
\end{figure}

In Figure 5 we have shown the fractional ionizing emissivity predicted by the model 2 in Table 3 at $z=4.5$, where the fraction is relative to the emissivity computed for $M_{1450}\leq -18$. From this curve it is possible to see that for faint-end slopes of the LF as flat as   $\beta \sim 1.7$ and break magnitudes $M_{break}\sim -25.8$, AGNs brighter than $\leq -22,-23$ are providing $\sim 70,60$\% of the AGN emissivity computed down to $M_{1450}= -18$, respectively. This is the region where at $z\sim 4-4.5$ we are starting to obtain reliable results both concerning spectroscopic redshift information and direct measures of the ionizing escape fraction $f_{esc}$.
Fainter AGNs are not expected to be dominant contributors to the UV background, independently of their escape fraction. In our computation we assume a high escape fraction down to $M_{1450}\sim -18$. Although high escape fractions are found in local Seyferts  (e.g., \citet{stevans14}), high redshift AGNs as faint as found in our sample at $z\sim 4$ still lack  detailed $f_{esc}$ measurements.

The resulting 1450 {\AA}  and ionizing emissivities for the different fitted luminosity functions derived in the two redshift bins are shown in Table 3.

The photoionization rate per hydrogen atom $\Gamma_{-12}$ in units of $10^{-12}$ s$^{-1}$ was computed following \citet{lusso15} who provided results similar to \citet{madau15}. The values derived from our AGN sample are shown in Table 3 and Figure 6. In the derivation of the photoionization rate we have increased the values of the AGN emissivity by a factor 1.2 to include the contribution by radiative recombination in the IGM following the considerations by \citet{daloisio18}.
The uncertainties in the photoionization rates of our data are derived from the joint 68\% confidence region in the break magnitude and  normalization of the luminosity function, keeping fixed the two slopes. 

The  photoionization rate provided by the AGN population in our analysis appears consistent with that derived from the analysis of the intergalactic Ly$\alpha$ forest 
statistics up to $z\sim 5$ as shown in Figure 6 where the black points represent values derived by different datasets and different procedures. When comparing with the values derived  in a model-dependent way from the IGM absorption statistics it is important to note that different methods for example related to the mean Ly$\alpha$  flux decrement in QSO spectra or to the proximity effect at the highest redshifts involve different systematic errors and different assumptions on the IGM ionization history (e.g., different average temperatures in the IGM low density regions). 

At $5\lesssim z\lesssim 6$ the redshift evolution of the photoionization rate predicted by the AGN population depends on the unknown shape evolution of the luminosity function.  In model 3, where a strong change in the shape  of the LF is envisaged with respect to that found at $z\sim 4.5$,  the photoionization rate predicted by the AGN population is $\sim 20$\% of that derived by the IGM ionization level ($\Gamma _{IGM}\sim 0.3$).  This is a minor but not negligible average fraction  which becomes more important in a scenario of inhomogeneous ionization by few clustered sources like AGNs. We have discussed however in the previous section the biases potentially involved for this solution.  The redshift evolution of the photoionization rate can be more similar to the one derived by the IGM if the shape of the AGN luminosity function does not change dramatically with respect to that  derived at $z\sim 4.5$ (model 4). Under this assumption, galaxies hosting active galactic nuclei at $z\sim 5.6$  can give a significant ($> 40$\%) contribution to the UV background near the reionization epoch,  competing with possible other classes of ionizing galaxies (see also \citet{finkelstein19}).

\begin{figure}
\epsscale{1.2}
\plotone{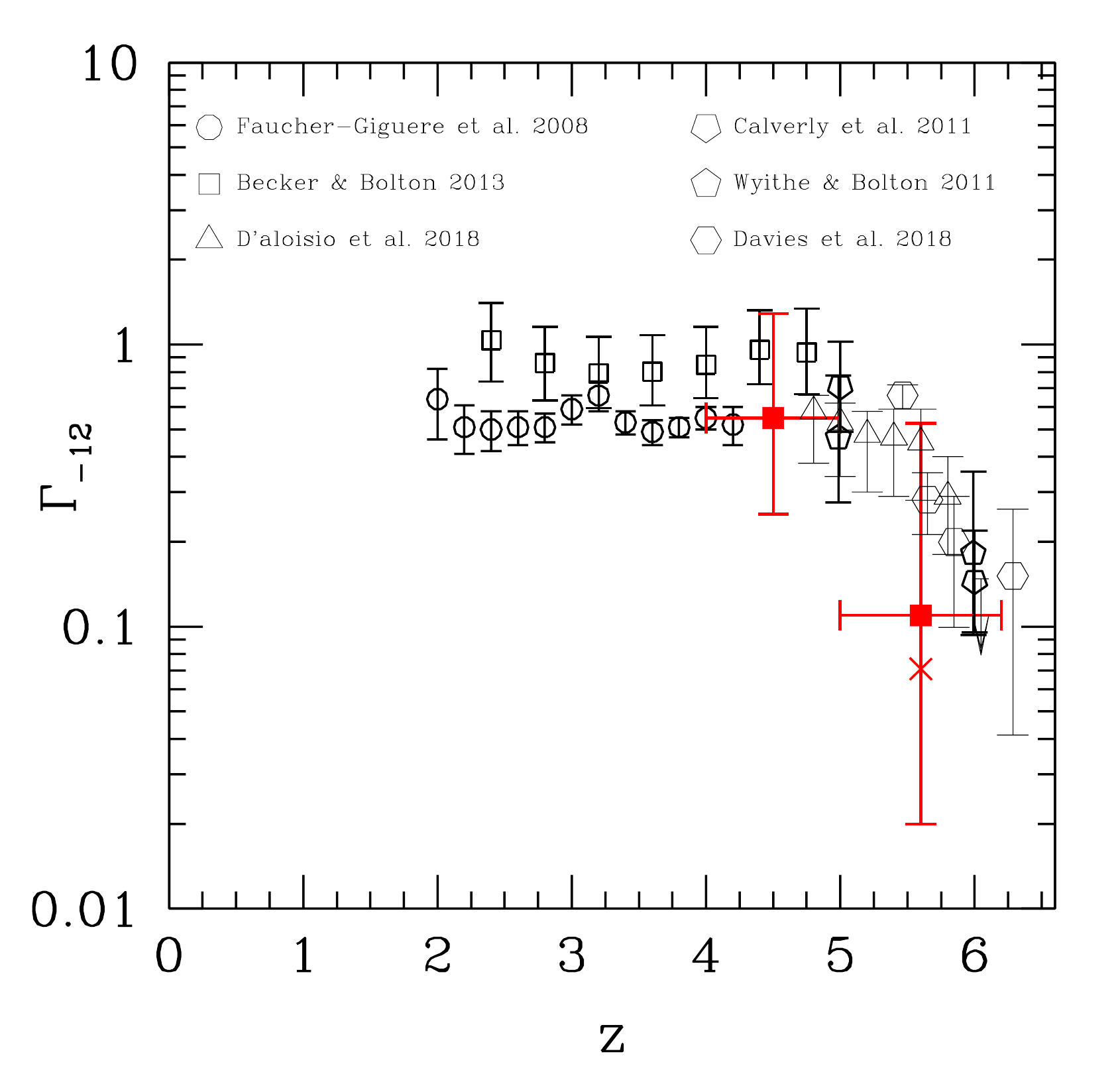}
\caption{Cosmic photoionization rate $\Gamma_{-12}$ in units of $10^{-12}$ s$^{-1}$ produced by AGNs as a function of redshift assuming $\langle f\rangle=1$. Red filled squares represent the predicted contribution at $z= 4.5$ and $z=5.6$ by the global AGN luminosity functions shown in table 3 as models 2 and 4, respectively. Red cross shows the prediction by the AGN model 3. Other open symbols are the values inferred from the ionization status of the IGM as derived from the Lyman $\alpha$ forest analysis in high $z$ QSO spectra.}
\end{figure}

\section{Discussions and Conclusions}

Motivated by the recent discovery (\cite{boutsia18}) of relatively high densities of $z\sim 4$ AGNs with magnitudes $-25<M_{1450}<-23$,  which support previous results obtained by \citet{glikman11},  we have updated and enlarged our original sample of fainter $z>4$ AGNs (G15) using the new deeper 7Msec Chandra X-ray images in GOODS-South coupled with shallower Chandra images in GOODS-North and EGS fields. As in G15 we have selected AGN candidates  starting from the H band selection of  galaxies whose photometric redshift probability distributions suggest a very high redshift $z>4$. For these galaxy candidates the NIR selection corresponds to a rest-frame UV selection. The AGN candidates have been derived from this parent sample thanks to the X-ray detection at the H-band position in the three mentioned fields.

We have revised our original sample (G15) taking into account both the X-ray association and uncertainties in photometric redshifts due also to different estimates made by various authors. While according to \citet{parsa18} we have removed a few sources from the sample on the basis of possible X-ray contamination by close interlopers, we have re-analyzed the low photometric redshifts derived  by \citet{parsa18} adopting pure AGN SEDs coupled with high dust extinction. We found on average a very good agreement between photometric redshifts derived by galaxy or AGN templates. However a fraction of 20\% of sources is found at significantly lower redshifts showing  AGN templates with strong reddening. We argued that these low-$z$ solutions are unlikely since imply a pure AGN template in sources with low luminosities  ($M_B>-18$ down to $M_B\sim -15$) without assuming any contamination by substantial optical/NIR emission by the host galaxy.
For this reason we kept our high redshift solutions based on galaxy templates.

Another source of uncertainty concerns the possible contamination of the observed  UV luminosity by the AGN host galaxy. In this respect
we have compared the distribution of our sources in the $L_X-L_{UV}$ plane to the  relation measured by \citet{lusso10} for the  type 1, COSMOS AGNs. It appears
that our sample simply follows the fainter end  of the COSMOS $L_X-L_{UV}$ correlation, suggesting  a dominant contribution to the UV luminosity by the AGN emission.

The new volume densities of the CANDELS sample are $\phi\sim 10^{-5}$  Mpc$^{-3}$ mag$^{-1}$ in the magnitude interval $-21.5\lesssim M_{1450}\lesssim -18.5$ at $z\sim 4.5$.  We have checked moreover that these volume densities are not strongly affected by the uncertainties in the derived photometric redshifts.
To this aim we have extracted randomly by a Monte Carlo technique the photometric redshift for each source according to its PDF(z) producing 1000 simulated catalogs.
The resulting  average simulated LFs appear consistent with that derived from the best fit photometric redshifts. Thus the broad PDF(z) distributions present in few sources of our sample do not significantly bias the estimate of the luminosity function.

Fitting a single power-law to the CANDELS AGNs  provides a rather flat slope $\sim 1.7$ for the faint end LF.  Adding to the analysis our brighter ($M_{1450}\sim -24$) spectroscopic  sample in the COSMOS field  (\citet{boutsia18}) and  the very bright ($M_{1450}<-27$) SDSS QSO sample  we derive  a double power-law shape  of the UV luminosity function with a break magnitude $M_{1450}\sim -25.8$ between a faint-end slope $\sim 1.7$ and a steep bright-end slope $\sim 3.7$.

Given the scanty data we have at $z>5$ we can not derive with a similar accuracy a global shape for the luminosity function and the results obtained combining the CANDELS data (statistically significant only at $M_{1450}>-20$) with the SDSS sample (at $M_{1450}<-27$) give a steeper faint-end slope $\sim 1.9$ and a shallower bright-end slope $\sim 3.1$. Assuming on the contrary that the  global shape of the LF does not change dramatically from $z\sim 4.5$ to  $z\sim 5.6$ we  fixed both the faint-end and bright-end slopes to the values found at $z\sim 4.5$.   Of course this assumption implies average densities at $M_{1450}\sim -23$ about 5-10 times higher with respect to the ones derived from  optical color selected surveys. It is interesting however to note in this respect that a similar discrepancy  is also present  at $z\sim 4-4.5$ between color selected (e.g. \citet{akiyama18}) and multiwavelength selected spectroscopic surveys (\citet{boutsia18} raising some concern about the completeness level of the faint color selected QSO samples  at $z>5$.

Our global luminosity functions have then been  used to predict ionizing emissivities and photoionization rates from the global AGN population at $4\lesssim z\lesssim 6$ assuming  full escape fraction due to the action of AGN radiative and/or mechanical outflows in the host galaxies. The latter assumption is supported by the measure of high escape fractions of ionizing photons in AGNs of intermediate luminosity ($M_{1450}\lesssim -23$) which are expected to provide most of the AGN ionizing emissivity (\citet{grazian18}).

The resulting  hydrogen photoionization rate  in the redshift interval $4<z<5$ is  in good agreement with the values derived from the statistical analysis of the IGM Ly$\alpha$ absorption  suggesting that AGN driven outflows in galaxies can play a crucial role in allowing the escape of  the required ionizing photons from the host galaxy into the IGM.

Extrapolation to  $z\sim 5.6$  suggests a decline in ionizing emissivity $\epsilon_{912}$ by a factor $\sim 3-5$   (model 4,3 respectively) depending on the assumed shape of the LF. In particular if the shape does not change dramatically between the two redshift bins then the AGN population sampled by our data can provide a fraction $\gtrsim 50$\% of the global photoionization rate derived from the ionization level of the IGM. This fraction could increase if the present color surveys are still affected by significant incompleteness at $z>5$.

Of course  a significant contribution to the IGM ionization by $z>4$ AGNs would produce an emission of hard UV photons able to produce a significant HeII ionization at $z>4$. This appears not in contrast  with  HeII QSO absorption spectra where an extended ionization period starting at $z>4$ has been inferred (\citet{worseck16}), especially in a  patchy reionization history (\citet{chardin17}). 

Moreover, in a scenario where AGN outflows are the main mechanism allowing significant escape fraction of ionizing photons, the spectral hardness of the escaping UV ionizing radiation will depend on the ratio between the AGN and host galaxy escaping ionizing flux. Thus for progressively fainter AGNs some softening of the ionizing UV spectrum could be expected depending on the ratio between the black hole and the host galaxy mass as well as on the outflow physics.

An earlier HeII ionization would produce a thermal heating of the IGM and higher temperatures are expected after the reionization epoch at $z\sim 3-5$. However, the knowledge of the thermal history of the IGM is a challenging, highly model dependent issue. Recent values for the average temperature derived at each redshift as a function of density can  differ by factors 2-3 depending on the procedures for the data analysis and model prescriptions (e.g., \citet{lidz10}, \citet{becker11}, \citet{garzilli12}, \citet{hiss18}, \citet{puchwein18}). Thus further investigation is needed to better constrain the ionizing spectral shape of the sources responsible for the reionization.

In summary, a significant contribution to the ionizing UV background by the AGN population fits well in the scenario of  late reionization  suggested by the Planck data  which put the reionization process  in place only at $z\sim 7.7\pm 0.7$ as derived from the estimate of the Thomson optical depth $\tau\simeq 0.054$ (\citet{aghanim18}). More interesting constraints however are coming from the recent spectral analysis of very high redshift bright QSOs and star forming galaxies. Indeed,  an almost neutral IGM is emerging from the analysis of the Ly$\alpha$ absorption damping wings in two bright QSOs 
($x_{HI}\gtrsim 0.5$ at $z\sim 7.1-7.5$, \citet{davies18}) and especially from the absence of Ly$\alpha$ emission in 68 star forming galaxies at $z\sim 7.6$ ($x_{HI}\sim 0.9$ \citet{hoag19}). Considering that the reionization could end at redshifts as low as $z\sim 5.5$ (\citet{becker15}) or even lower ($z\sim 5.2$ \citet{keating19}), most of the effective reionizing photons should be spread-out into the IGM in $\sim 3-4\times 10^8$ yr.

In this  "accelerating reionization" at relatively late times AGN activity in star forming  galaxies could add the required boost of ionizing photons escaping into the IGM once supermassive black holes have had time to grow. In this scenario outflows could play a significant role in driving and delaying to late cosmic times the ionization history of the IGM.

\acknowledgments
{We thank the anonymous referee for her/his constructive comments which improved the robustness of our results and the clarity of the paper. We acknowledge support  from INAF under the contract PRIN-INAF-2016 FORECAST, and ASI/INAF contract I/037/12/0. This work is based on observations taken by the CANDELS Multi-Cycle Treasury Program with the NASA/ESA HST, which is operated by the Association of Universities for Research in Astronomy, Inc., under NASA contract NAS5-26555. This work is based in part on observations made with the Spitzer Space Telescope, which is operated by the Jet Propulsion Laboratory, California Institute of Technology under a contract with NASA. Support for this work was provided by NASA through an award issued by JPL/Caltech.}

\begin{appendix}

\section{Spectral energy distributions and redshift probability distribution functions}

In this section we show the spectral energy distributions (SEDs) and the redshift probability distribution functions  (PDFs) for each AGN candidate derived from a new CANDELS analysis (Kodra et al. 2018, in preparation). Figure 7 left panels: spectral energy distributions. Continuous curves show theoretical SEDs from Bruzual \& Charlot models at the phot.z best fit or spectroscopic redshift. They are not the results of the best fitting procedure but are shown only as a representative model solution. Right panels: Probability distribution functions PDFs of photometric redshifts. Most of the candidates in Figure 7 show PDFs confined at $z>4$ with only small wings at $z<4$. 

{\it Notes on individual objects.} The very high photometric redshift estimated for GDS29323 in the previous CANDELS redshift catalog (\citet{dahlen13},      \citet{santini15}, G15) was due to SED artifacts (see \citet{cappelluti16}).  For three objects, GDS2527,  GDS6131 and GDS11847, we have measured new HST colors in a smaller aperture to avoid faint contamination  by close sources. The new SEDs are shown in Figure 6. 

GDN12027 (RA=189.17539633 DEC=62.22540103) which was included in the GOODS-north AGN X-ray catalog  with a spectroscopic redshift of $z_{spec}=4.42$ (\citet{waddington99}) has then been put at $z_{spec}=2.018$ by \citet{murphy17} based on MOSFIRE NIR spectra and it has been removed from the GOODS-North catalog of $z>4$ X-ray sources.

The redshift difference between estimates based on galaxy (BC) and AGN templates are shown in Figure 8. The average offset is small $\langle z(BC)-z(agn)\rangle = 0.04$ with 1$\sigma =0.28$ after iterative $\sigma$ clipping. A fraction of $\sim 20$\% of objects show low redshift solutions by AGN templates with strong reddening.

The associated multiband optical/NIR images are shown in Figures 9,10,11 for each candidate in the three fields.

Finally we note that GDS4356, GDS5375, GDS8687, GDS19713, GDS20765, GDS23757, GDS29323, GDS33160 are in common with the Parsa et al. 2018 catalog who however provided appreciably lower redshift solutions. In all the cases except one this is due to best fits for the photometric redshifts obtained with the adoption of AGN templates coupled with large dust reddening (See the main text for a discussion on the critical issues associated with these discrepant low redshift estimates). 

In Figures 12, 13, 14 we show X-ray contours  overlaid with the HST H-band images in the GDS, GDN and EGS fields, respectively. The H-band AGN candidates are in the center of a circle with a radius of 2 arcsec.
For most of the 14 new sources the offsets between the H-band position and the X-ray centroid are smaller than 1 arcsec.
For  GDS8884, GDS23757, GDS28476 and  EGS23182 the difference is slightly larger. In particular GDS23757 is almost equidistant from another fainter source GDS23751 which however is at the same photometric redshift thus we assigned the X-ray detection to the brighter GDS23757. 
A complex morphology appears also for GDS9945,  a source with three components (the central point-like)  just north of the brightest galaxy.
We also remark that offsets up to $\sim 1.5$ arcsec are common when detecting the X-ray counterparts of HST optical/NIR sources, especially when the X-ray detection involves few tens of X-ray counts even at small off-axis angles (see e.g., \citet{luo17}).

We have also stacked the X-ray emission of the new 14 sources found by our selection criterion, namely GDS8687, GDS9945, GDS11287, GDS11287, GDS11847, GDS23757, GDS28476, GDS33160, GDN24110, GDN28055, EGS7454, EGS8046, EGS20415, EGS23182. The stacked image is shown in Figure 15. We have found 369 counts in the 0.8-3 keV energy band and in a region of 64 pxl$^2$ corresponding to a circle of $\sim 2$ arcsec in radius. The associated background in the same area amounts to 212 counts corresponding to a S/N ratio of about 10. The S/N reduces to 8 when considering the 0.5-2 keV band.

\begin{figure}
\centering
\scalebox{0.95}[0.95]{\hspace{-0.5cm}\includegraphics[width=\hsize]{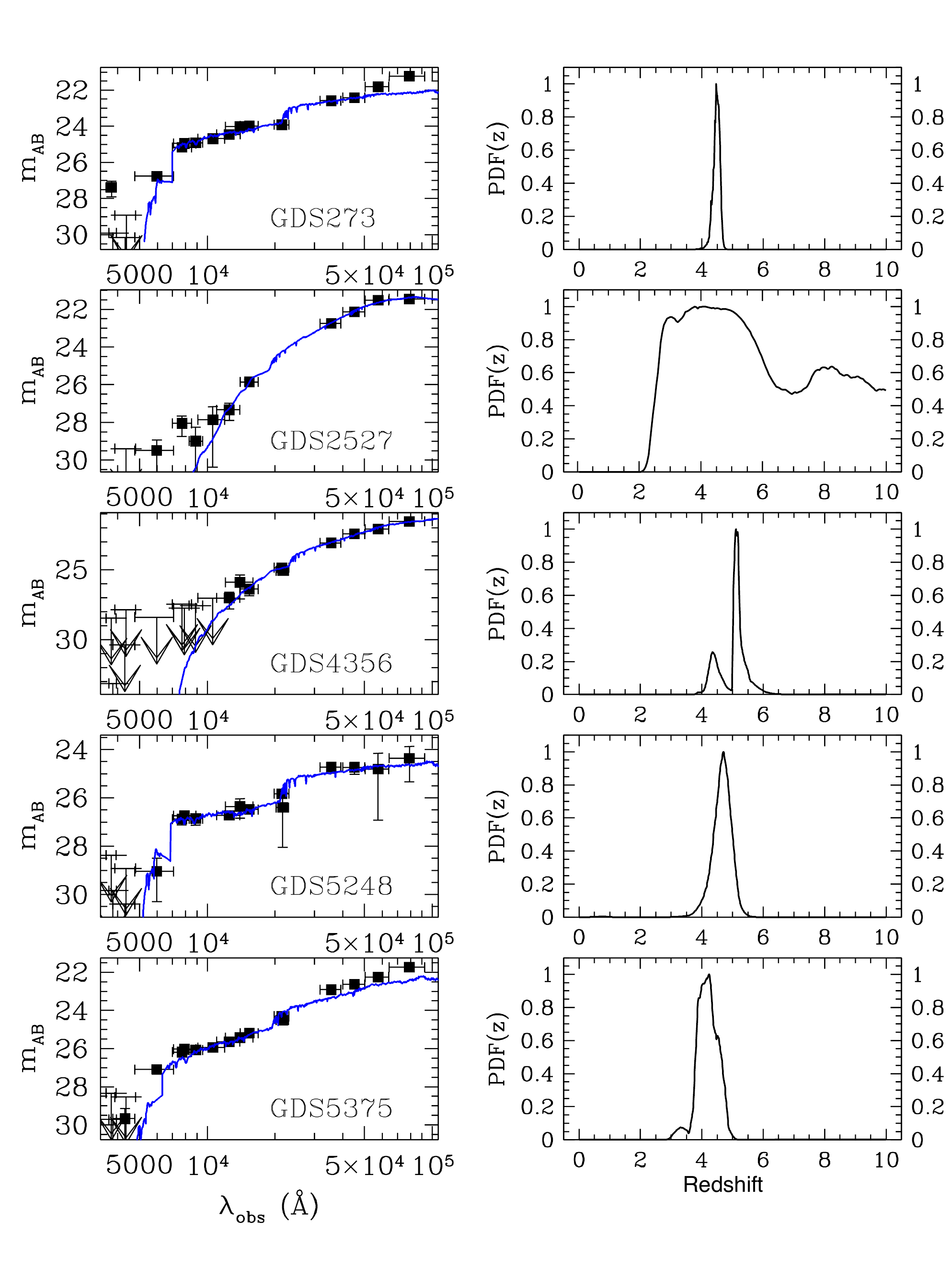}}
\caption{Spectral energy distributions and probability distribution functions of photometric redshifts for all the AGN candidates shown in Table 1. 1-$\sigma$ upper limits are also shown as downward arrows. The blue SEDs represent  the best fit galaxy templates from the \cite{bruzual03} library computed at the photometric redshift for each object  for illustrative purposes only. Spectroscopic redshifts are adopted where available in Table 1.}
\end{figure}

\begin{figure}
\centering
\scalebox{1}[1]{\hspace{-0.5cm}\includegraphics[width=\hsize]{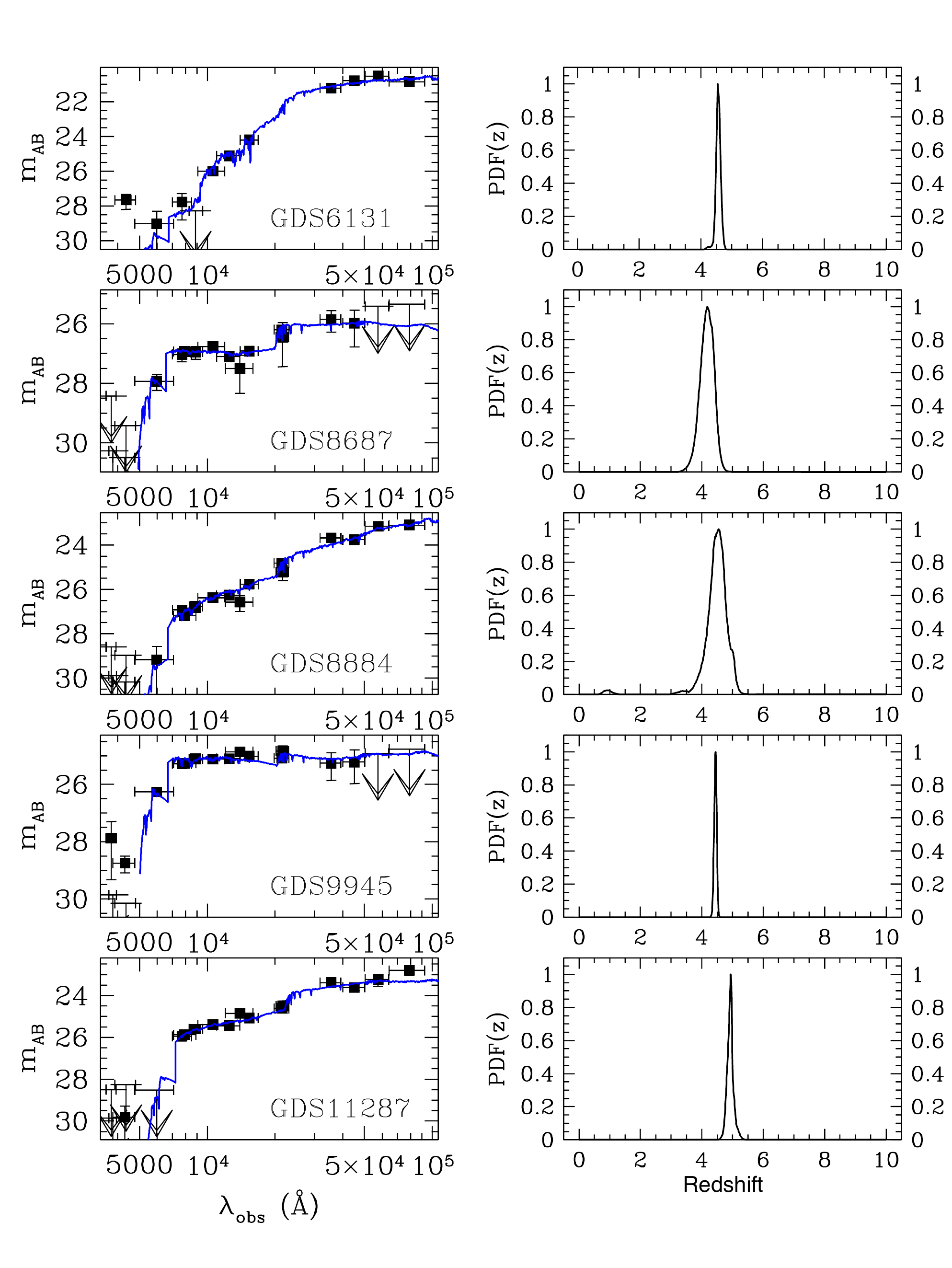}}
Fig. 7 Continued
\end{figure}

\begin{figure}
\centering
\scalebox{1}[1]{\hspace{-0.5cm}\includegraphics[width=\hsize]{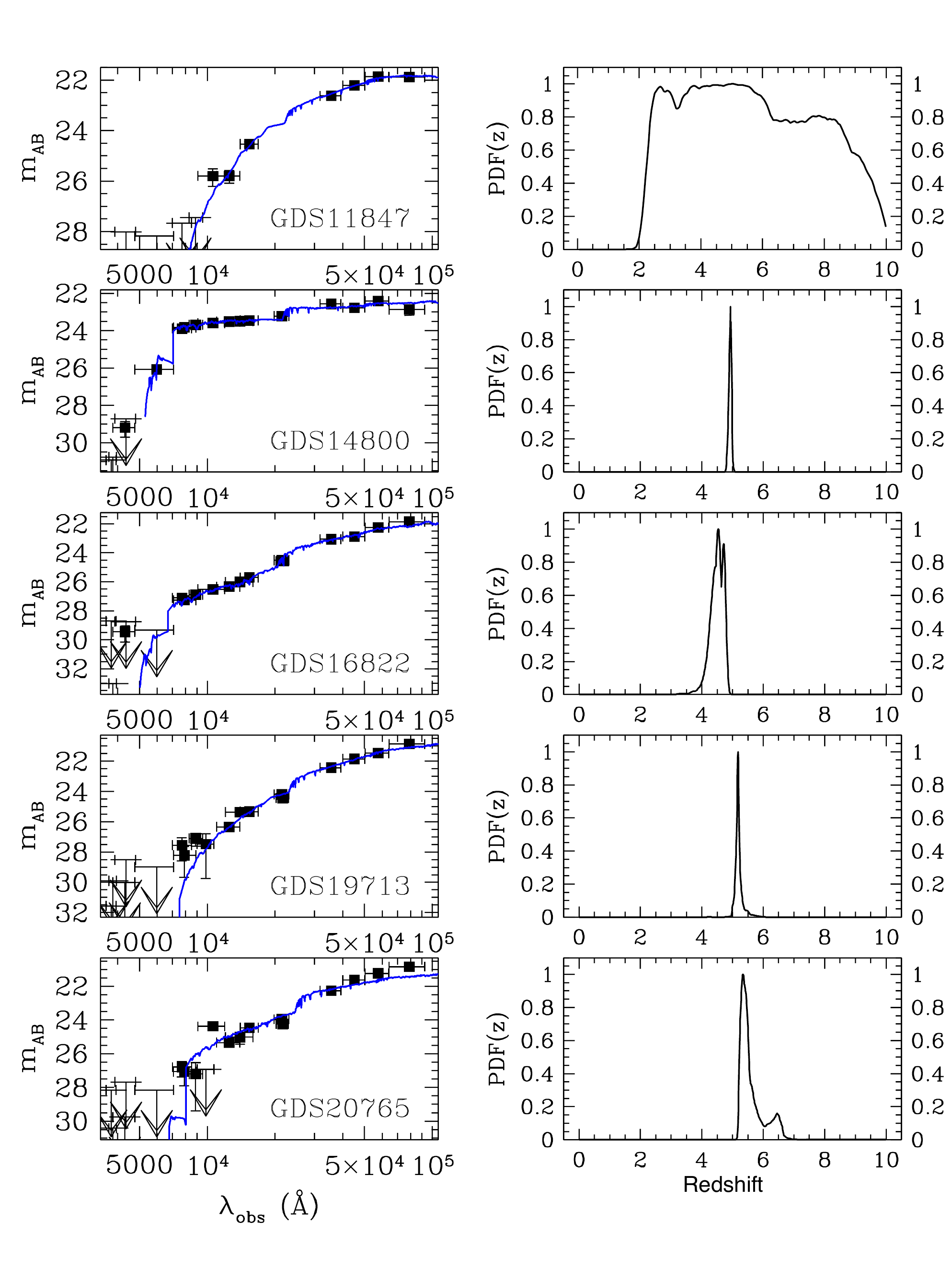}}
Fig. 7 Continued
\end{figure}

\begin{figure}
\centering
\scalebox{1}[1]{\vspace{-1cm}\includegraphics[width=\hsize]{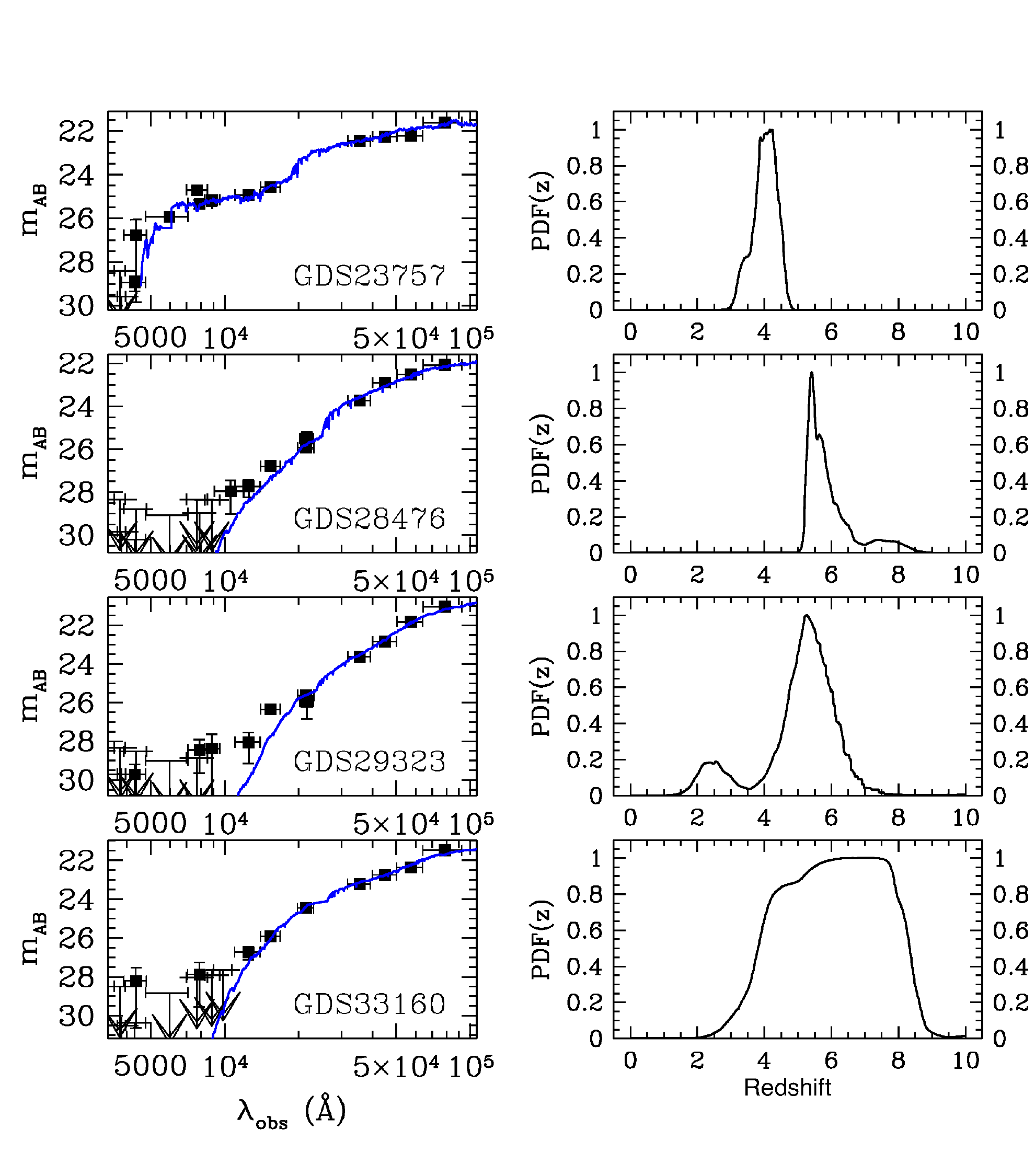}}
Fig. 7 Continued
\end{figure}

\begin{figure}
\centering
\scalebox{1}[1]{\hspace{-0.5cm}\includegraphics[width=\hsize]{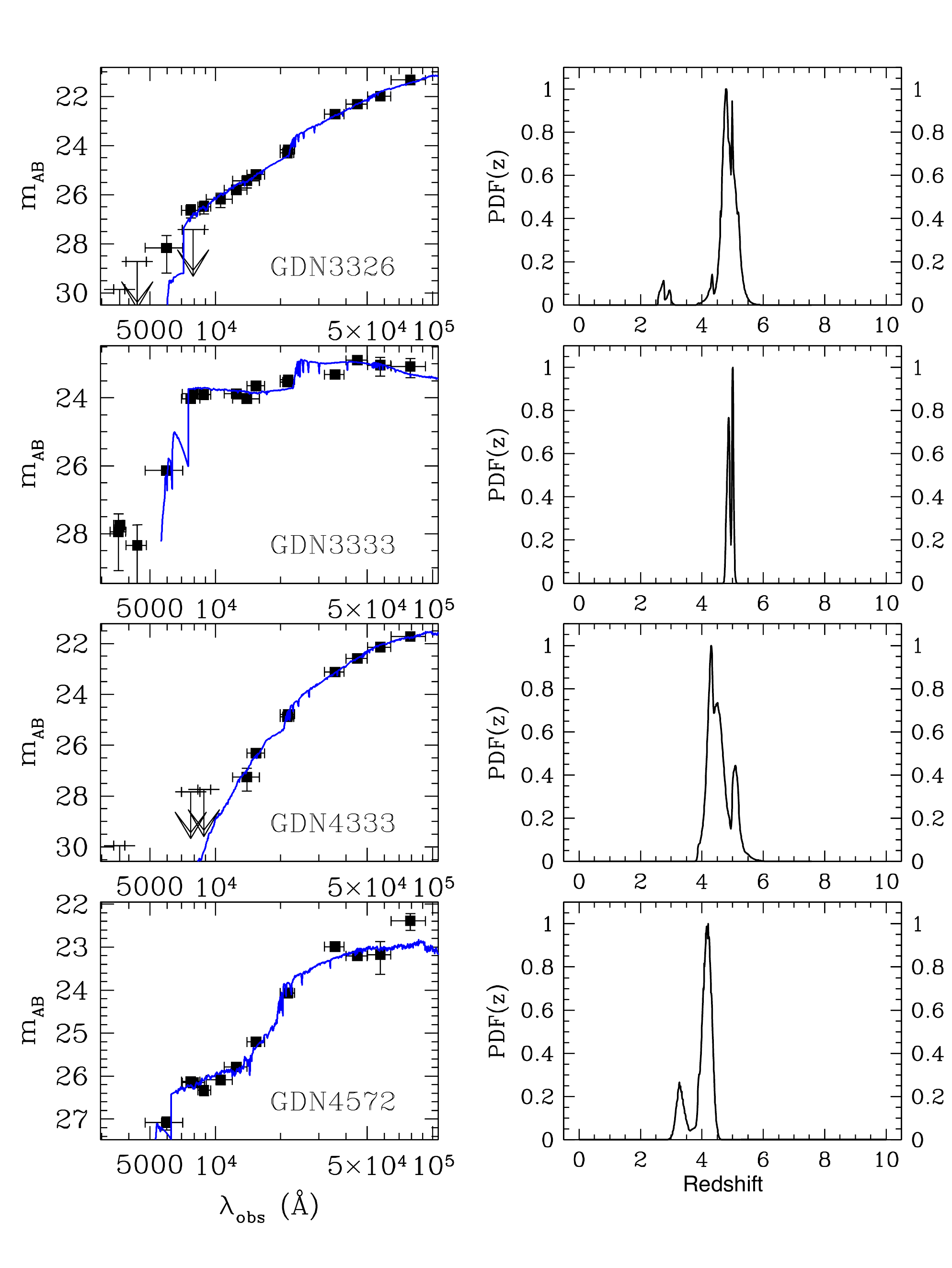}}
Fig. 7 Continued
\end{figure}

\begin{figure}
\centering
\scalebox{1}[1]{\hspace{-0.5cm}\includegraphics[width=\hsize]{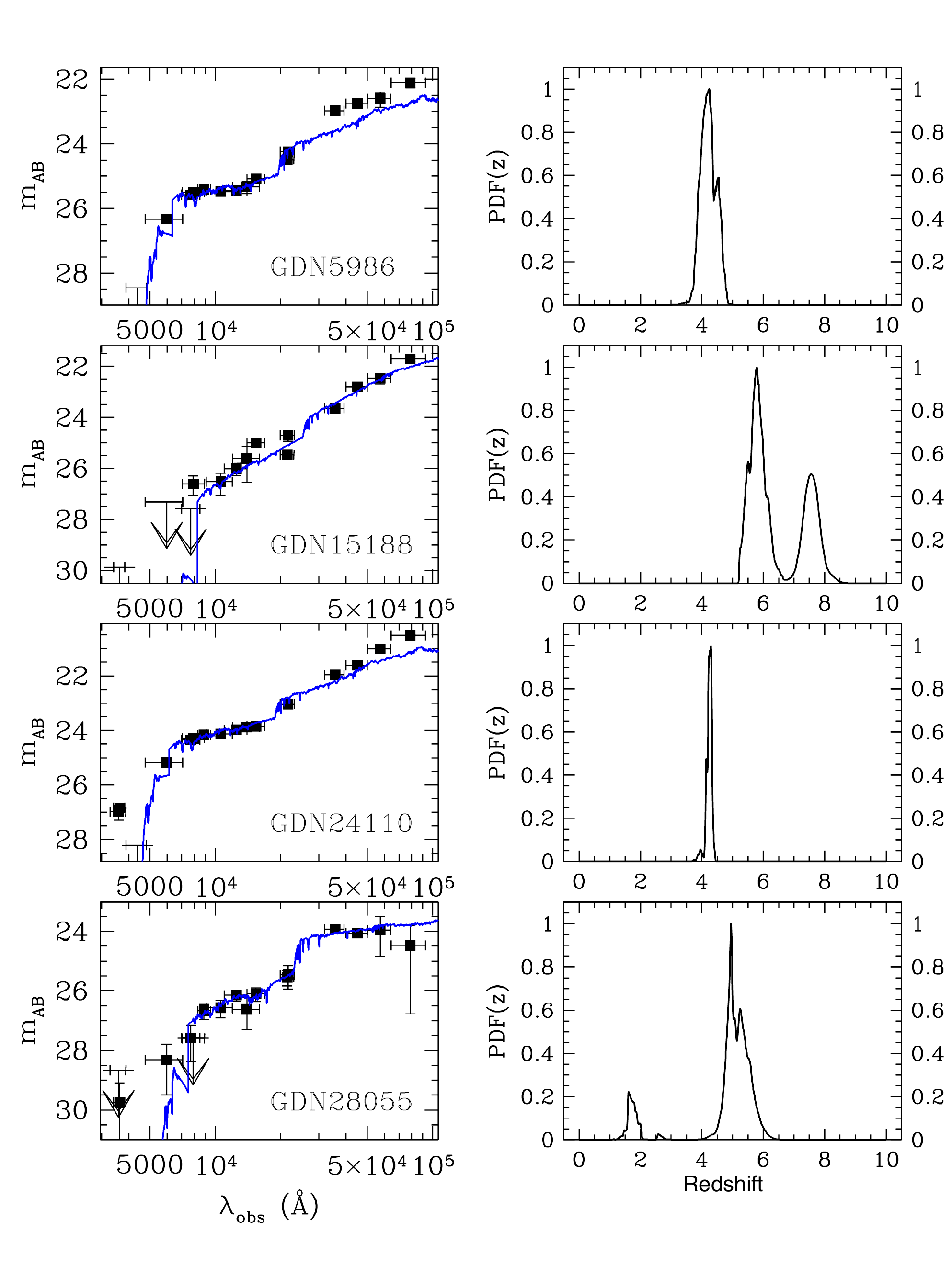}}
Fig. 7 Continued
\end{figure}

\begin{figure}
\centering
\scalebox{1}[1]{\hspace{-0.5cm}\includegraphics[width=\hsize]{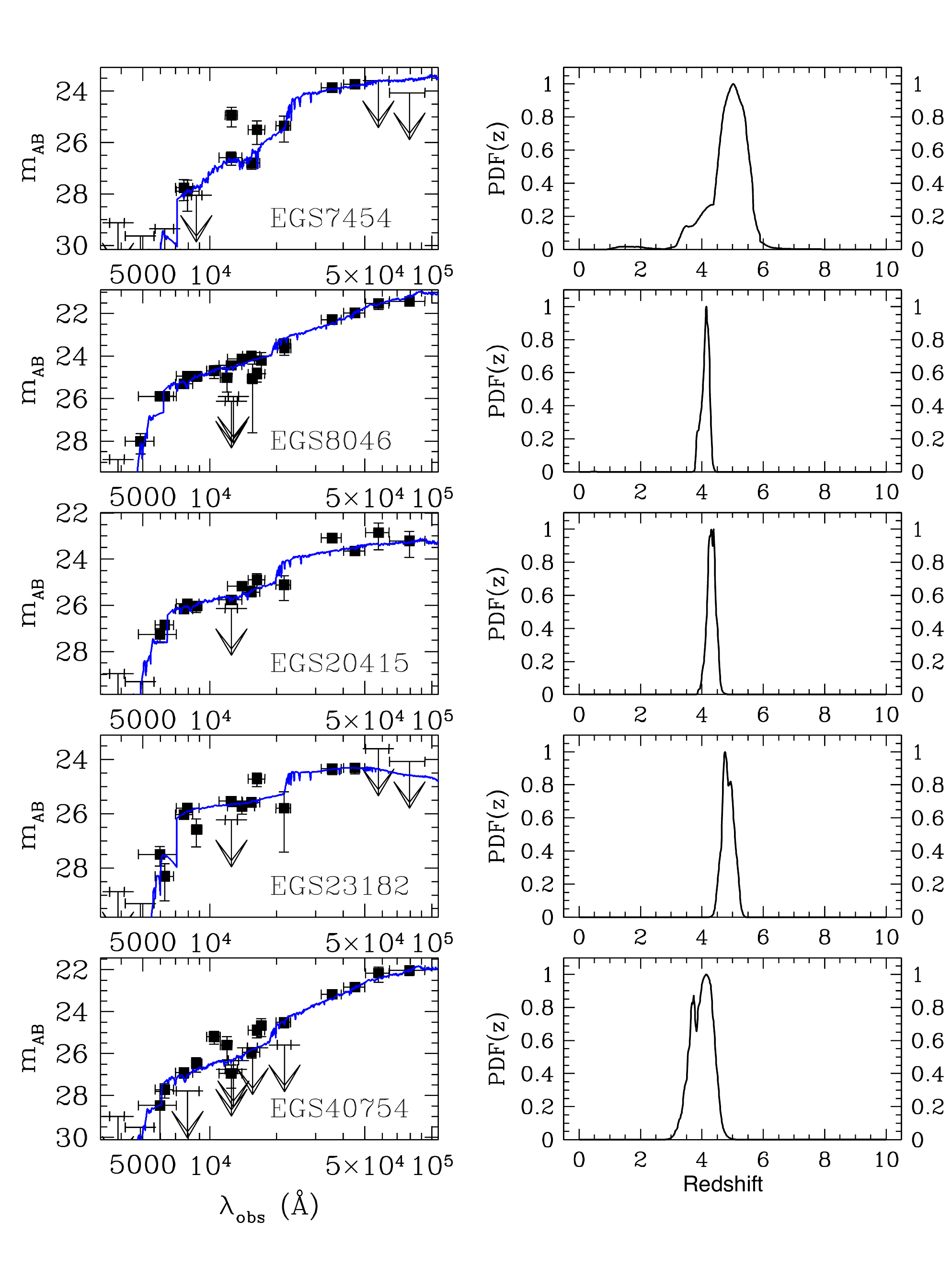}}
Fig. 7 Continued
\end{figure}

\clearpage

\begin{figure}
\centering
\scalebox{1}[1]{\hspace{-0.5cm}\includegraphics[width=\hsize]{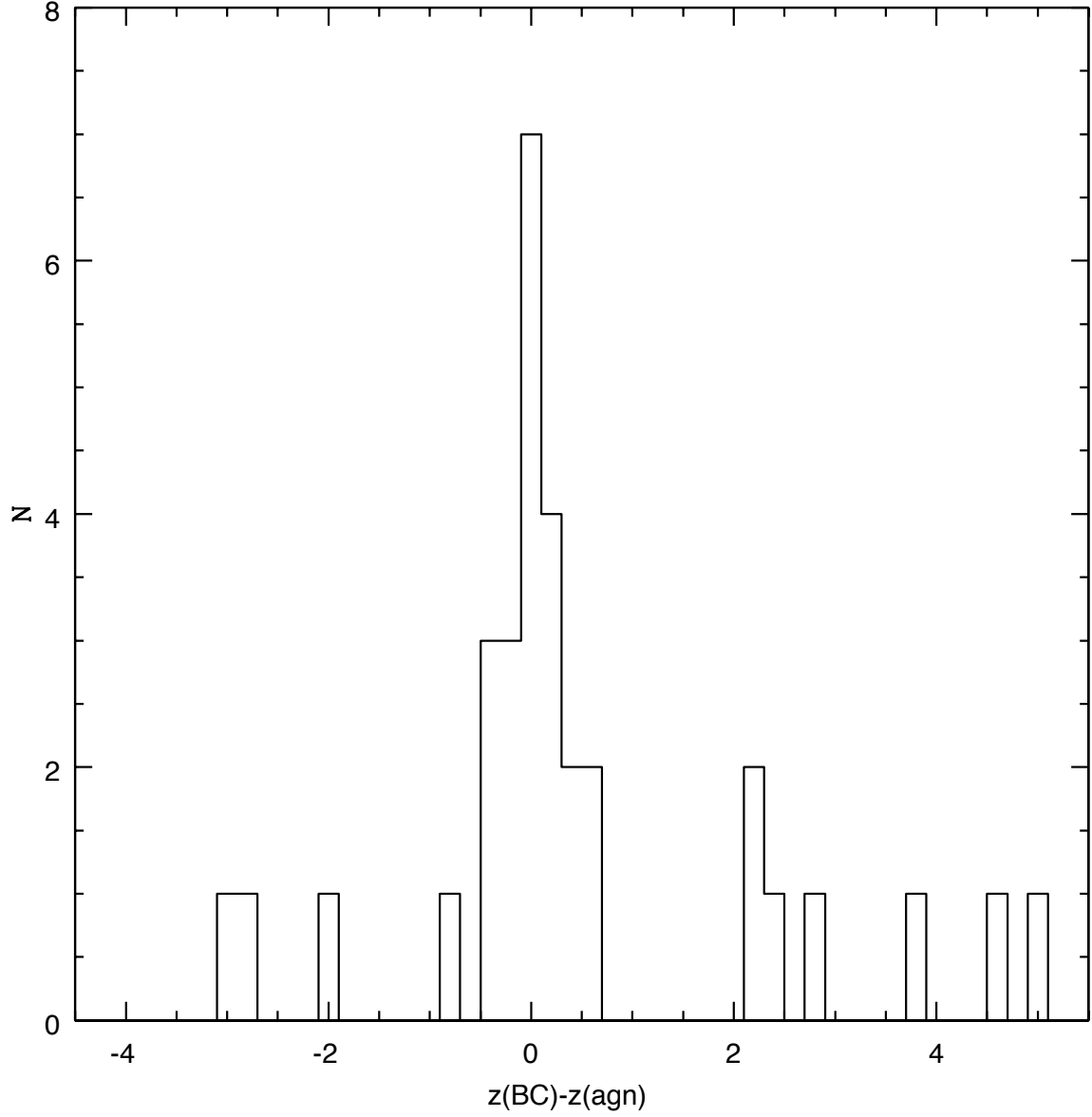}}
\caption{ Histogram of redshift difference between estimates based on galaxy templates (BC) vs. AGN templates. Dust reddening is also included in the libraries.
Small magellanic clouds extinction curves are adopted for AGN templates. The Calzetti law is also added in galaxy templates. }
\end{figure}

\begin{figure}
\centering
{\bf GDS 273}\\
\scalebox{1}[1]{\includegraphics{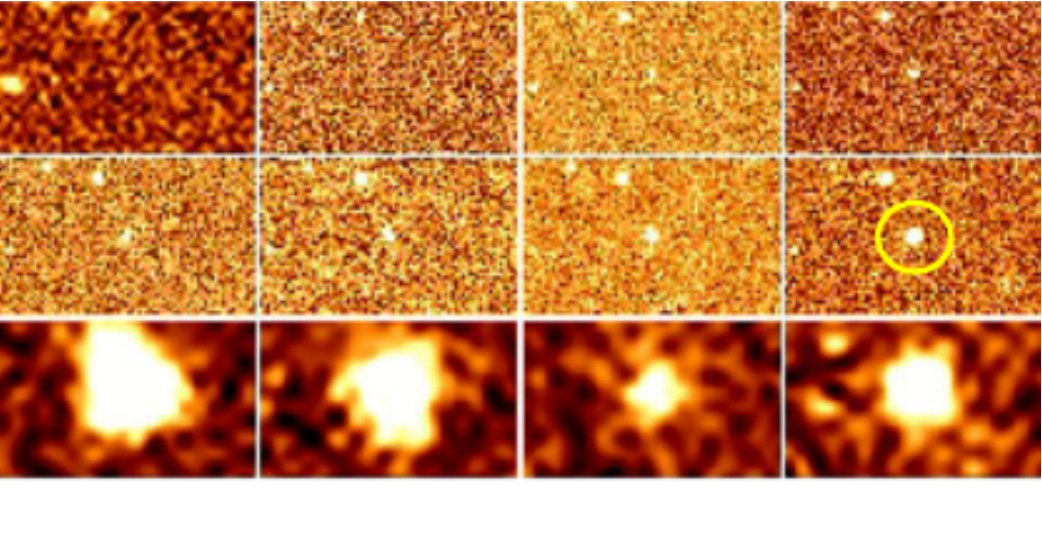}}

{\bf GDS 2527}\\
\scalebox{1}[1]{\includegraphics{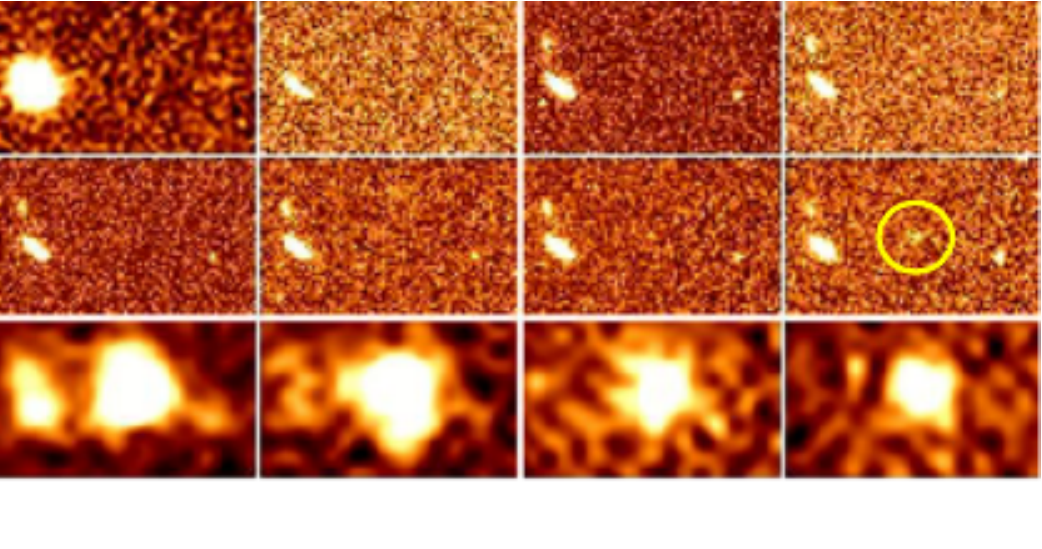}}

{\bf GDS 4356}\\
\scalebox{1}[1]{\includegraphics{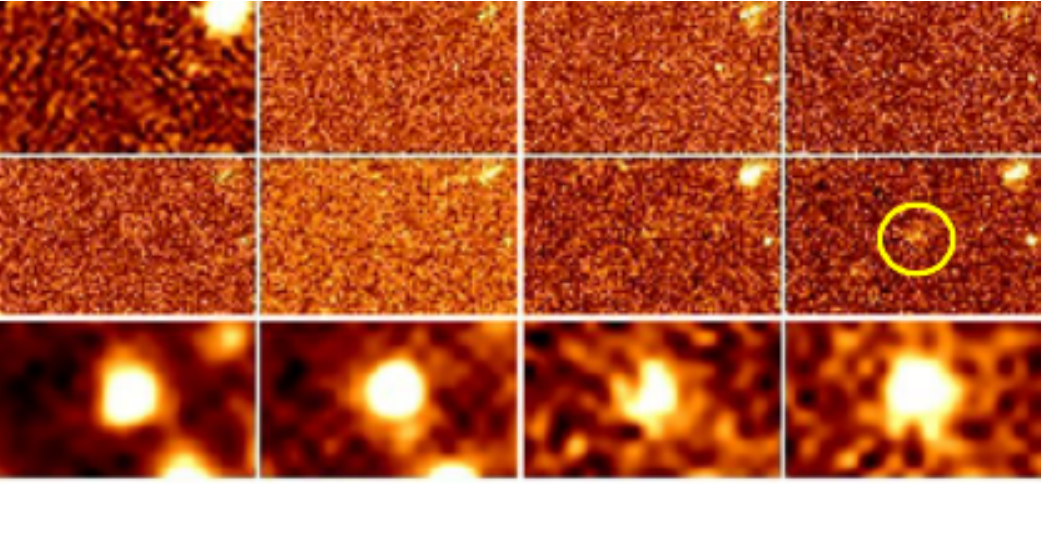}}

{\bf GDS 5248}\\
\scalebox{1}[1]{\includegraphics{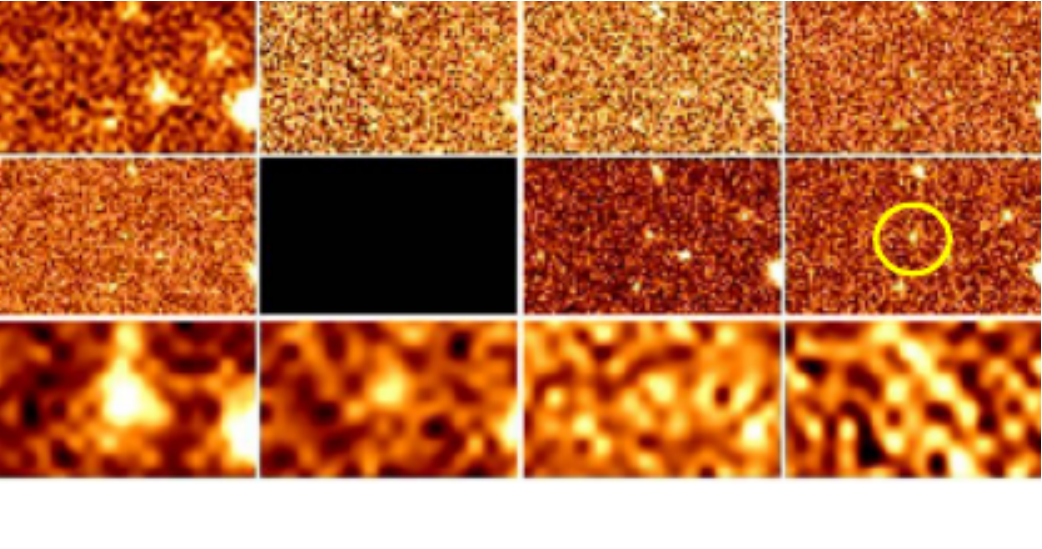}}

\end{figure}

\clearpage

\begin{figure}
\centering
{\bf GDS 5375}\\
\scalebox{1}[1]{\includegraphics{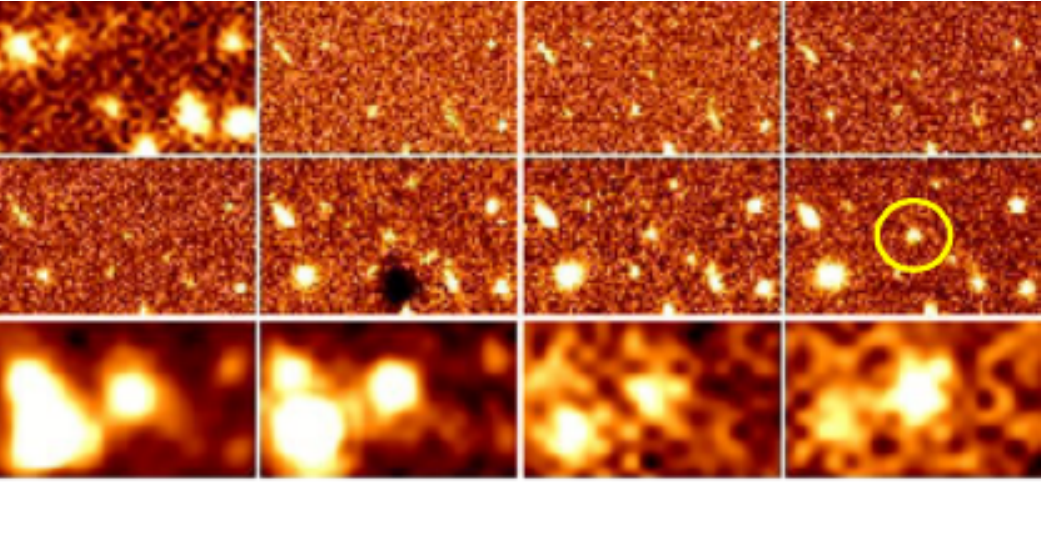}}

{\bf GDS 6131}\\
\scalebox{1}[1]{\includegraphics{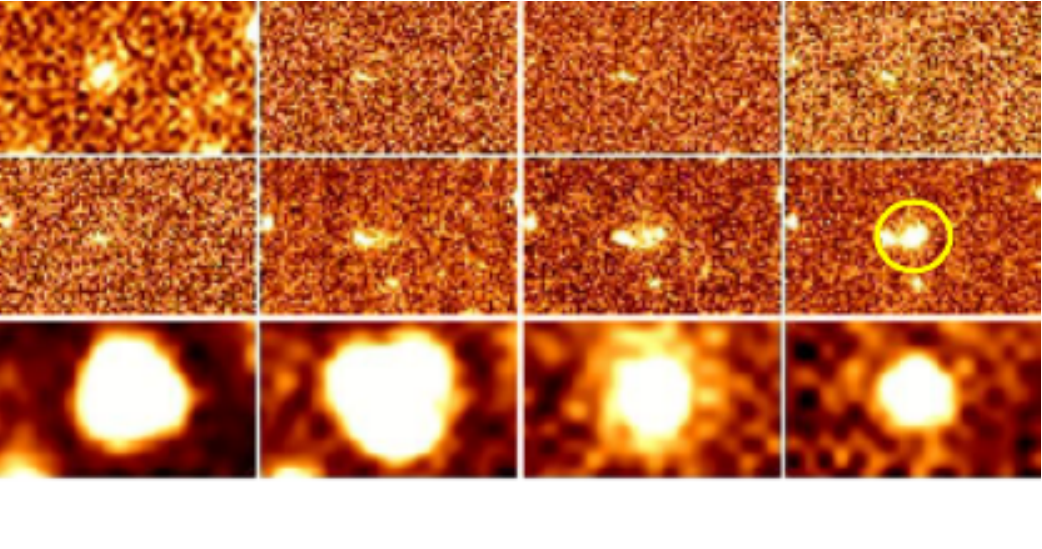}}

{\bf GDS 8687}\\
\scalebox{1}[1]{\includegraphics{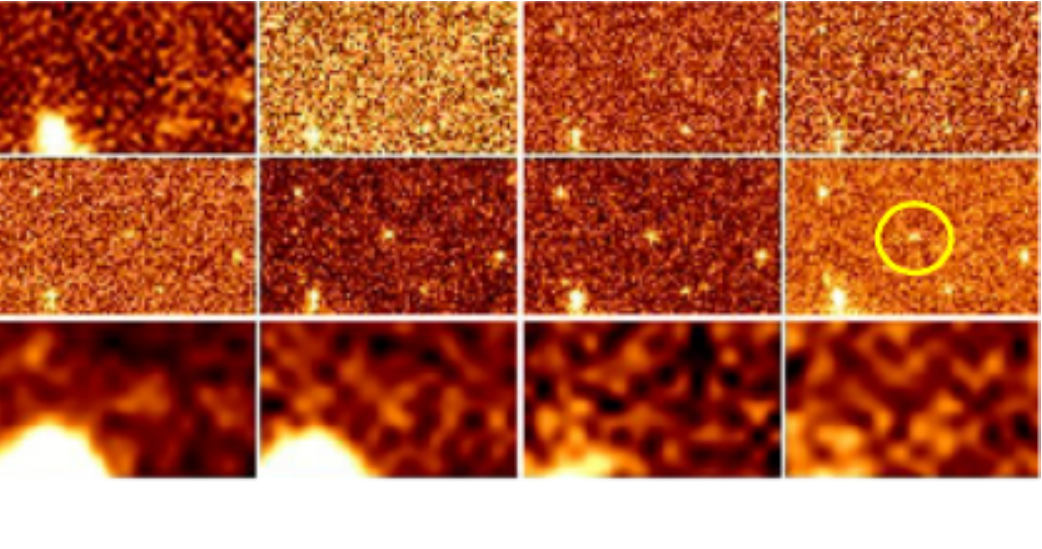}}

{\bf GDS 8884}\\
\scalebox{1}[1]{\includegraphics{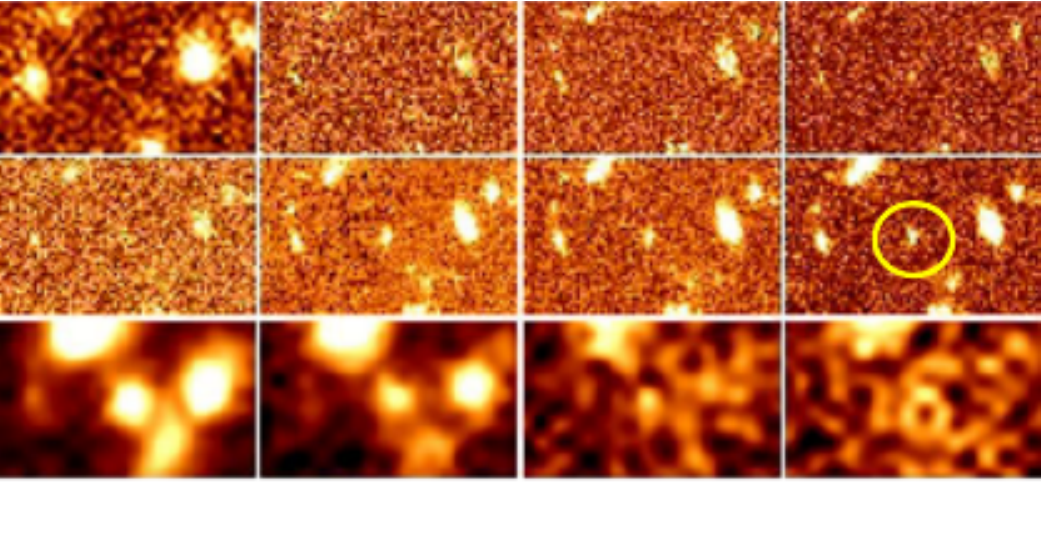}}
\end{figure}

\clearpage

\begin{figure}
\centering
{\bf GDS 9945}\\
\scalebox{1}[1]{\includegraphics{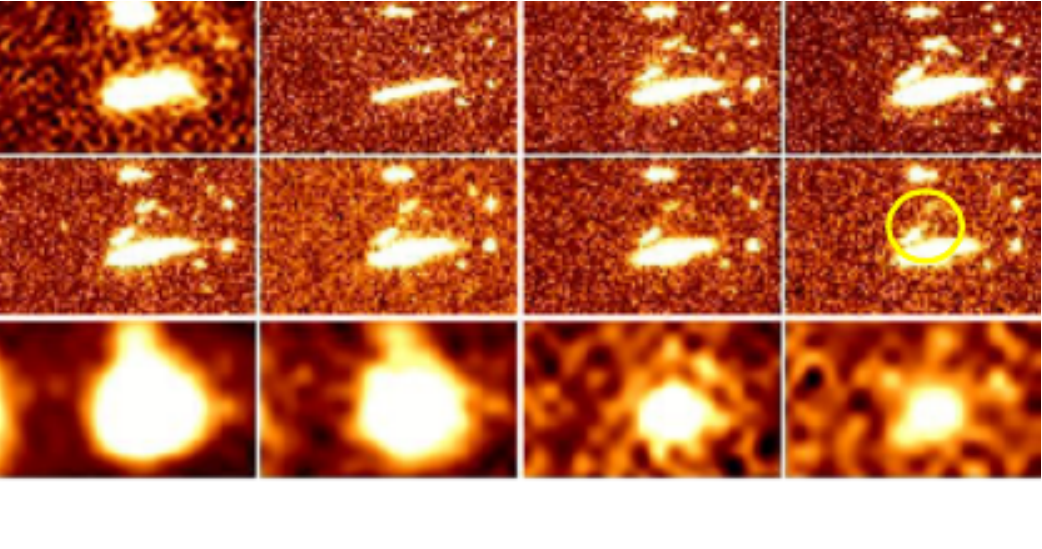}}

{\bf GDS 11287}\\
\scalebox{1}[1]{\includegraphics{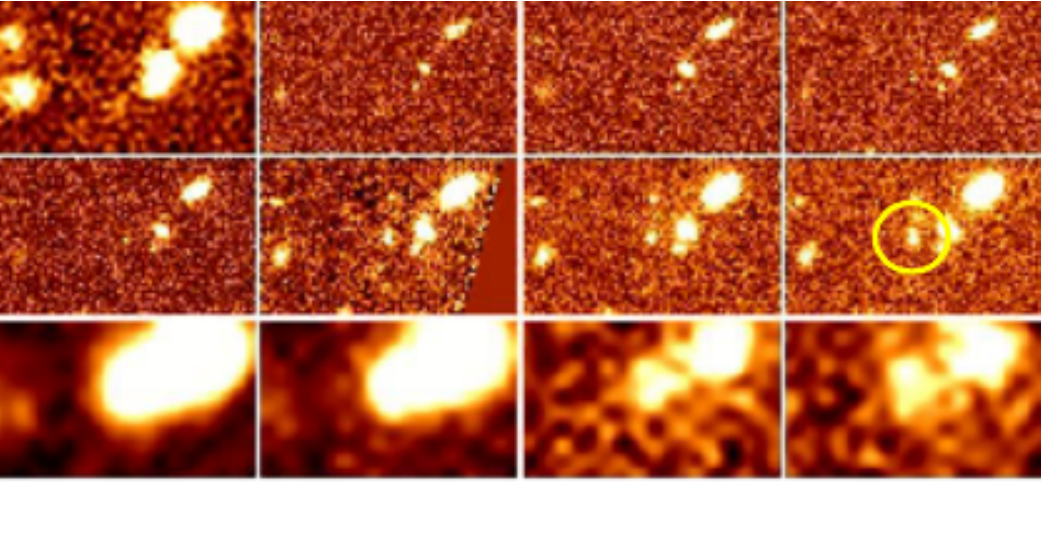}}

{\bf GDS 11847}\\
\scalebox{1}[1]{\includegraphics{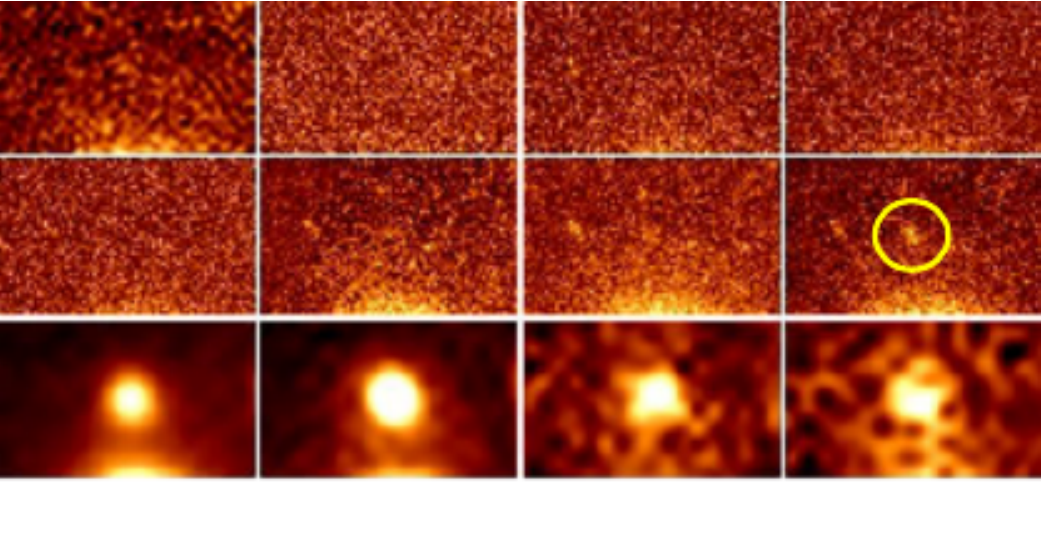}}

{\bf GDS 14800}\\
\scalebox{1}[1]{\includegraphics{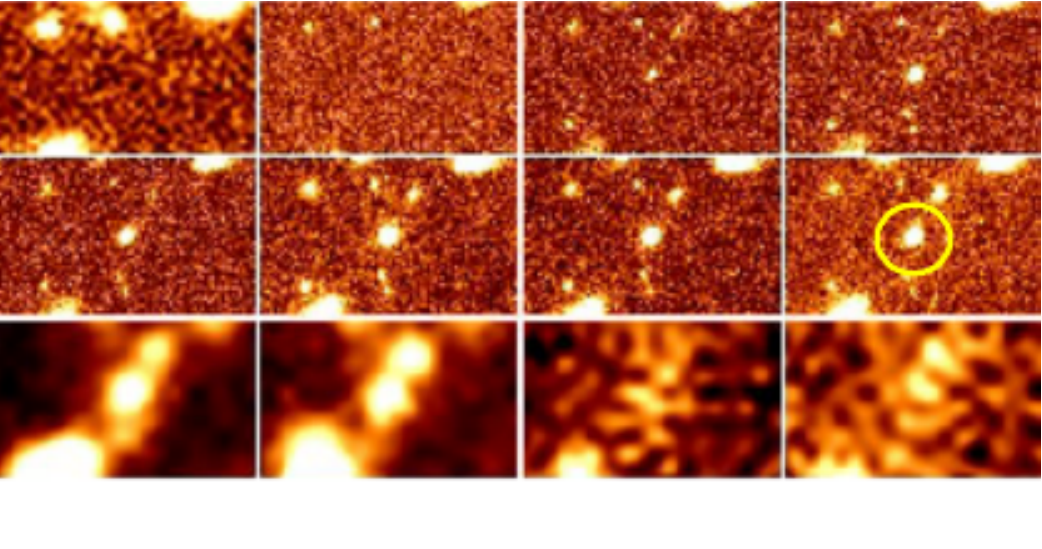}}
\end{figure}

\clearpage

\begin{figure}
\centering
{\bf GDS 16822}\\
\scalebox{1}[1]{\includegraphics{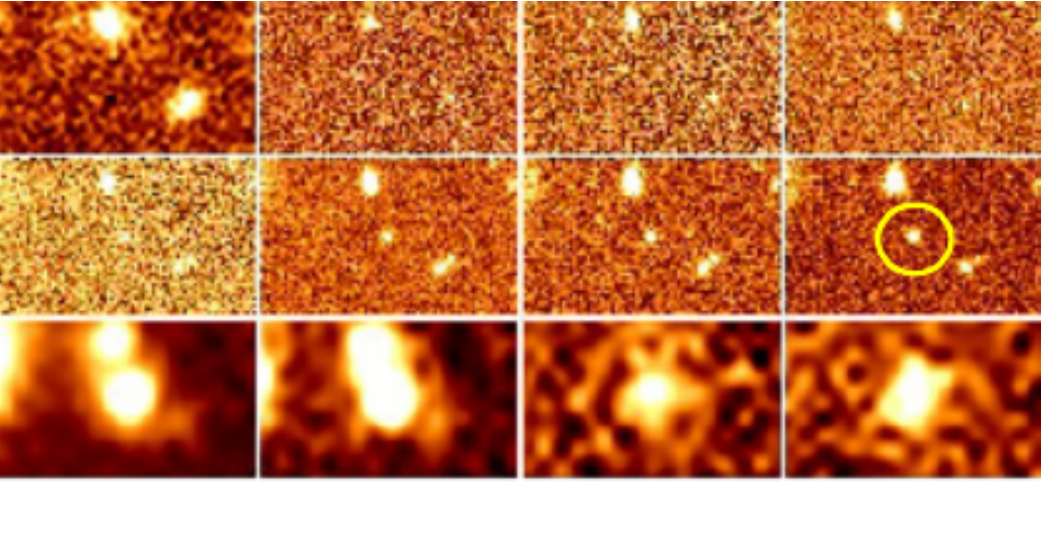}}

{\bf GDS 19713}\\
\scalebox{1}[1]{\includegraphics{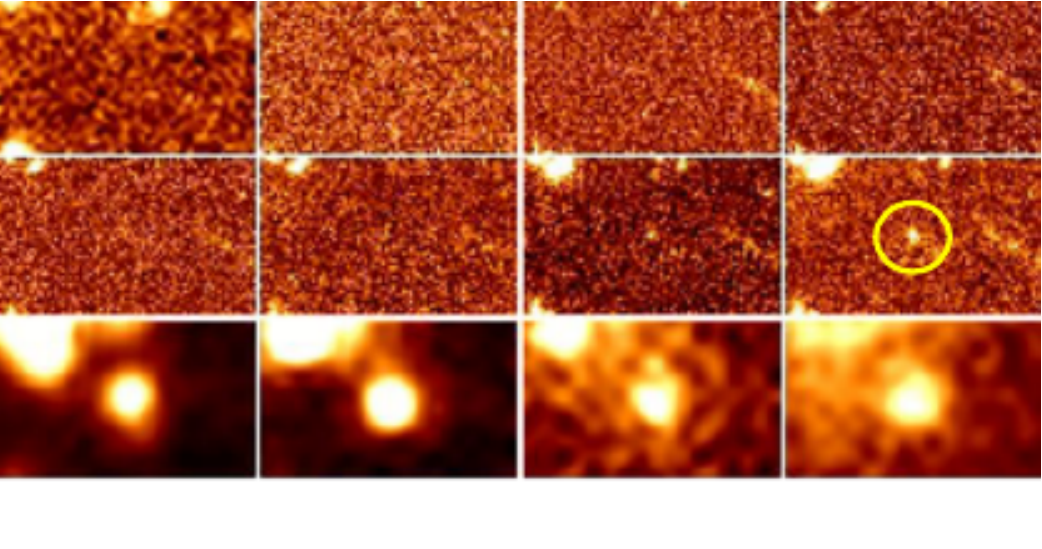}}

{\bf GDS 20765}\\
\scalebox{1}[1]{\includegraphics{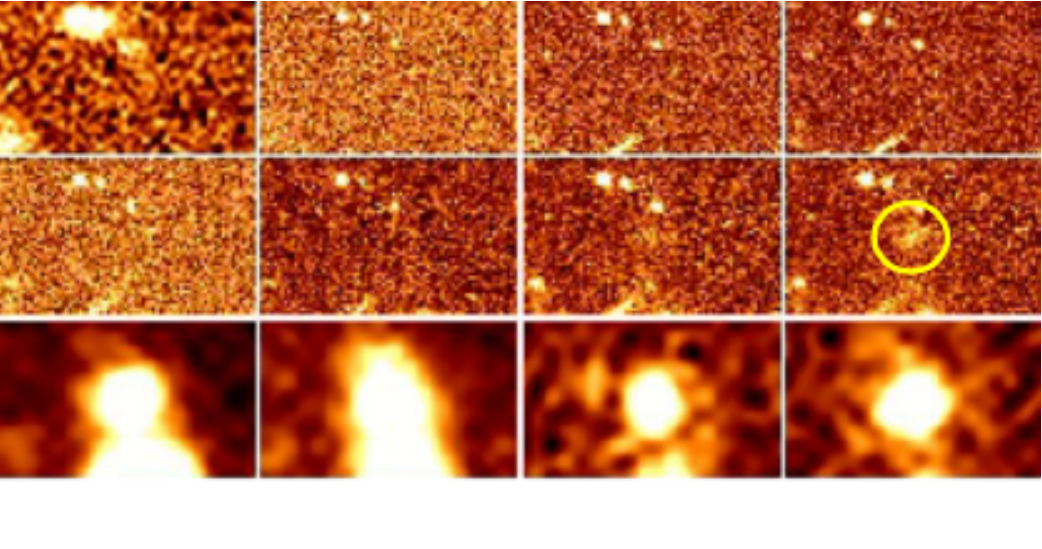}}

{\bf GDS 23757}\\
\scalebox{1}[1]{\includegraphics{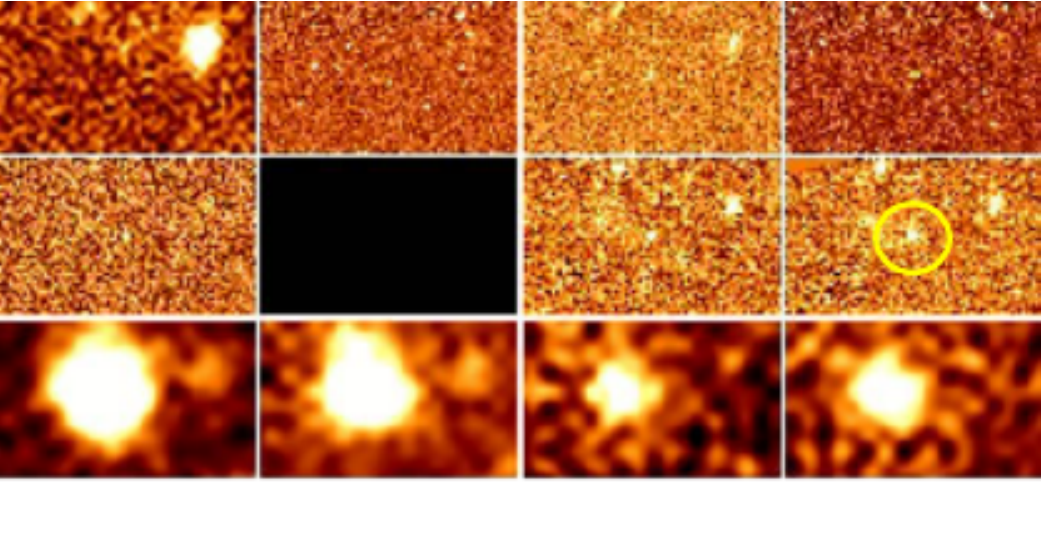}}
\end{figure}

\clearpage

\begin{figure}
\centering
{\bf GDS 28476}\\
\scalebox{1}[1]{\includegraphics{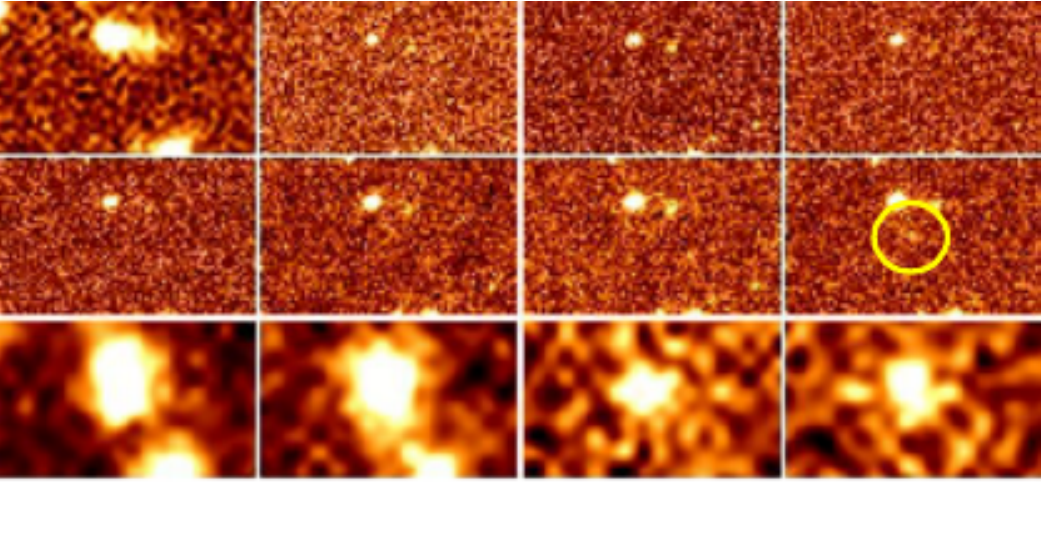}}

{\bf GDS 29323}\\
\scalebox{1}[1]{\includegraphics{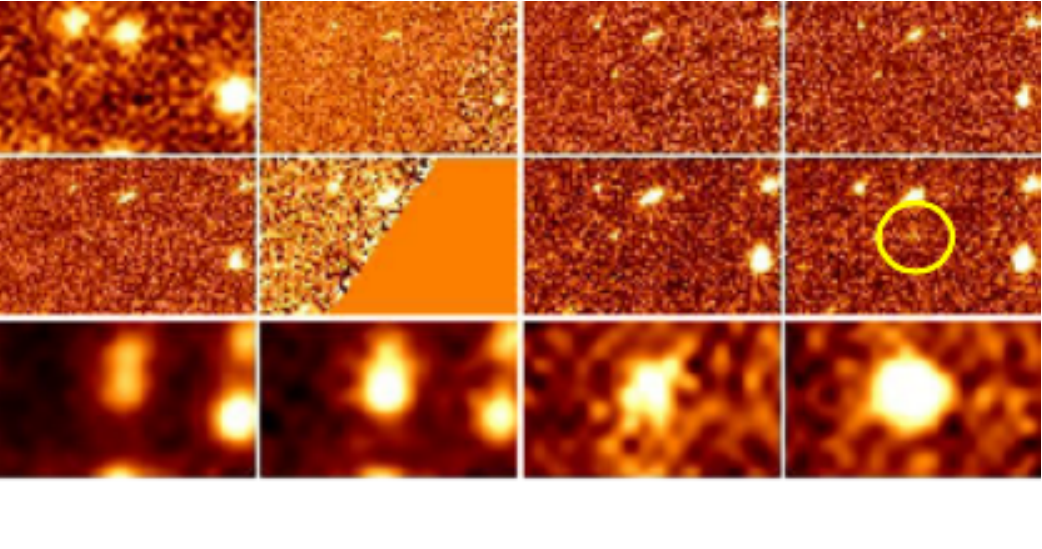}}

{\bf GDS 33160}\\
\scalebox{1}[1]{\includegraphics{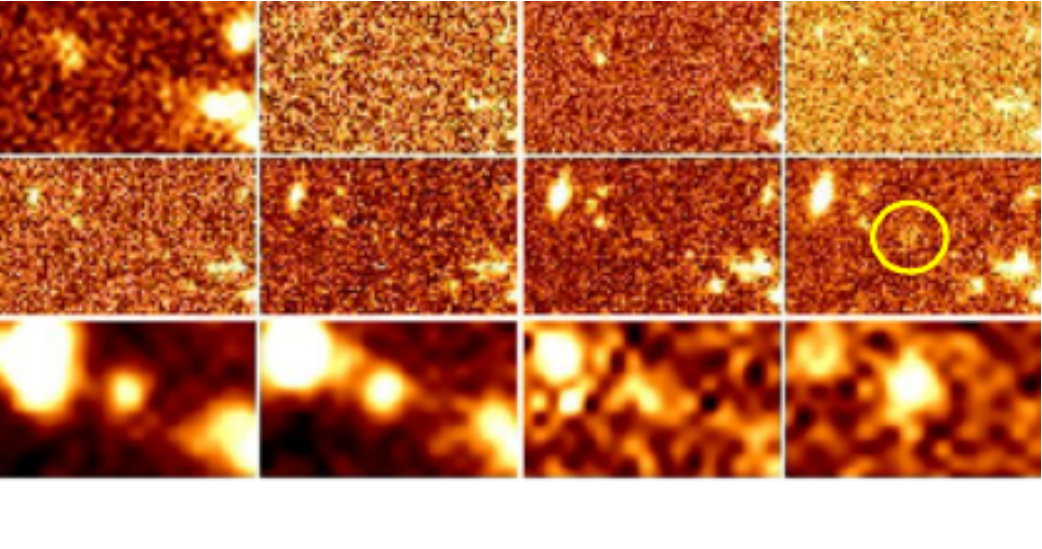}}
\caption{Multiwavelength UV-NIR distribution of all the GOODS-South candidates. From top left to bottom right the VIMOS U, HST (B,V,I,Z,Y,J,H), IRAC (3.6,4.5,5.8,8 $\mu$m)  images are shown. The sizes are $\sim 9\times 6$ arcsec$^2$. The targets are in  the center of the circle in the H-band image.}
\end{figure}

\clearpage

\begin{figure}
\centering

{\bf GDN 3326}\\
\scalebox{1}[1]{\includegraphics[trim=0 0 0 0]{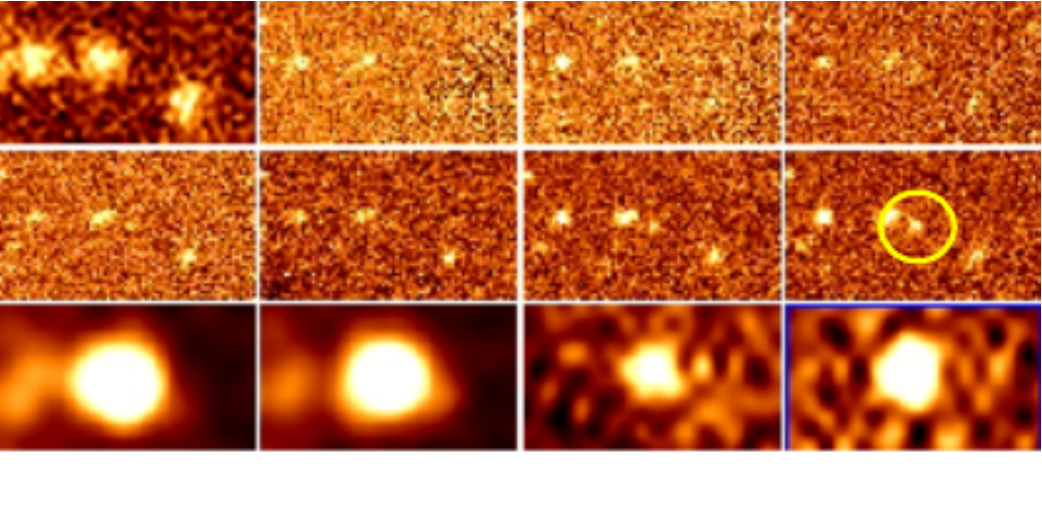}}

{\bf GDN 3333}\\
\scalebox{1}[1]{\includegraphics{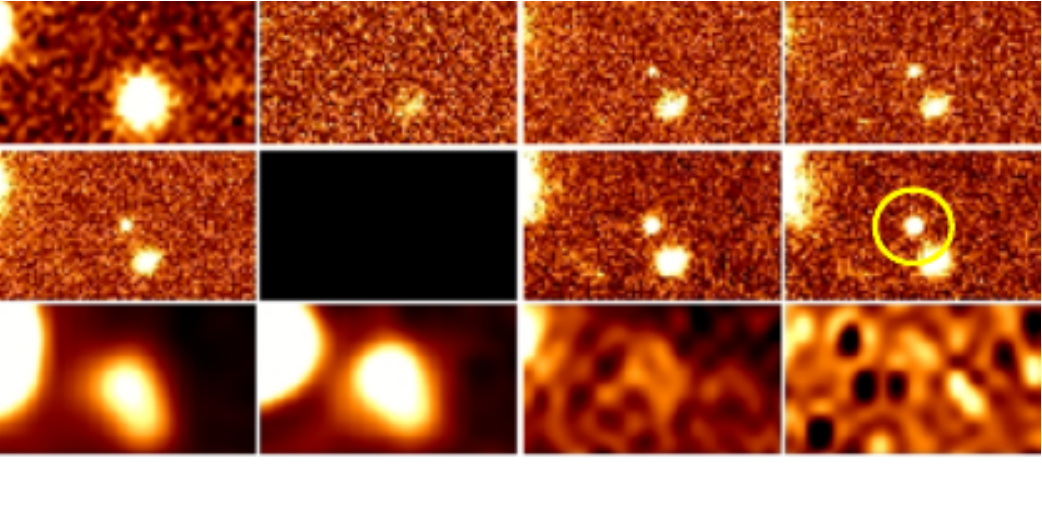}}

{\bf GDN 4333}\\
\scalebox{1}[1]{\includegraphics{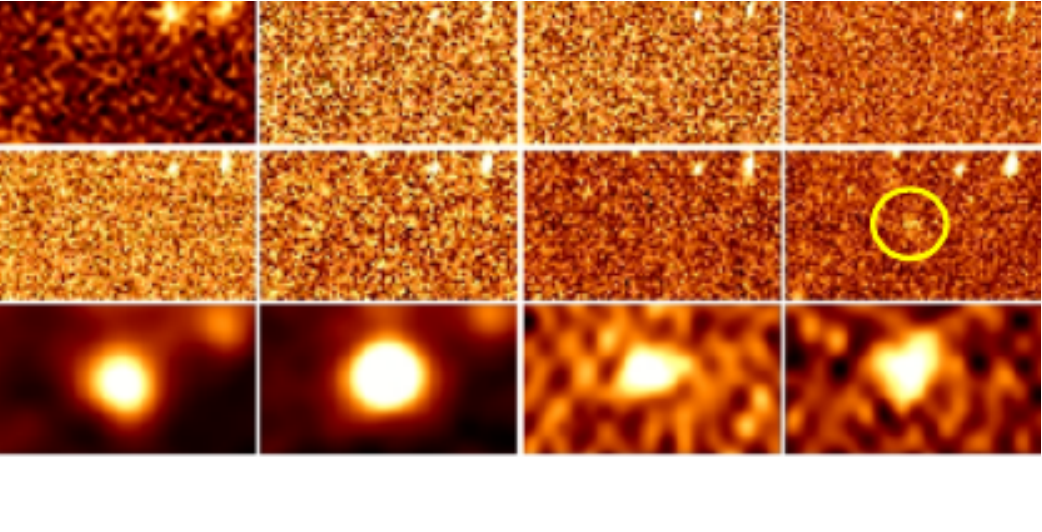}}

{\bf GDN 4572}\\
\scalebox{1}[1]{\includegraphics{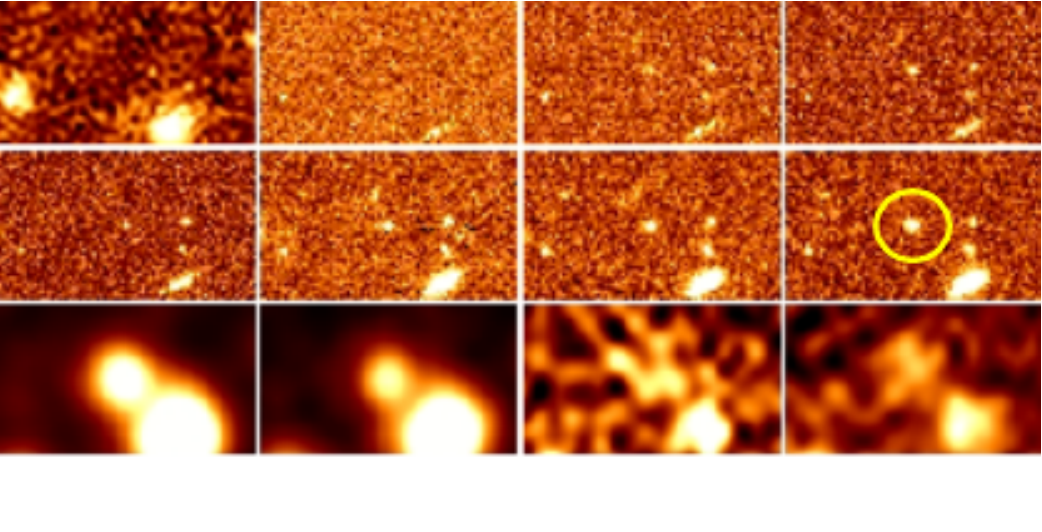}}
\end{figure}

\clearpage 

\begin{figure}
\centering

{\bf GDN 5986}\\
\scalebox{1}[1]{\includegraphics{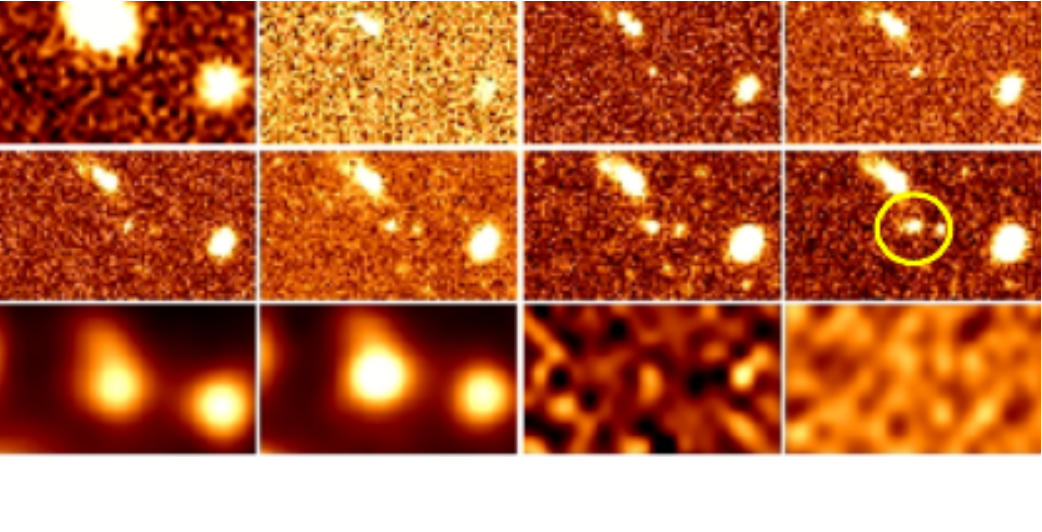}}

{\bf GDN 15188}\\
\scalebox{1}[1]{\includegraphics[trim=0 -0.8cm 0 0]{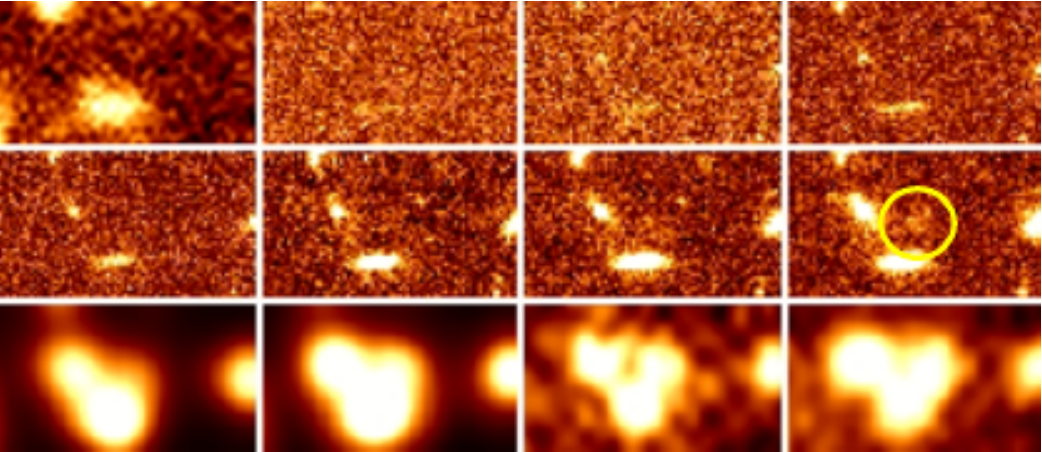}}

{\bf GDN 24110}\\
\scalebox{1}[1]{\includegraphics{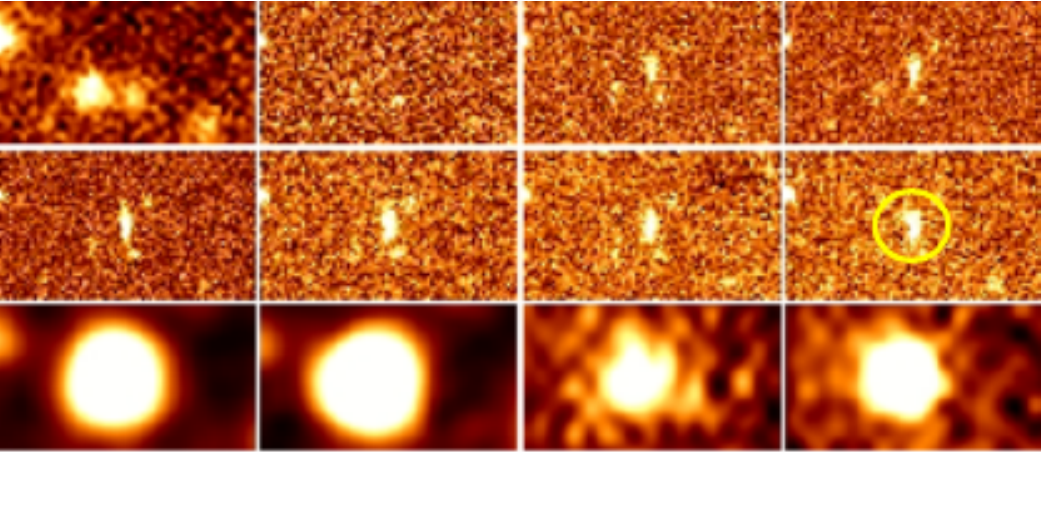}}

{\bf GDN 28055}\\
\scalebox{1}[1]{\includegraphics{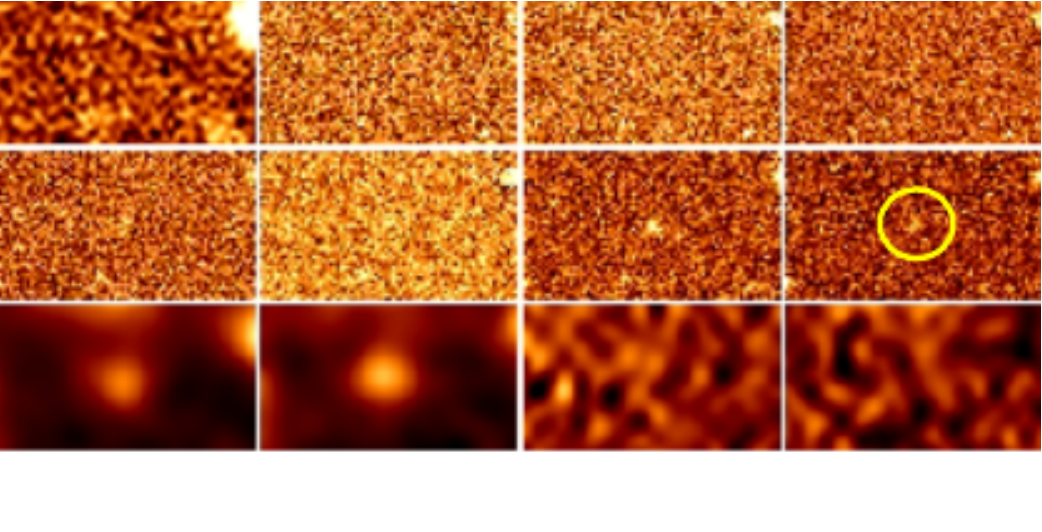}}
\caption{Multiwavelength UV-NIR distribution of all the GOODS-North candidates. From top left to bottom right the LBT U, HST (B,V,I,Z,Y,J,H), IRAC (3.6,4.5,5.8,8 $\mu$)  images are shown. The sizes are $\sim 9\times 6$ arcsec$^2$. The targets are in  the center of the circle in the H-band image.}
\end{figure}

\clearpage

\begin{figure}
\centering

{\bf EGS 7454}\\
\scalebox{1}[1]{\includegraphics{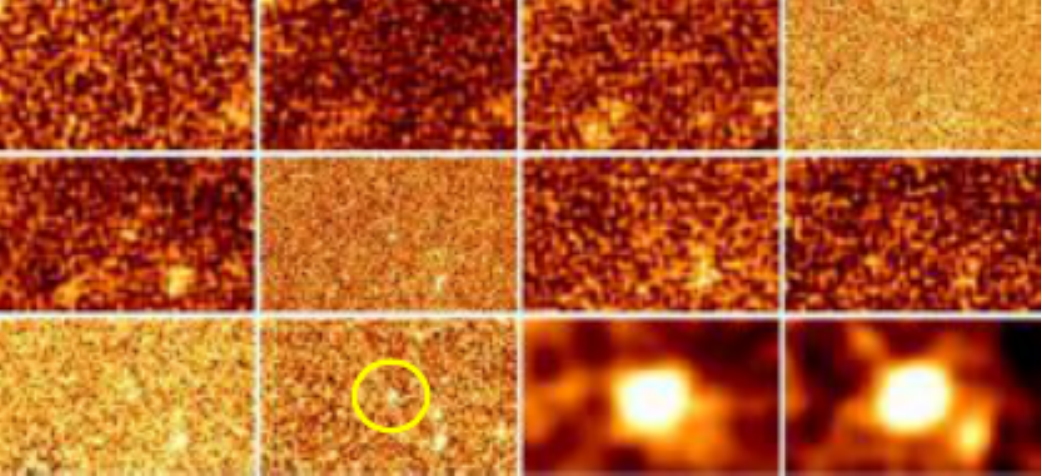}}

{\bf EGS 8046}\\
\scalebox{1}[1]{\includegraphics{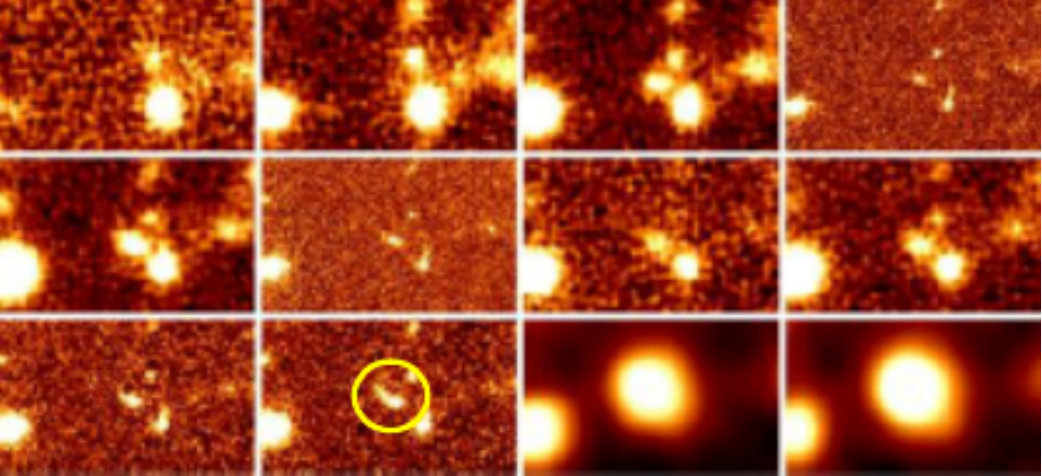}}

{\bf EGS 20415}\\
\scalebox{1}[1]{\includegraphics{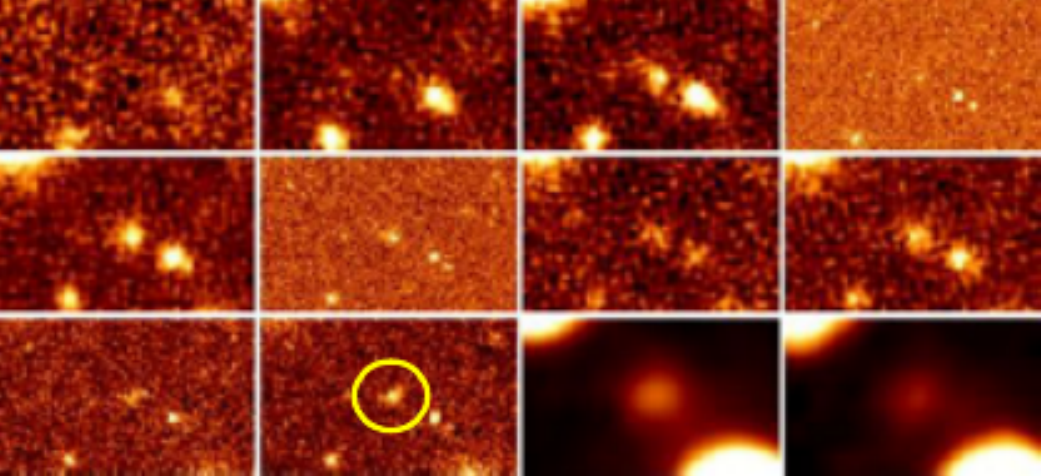}}

{\bf EGS 23182}\\
\scalebox{1}[1]{\includegraphics{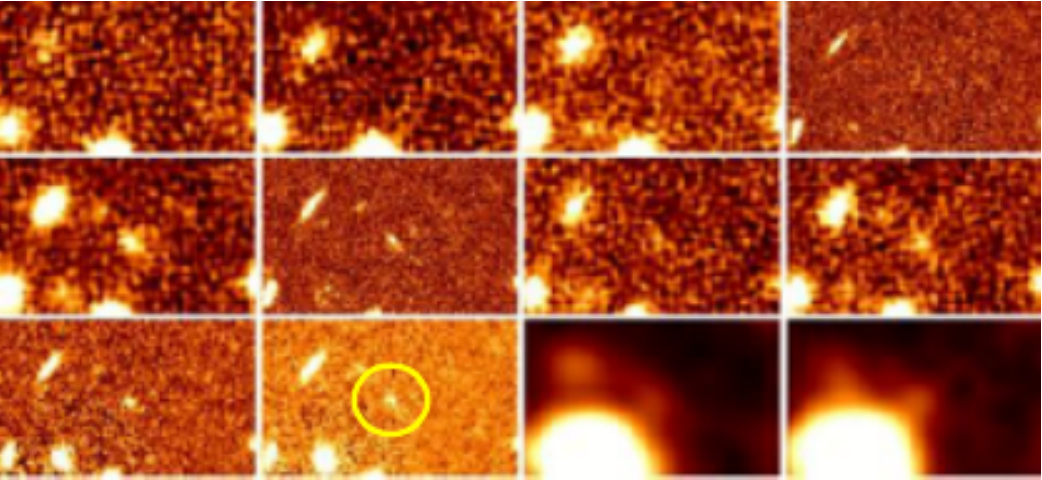}}

\end{figure}

\clearpage

\begin{figure}
\centering
{\bf EGS 40754}\\
\scalebox{1}[1]{\includegraphics{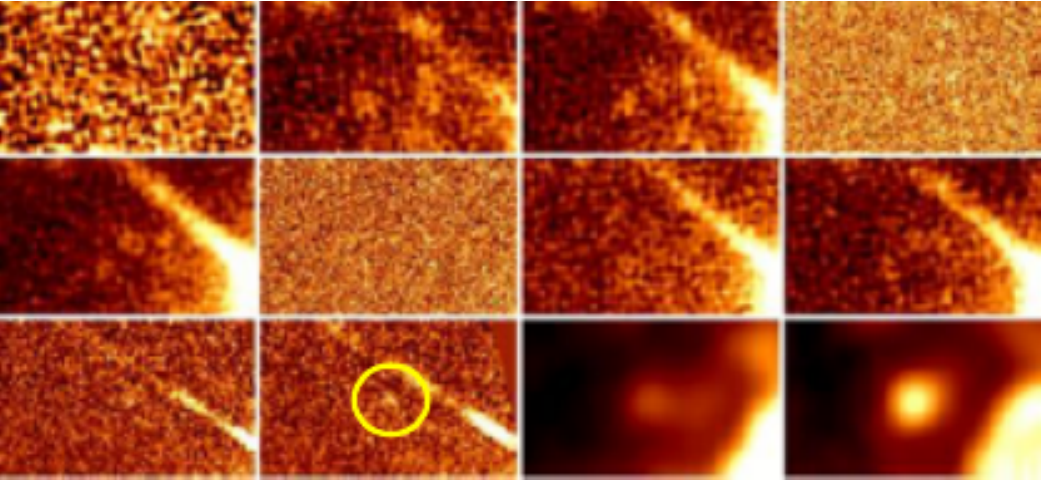}}
\caption{Multiwavelength UV-NIR distribution of all the EGS candidates. From top left to bottom right the CFHT (U,G,R), HST V606, CFHT I, HST I814, CFHT (Z,Y) HST (J,H),
IRAC (3.6, 4.5 $\mu$m)  images are shown. The sizes are $\sim 9\times 6$ arcsec$^2$. The targets are in  the center of the circle in the H-band image.}
\end{figure}

\clearpage

\begin{figure}
\centering
\scalebox{0.7}[0.7]{\includegraphics[width=\hsize]{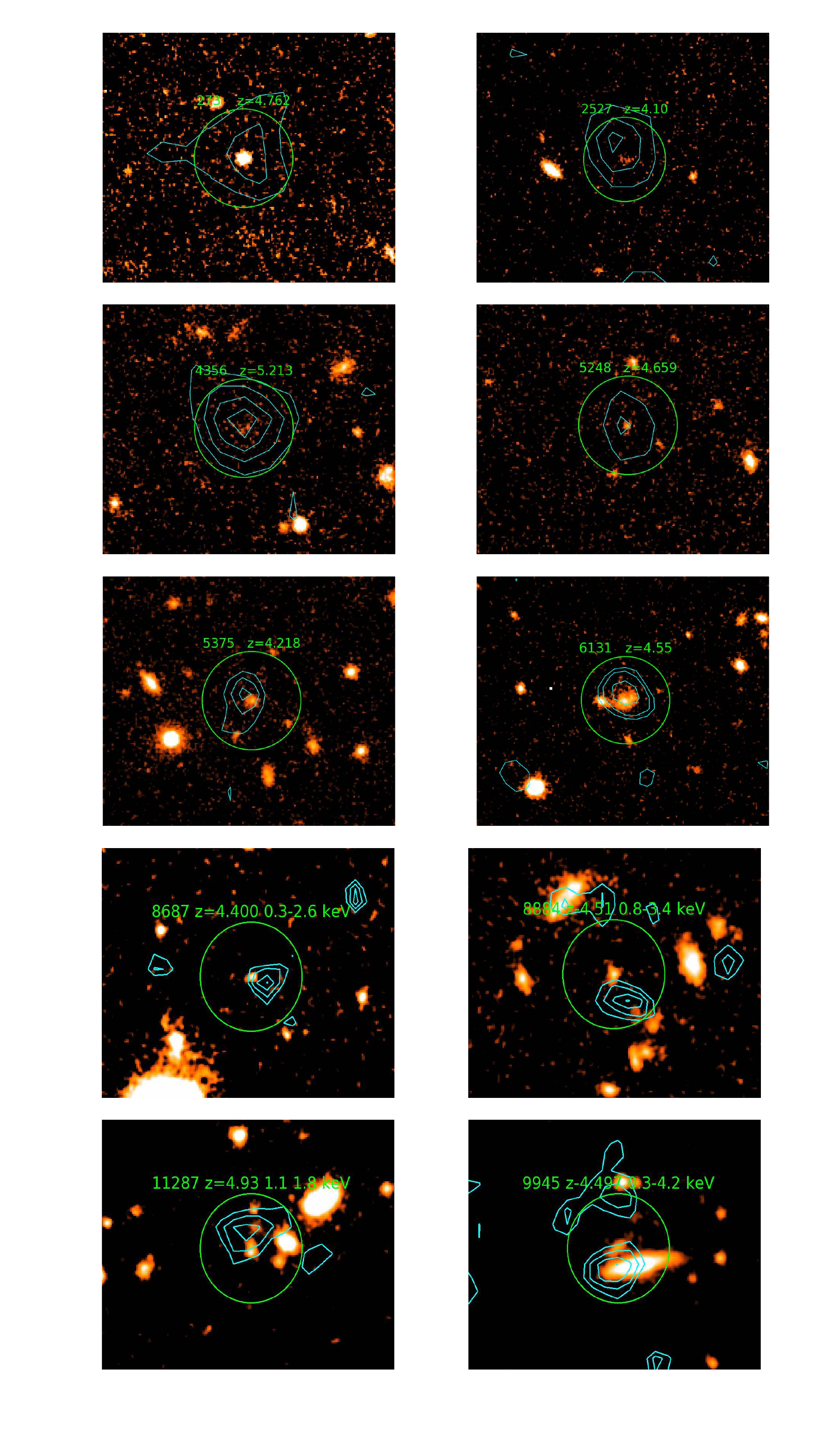}}
\caption{X-ray contours of the AGN candidates in the GOODS-South field overlaid with the HST H-band image. The  AGN candidate is at the center of the H-band image. The green circle radius is 2 arcsec. The X-ray energy bands adopted for the detection are  also shown  for the new sources. For the other known sources the energy band is 0.5-2 keV. Intensity contours are on a linear scale with a factor 2 of dynamical range for the new sources. For GDS273, GDS2527, GDS4356, GDS5248, GDS5375, GDS14800 GDS19713, GDS20765 the intensity contours are with a factor 3 of dynamical range. For GDS 6131, GDS16822, GDS29323 we adopted intensity contours with a square root scale and a factor 10 of dynamical range. }
\end{figure}

\clearpage

\begin{figure}
\centering
\scalebox{0.7}[0.7]{\includegraphics[width=\hsize]{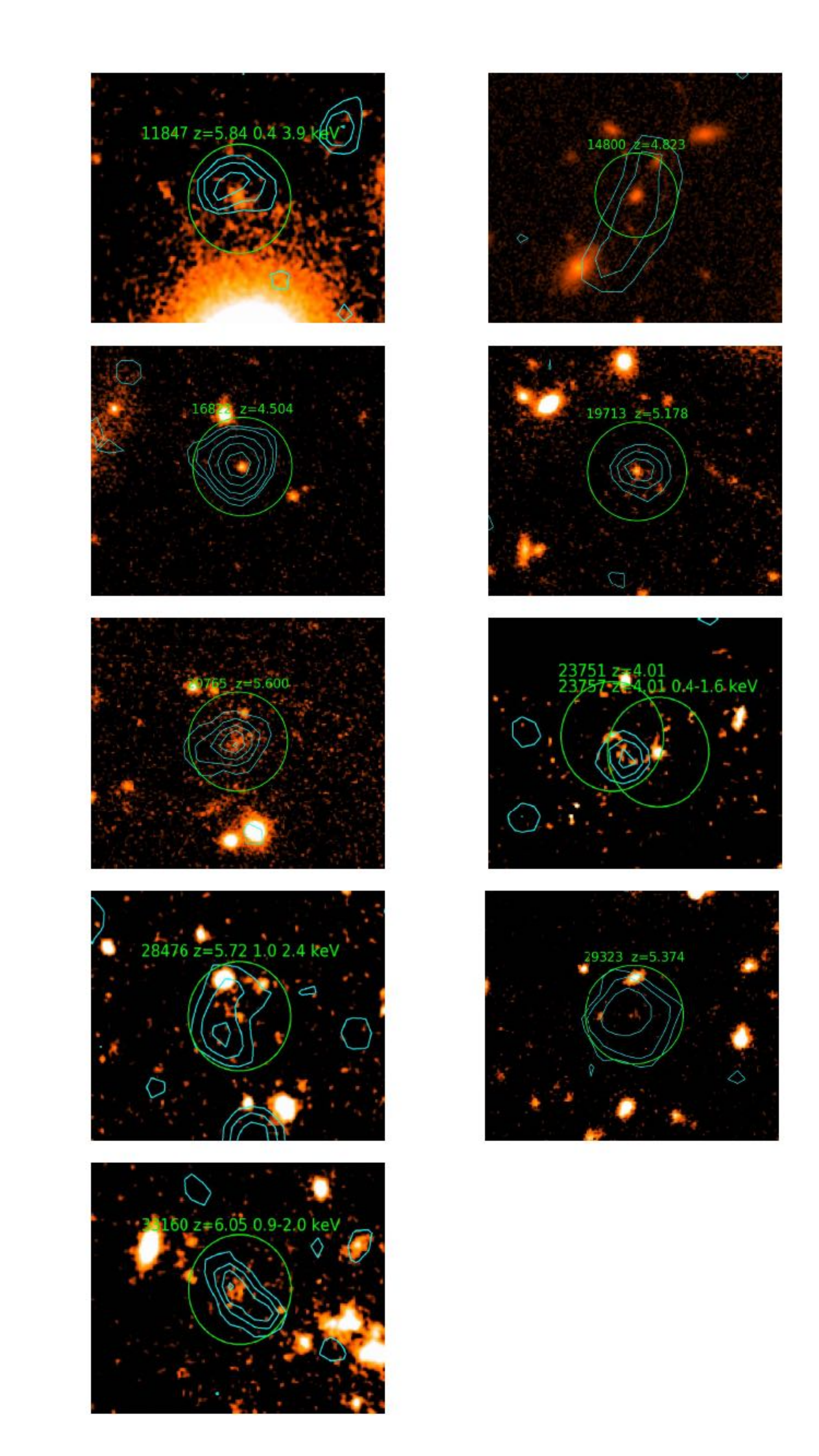}}
\centering
Figure 12 Continued
\end{figure}

\clearpage

\begin{figure}
\centering
\scalebox{0.7}[0.7]{\includegraphics[width=\hsize]{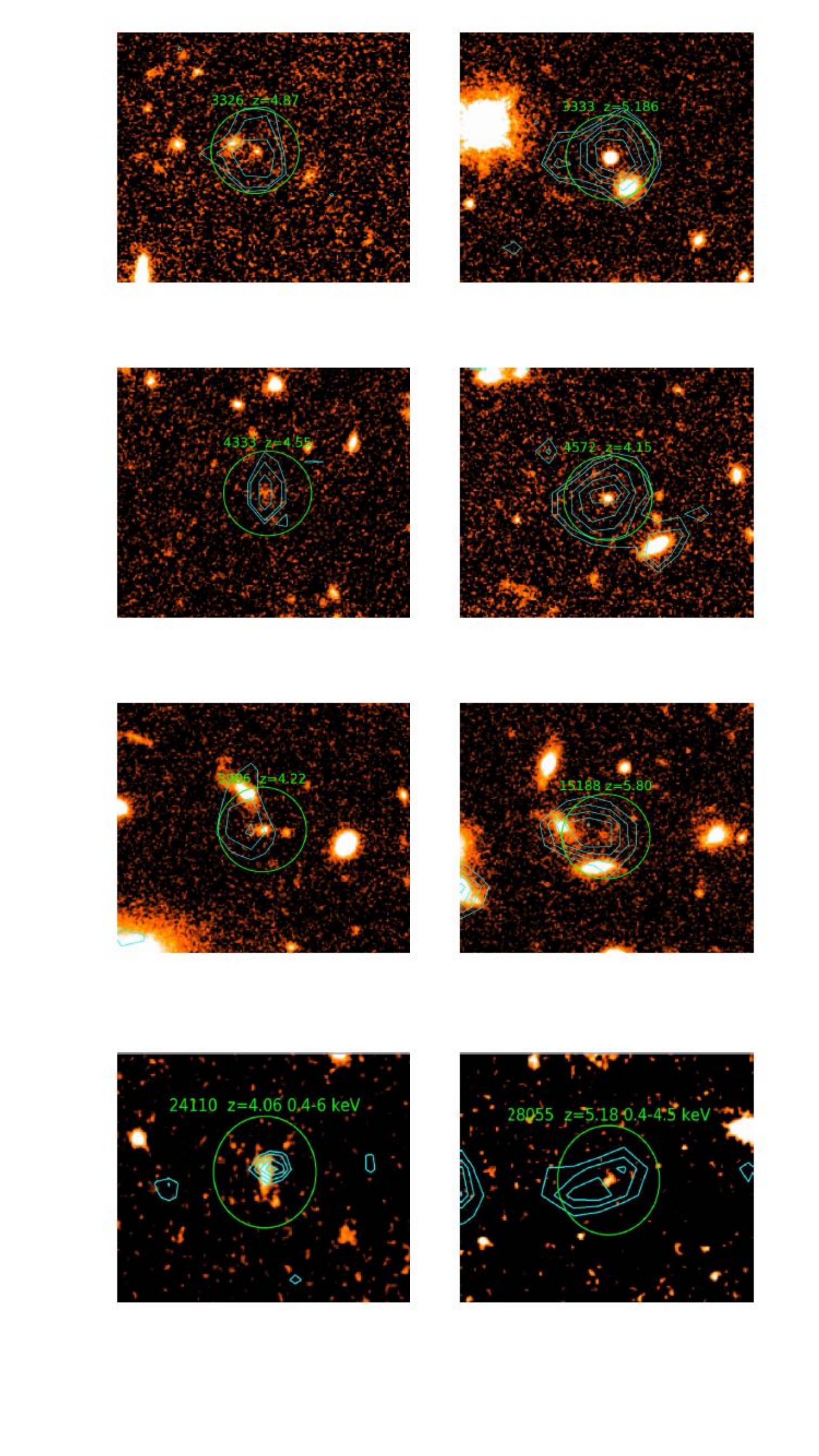}}
\centering
\caption{Figure similar to the previous one for the X-ray GOODS-North sources. The X-ray energy bands adopted for the detection are  also shown  for the new sources. For the other known sources the energy band is 0.5-2 keV. Intensity contours are on a linear scale with a factor 2 of dynamical range for the new sources.  For GDN5986, GDN15188  the intensity contours are with a factor 3 of dynamical range. For  GDN3326, GDN3333, GDN4333, GDN4572 we adopted intensity contours with a square root scale and a factor 10 of dynamical range. }
\end{figure}

\clearpage

\begin{figure}
\centering
\scalebox{0.7}[0.7]{\includegraphics[width=\hsize]{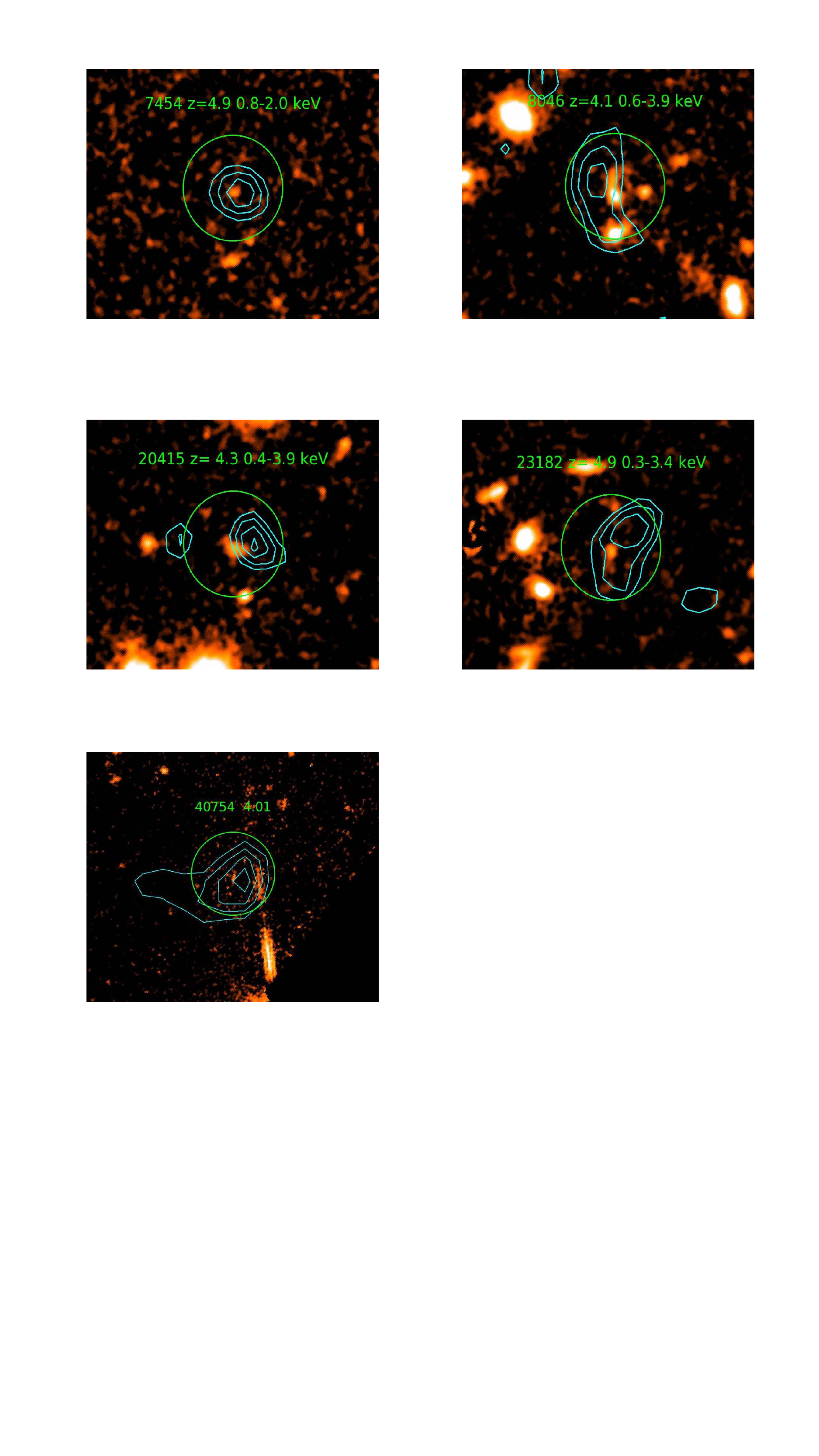}}
\centering
\caption{Figure similar to the previous one for the X-ray EGS sources. The X-ray energy bands adopted for the detection are  also shown  for the new sources. For the other known sources the energy band is 0.5-2 keV. Intensity contours are on a linear scale with a factor 2 of dynamical range for the new sources.  For EGS40754  the intensity contours are with a factor 3 of dynamical range. }
\end{figure}

\clearpage

\begin{figure}
\centering
\scalebox{0.5}[0.5]{\includegraphics[width=\hsize]{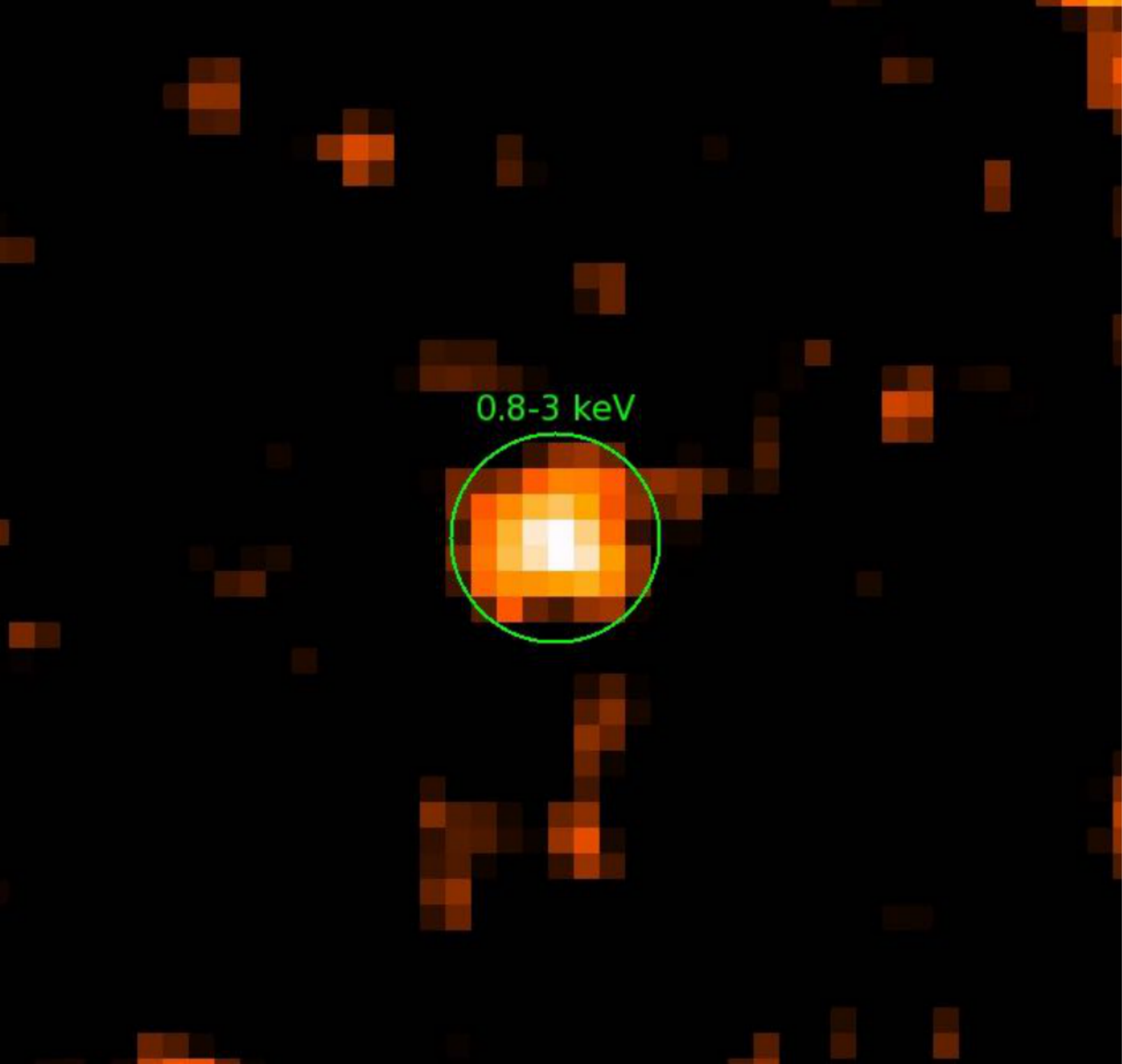}}
\caption{X-ray stack image in the 0.8-3 keV energy band of the new 14 sources found in the present work. The photometric area is shown by a circle with  a radius of $\sim 2$ arcsec. See the Appendix for more details.}
\end{figure}

\end{appendix}
\end{document}